\documentstyle[11pt,epsf]{article}
\makeatletter
\newcommand{\artsectnumbering}{%
\@addtoreset{equation}{section}
\renewcommand{\theequation}{\thesection.\arabic{equation}}}
\makeatother
\newcommand{\al}{\alpha}
\newcommand{\bt}{\beta}

\newcommand{\de}{\delta}

\newcommand{\fr}{\frac}
\newcommand{\ga}{\gamma}
\newcommand{\hb}{\hbar}

\newcommand{\La}{\Lambda}
\newcommand{\la}{\lambda}
\newcommand{\lb}{\label}

\newcommand{\om}{\omega}
\newcommand{\Om}{\Omega}
\newcommand{\rh}{\rho}

\newcommand{\th}{\theta}

\newcommand{\be}{\begin{equation}}
\newcommand{\ee}{\end{equation}} 
\newcommand{\eei}{\end{equation}\indent\indent}
\newcommand{\bc}{\begin{center}}
\newcommand{\ec}{\end{center}}
\newcommand{\ber}{\begin{eqnarray}}
\newcommand{\ear}{\end{eqnarray}}
\newcommand{\ba}{\begin{array}}
\newcommand{\ea}{\end{array}}

\newcommand{\p}{\partial}

\def\case#1/#2{\textstyle\frac{#1}{#2} }
\begin{document}
\title{Vacuum Energy}
\author{Mark D. Roberts, \\\\
117 Queen's Road, Wimbledon, London SW19 8NS,  Email:mdr@ic.ac.uk\\
http://cosmology.mth.uct.ac.za/$\sim$roberts\\}
\date{\today}
\maketitle
\bc Eprint: hep-th/0012062 \ec
\bc Comments:\ec 
A comprehensive review of Vacuum Energy,  
which is an extended version of a poster presented at L\"{u}deritz (2000).
This is not a review of the cosmological constant {\it per se},
but rather vacuum energy in general,  my approach to the cosmological constant
is not standard.
Lots of very small changes and several additions 
for the second and third versions:  constructive feedback still welcome,
but the next version will be sometime in coming 
due to my sporadiac internet access.
\bc First Version 153 pages,  368 references.\ec
\bc Second Version 161 pages,  399 references.\ec
\bc Third Version 167 pages,  412 references.\ec
The 1999 PACS Physics and Astronomy Classification Scheme:\\ 
http://publish.aps.org/eprint/gateway/pacslist\\ 
11.10.+x,  04.62.+v,  98.80.-k,  03.70.+k;\\
The 2000 Mathematical Classification Scheme:  
http://www.ams.org/msc\\
81T20,  83E99,  81Q99,  83F05.\\
\bc 3 KEYPHRASES: \ec
\bc Vacuum Energy,  Inertial Mass,  Principle of Equivalence.\ec
\newpage
\begin{abstract}
There appears to be {\em three},  perhaps related,  ways of approaching the
nature of vacuum energy.   
The {\em first} is to say that it is just the 
lowest energy state of a given,  usually quantum, system.   
The {\em second} is to equate vacuum energy with the Casimir energy.   
The {\em third} is to note that an energy difference 
from a complete vacuum might have some long range effect,  
typically this energy difference is interpreted as 
the cosmological constant.   
All three approaches are reviewed, with an emphasis on recent work. 
It is hoped that this review is comprehensive in scope.
There is a discussion on whether there is a relation between vacuum energy
and inertia.   
The solution suggested here to the nature of the vacuum is that Casimir 
energy can produce short range effects because of boundary conditions,  
but that at long range there is no overall effect of vacuum energy,
unless one considers lagrangians of higher order than Einstein's as vacuum 
induced.   No original calculations are presented in support of this position.
\end{abstract}
\artsectnumbering
{\small\tableofcontents}
\section{Introduction.}
\subsection{Forward.}
In physics there are a great variety of views of what vacuum energy is.
It is understood differently in many formalisms and
domains of study,  and ideas from one area are sometimes used in another;
for example a microscopic origin for the cosmological constant is often used to
justify its inclusion in cosmology.  
A {\em narrow} view is that vacuum energy is synonymous with the 
cosmological constant;  why this is the case is explained in \S2.1\P2.
A {\em wide} view is that vacuum energy is just the lowest energy of any
system under consideration.
Finkelstein \cite{fink} (1991) takes an {\em extreme} 
view of what the vacuum is:
\begin{quote}
The structure of the vacuum is the central problem of physics today:  
the fusion of the theories of gravity and the quantum is a subproblem.
\end{quote}
In general relativity gravitational energy is hard to calculate,
and often ambiguous,  especially in non-asymptotically flat spacetimes.
Now how vacuum energy fits into gravitation is not straightforward
as the gravitational field might have its own vacuum energy,
this is discussed in \S6.
Einstein \cite{einstein} (1924) discusses the effect of ``ether'',
perhaps the ether can be thought of as vacuum energy.
Vacuum energy,  like any other energy,  can be thought of as having an 
equivalent mass via $E=mc^2$.
Moving masses have inertia,  and this is one way of looking at how inertia is
related to vacuum energy,  this is discussed in \S7.
The principle of equivalence,  \S5,  can be formulated in a way which says
that the laws of physics are the same in all inertial frames.
An inertial frame is thus a primitive concept,  and one way of thinking of
it is of it occurring because of some property of the vacuum.
\subsection{Historical Background.}
The presentation here of the history of zero-point energy largely
follows Sciama \cite{sciama} (1991),  the is also a historical
discussion in Lima and Maia \cite{limamaia} (1995).
\subsubsection{The Zero-point Energy of a Harmonic Oscillator.}
Zero-point energy was introduced into physics by Max Planck in
1911.   
In a renewed attempt to understand the interaction between matter 
and radiation and its relationship to the black body spectrum Planck
put forward the hypothesis that the absorption of radiation occurs continuously
while its emission is discrete and in energy units of $h\nu$,  
On this 
hypothesis the average energy $\bar{\varepsilon}$ of a harmonic oscillator
at temperature $T$ would be given by
\be
\bar{\varepsilon}=\fr{1}{2}h\nu+B(\nu,T)
\lb{eq:1}
\ee
where
\be
B(\nu,T)\equiv\fr{h\nu}{e^{h\nu/kT}-1}.
\lb{defB}
\ee
Thus the oscillator would have a finite zero-point energy $\fr{1}{2}h\nu$ 
even at the absolute zero of temperature.
When it comes to quantum field theory,
there can be thought of as being a quantum 
oscillation at each point in spacetime,
this leads to the zero-point energy giving an overall infinite energy
contribution,  so that there is a problem in how to adjust this 
contribution in order to allow quantum field theory to be well defined.

Planck abandoned his discrete emission of energy,  in units of $h\nu$,  
hypothesis in 1914 when Fokker showed that
an assembly of rotating dipoles interacting classically with
electromagnetic radiation would posses statistical properties 
(such as specific heat) in conflict with observation.   
Planck then became convinced that no classical discussion 
could lead in a satisfactory manner to a derivation of his 
distribution for black body radiation,
in other words the distribution was fundamental and $h$
could not be derived from previously know properties.
Nevertheless according to Sciama \cite{sciama} (1991) the idea of 
zero-point motion for a quantum harmonic oscillator 
continued to intrigue physicists,  
and the possibility was much discussed before its existence was definitely
shown in 1925 to be required by quantum mechanics,  as a direct consequence
of Heisenberg's uncertainty principle.

The Einstein-Hopf (1910) derivation of the black body radiation spectrum,
requires an expression for zero-point motion.   
That derivation involved 
a study of the interchange of energy and momentum between a harmonic oscillator
and the radiation field.   It led to the Rayleigh-Jeans distribution rather 
than to Planck's because of the classical assumptions which were originally
used for the harmonic oscillators and the radiation field.   In the 
Einstein-Hopf calculation the mean square momentum of an oscillator 
was found to be proportional to the mean energy of the oscillator and to
the mean energy density of the radiation field.   Einstein and Stern 
now included the zero-point energy of the oscillator in its mean energy,
in the hope of deriving the Planck rather than the Rayleigh-Jeans distribution,
but found that they could do so only if they took $h\nu$ rather than 
$\fr{1}{2}h\nu$ for the zero-point energy.

According to Sciama \cite{sciama} (1991) the reason for this difficulty 
is their neglect of the zero-point energy of the
radiation field itself.   It is not a consistent procedure to link together 
two physical systems only one of which possesses  zero-point fluctuations in
a steady state.   These fluctuations would simply drive zero-point fluctuations
in the other system or be damped out,  depending on the relative number of 
degrees of freedom in the two systems.   
Nerst (1916),  first proposed that 'empty' space was everywhere
filled with zero-point electromagnetic radiation.
According to Sciama \cite{sciama} (1991)
the main considerations of Einstein and Stern are wrong;
when one takes the classical limit $kT>>h\nu$ for the Planck distribution,
one finds
\be
B(\nu,T)\rightarrow kT-\fr{1}{2}h\nu
+h\nu\times O\left(\fr{h\nu}{kT}\right),
\ee
whereas
\be
\fr{1}{2}h\nu+B(\nu,T)\rightarrow kT
+h\nu\times O\left(\fr{h\nu}{kT}\right).
\ee
Thus the correct classical limit is obtained 
only if the zero-point energy is included.

Einstein and Stern gave two independent arguments in favour of retaining the 
zero-point energy,  the {\it first} involved rotational specific heats 
of molecular gases and the {\it second} from an attempted,  
derivation of the Planck distribution itself.
In the first argument it was assumed that a freely rotating molecule would
possess a zero-point energy,
this assumption is now known not to be compatible with quantum mechanics.

Nogueira and Maia \cite{NM2} (1995) discuss the possibility 
that there might be no zero-point energy.
\subsubsection{The Experimental Verification of Zero-point Energy.}
For a single harmonic oscillator,  the existence of a zero-point energy
would not change the spacing between the various oscillator levels and
so would not show up in the energy spectrum.   It might,  of course,  alter
the gravitational field produced by the oscillator.   
For the experimental physicists
influenced by Planck and also by Einstein and Stern,  
demonstrating the existence
of zero-point energy amounted to finding a system in which the 
{\it difference} in this energy for different parts of the system could
be measured.   According to Sciama \cite{sciama} (1991)
it was soon realized that a convenient way to do this was to
look for isotope effects in the vibrational spectra of molecules.
The small change in mass associated with an isotopic replacement would
lead to a small change in the zero-point energy,  and in the energies of all
vibrational levels.   These changes might then show up in the vibrational
spectrum of the system.

Many attempts were made to find this effect,  but the first conclusive
experiment was made by Mulliken in 1925,   in his studies of the band spectrum
of boron monoxide.   This demonstration,  made only a few months before
Heisenberg (1925) first derived the zero-point energy for a harmonic 
oscillator from his new matrix mechanics,  provided the first experimental 
verification of the new quantum theory.
New quantum theory here meaning Heisenberg-Schr\"odinger theory as opposed
to the old Bohr-Sommerfeld quantum theory.

Since then,  zero-point effects have become commonplace in quantum physics,  
{\bf three} examples are in spectroscopy,  in chemical reactions,  
and in solid-state physics.   
Perhaps the most dramatic example is their role in maintaining 
helium in the liquid state under its own vapour pressure at 
(almost) absolute zero.
The zero-point motion of the atoms keep them sufficiently far apart 
on average so that the attractive forces between them are too weak 
to cause solidification.   This can be expressed in a rough way by defining 
an effective temperature $T_{eff}$ where $kT_{eff}$ is equal to the
zero-point energy per atom.  For helium $T_{eff}$ exceeds a classical 
estimate of the melting point under its own vapour pressure.
Thus,  even close to absolute zero helium is 'hot' enough to be liquid.
\subsection{Finding,  Creating and Exploiting Vacua.}
Perhaps related to the nature of vacuum energy is the nature of vacua in 
general.   Absolute vacua cannot be {\em created} or {\em found}.
Looking at ``found'' {\it first},  note that physical spacetime is never a 
vacuum,  for example interplanetary space is not a vacuum,  but has density
$\rho=10^{-29}{\rm g.cm.}^{-3}$ or $10^{-5}$ protons ${\rm cm.}^{-3}$,
the magnitude of the upper limit of the effective cosmological constant
is about $\rho_\Lambda=10^{-16}{\rm g.cm.}^{-3}$,  
see Roberts \S5.2 \cite{mdrse} (1998).
The Universe itself is not a vacuum but has various densities associated with
itself,  see \S4 below.   
A particle gas,  as used in kinetic theory,  can be thought of as 
billiard balls moving through space,  the space being a vacuum:
however such a theory is not fundamental,  whereas field theories
are,  and in them particles moving through space are replaced by
fields defined at each point,  if the fields are non-zero there
is no vacuum.
To look at ``created'' {\it second},  note that a particle
accelerator needs to work in as close to a vacuum as can be achieved.
A recent discussion of a particular approach to vacuum creation for 
particle accelerators is Collins {\it et al} \cite{CGNRSV} (2000).
They study
ion induced vacuum instability which was first observed in the 
Intersecting Proton Storage Rings (ISR) at CERN and in spite 
of substantial vacuum improvements, 
this instability remains a limitation of the
maximum beam current throughout the operation of the machine. 
Extensive laboratory studies and dedicated machine experiments 
were made during this period to understand the details of this
effect and to identify ways of increasing the limit to higher beam currents. 
Stimulated by the recent design work for the LHC vacuum system, 
the interest in this problem has been revived with a
new critical review of the parameters which determine 
the pressure run-away in a given vacuum system with high intensity beams.

Given the nature of the quantum vacuum and its 
fluctuations, it can be imagined that there is a lot of energy 
out there floating around in the vacuum: 
is there any way to get hold of it? 
The {\it second law of thermodynamics} might suggest not, but that law is not 
cognizant of quantum field theory (nor of gravity). 
There are occasional discussions on whether vacuum energy can be 
{\it exploited},  typically used to power electrical or mechanical devices,
an example is 
Yam \cite{yam} (1997) who concludes that zero-point energy probably 
cannot be tapped,  see also \S 3.7 where the exploitation of Casimir energy
is discussed.
Xue \cite{xue} (2000) presents and studies a possible mechanism 
of extracting energies from the vacuum by external classical fields. 
Taking a constant magnetic field as an example, 
he discusses why and how the vacuum energy can be released 
in the context of quantum field theories. 
In addition, he gives a theoretical computation showing 
how much vacuum energies can be released. 
He discusses the possibilities of experimentally
detecting such a vacuum-energy releasing. 
Scandurra \cite{scandurram} (2001)
extends the fundamental laws of thermodynamics 
and of the concept of entropy to the ground state fluctuations of quantum
fields. He critically analyzes a device to extract energy from the vacuum. 
He finds that no energy can be extracted cyclically from the vacuum. 

There are {\it three things} suggest a positive possibility 
of energy extraction.
{\it Firstly}, the theory of quantum evaporation from ``black holes'' 
can be thought of as a form of energy extraction from the quantum vacuum. 
{\it Secondly}, the inflationary universe can also be thought of as
extracting energy from the vacuum, see \S4.4. 
So there is the possibility that what is happening in
either of those cases could be made to happen in a controlled way in
a laboratory. 
The {\it third} things is an analogy: it is  said in physics text books
that you can not extract the vast amount of thermal energy in the sea,
because there is no lower temperature heat bath available; so it is
unavailable energy, despite there being so much of it there. 
Ellis \cite{gfre} (1979) pointed out that in
fact it can be extracted when you remember that the dark night sky 
acts as a heat sink at 3K (the temperature of the Cosmic background
radiation). 
In some ways this seems reminiscent of 
the energy bath that is the quantum vacuum. 
If some particles in that 
vacuum have an effective energy of greater than 3K, maybe one can 
radiate them off to the night sky and leave their partner behind.
\section{The Quantum Field Theory Vacuum.}
\subsection{The Vacuum State of QFT's in general.}
From equation \ref{eq:1} at zero temperature an harmonic oscillator
contributes $\fr{1}{2}h\nu$ to the energy.   
Quantum field theory (QFT) assumes that each point of a field can be treated as
a quantum system,  typically as a quantum harmonic oscillator,
thus each point contributes $\fr{1}{2}h\nu$ to the energy resulting in an
overall infinite energy.   Various ways have been contrived to deal with this
infinite overall energy, {\it two} examples are: {\it firstly} renormalization 
schemes which invoke the idea that it is only energy differences that are 
measurable and that attempts to subtract infinite energy from quantum 
vacuum energy,  see \S2.16 below,  and {\it secondly} supersymmetry where
the infinite vacuum energy of one field is cancelled out term by term by
its supersymmetric partner's opposite sign vacuum energy. see \S2.6 below.
Dmitriyev \cite{dmitriyev} (1992) describes elastic analogs of the vacuum.
Fields are used to describe matter and the relation of the vacuum 
to elementary matter is discussed in
St\"ocker {\it et al} \cite {SGH} (1997).

Fauser \cite{fauser} (1997)
proposes a geometric method to parameterize inequivalent vacua 
by dynamical data. Introducing quantum Clifford algebras with
arbitrary bilinear forms we distinguish isomorphic algebras 
--as Clifford algebras-- by different filtrations resp. induced gradings. 
The idea of a vacuum is introduced as the unique algebraic projection 
on the base field embedded in the Clifford algebra, which is however
equivalent to the term vacuum in axiomatic quantum field theory 
and the GNS construction in $C^*$-algebras. This approach is shown to
be equivalent to the usual picture which fixes one product 
but employs a variety of GNS states. The most striking novelty of the 
geometric approach is the fact that dynamical data fix uniquely 
the vacuum and that positivity is not required. 
The usual concept of a statistical
quantum state can be generalized to geometric meaningful but non-statistical, 
non-definite, situations. Furthermore, an algebraization of
states takes place. An application to physics is provided 
by an U(2)-symmetry producing a gap-equation which governs a phase transition.
The parameterization of all vacua is explicitly calculated 
from propagator matrix elements. 
A discussion of the relation to BCS theory and
Bogoliubov-Valatin transformations is given. 

Chen \cite{chen} (2001) provides a new explaination of the meaning of 
negative energies in the relativistic theory. 
On the basis of that he presents two new conjectures. 
According to the conjectures, particles have two sorts 
of existing forms which are symmetric. 
From this he presents a new Lagrangian density and a new
quantization method for QED. 
That the energy of the vacuum state is equal to zero is naturally obtained. 
From this he can easily
determine the cosmological constant according to experiments, 
and it is possible to correct nonperturbational methods which depend on
the energy of the ground state in quantum field theory. 

Grandpeix \& Lurcat \cite{GL} (2001)
first sketch the "frame problem": the motion of an isolated particle obeys 
a simple law in galilean frames, but how does the galilean
character of the frame manifest itself at the place of the particle? 
A description of vacuum as a system of virtual particles will help to
answer this question. For future application to such a description, 
they define and study the notion of global particle. To this end, a
systematic use of the Fourier transformation on the Poincare group 
is needed. The state of a system of n free particles is represented by a
statistical operator $W$, which defines an operator-valued measure on the 
$n-th$power of the dual of the Poincare group. The inverse
Fourier-Stieltjes transform of that measure is called the characteristic 
function of the system; it is a function on the $n-th$ power of the 
Poincare group. The main notion is that of global characteristic function: 
it is the restriction of the characteristic function to the diagonal
subgroup ; it represents the state of the system, 
considered as a single particle. The main properties of 
characteristic functions, and particularly of global characteristic functions,
are studied. A mathematical Appendix defines two functional spaces involved. 
They describe vacuum as a system of virtual particles, 
some of which have negative energies.  
Any system of vacuum particles is a part of a
keneme, i.e. of a system of $n$ particles which can, 
without violating the conservation laws, 
annihilate in the strict sense of the word (transform into nothing). 
A keneme is a homogeneous system, 
i.e. its state is invariant by all transformations of the invariance group. 
But a homogeneous system is not necessarily a keneme. 
In the simple case of a spin system, where the invariance group is $SU(2)$, a
homogeneous system is a system whose total spin is unpolarized; 
a keneme is a system whose total spin is zero. 
The state of a homogeneous system is described 
by a statistical operator with infinite trace (von Neumann), 
to which corresponds a characteristic distribution. 
The characteristic distributions of the homogeneous systems 
of vacuum are defined and studied. 
Finally they show how this description of
vacuum can be used to solve the frame problem posed in the first paper. 

The standard argument that a non-zero vacuum leads to a cosmological
constant is as follows.   Consider the action of a single scalar field 
with potential $V(\phi)$ is
\be
S=\int d^4x\sqrt{-g}\left[\fr{1}{2}g^{ab}(\p_a\phi)(\p_b\phi)-V(\phi)\right]
\lb{eq:simpleaction}
\ee
where the lagrangian is
\be
{\cal L}=-\fr{1}{2}\phi_a\phi^a+V(\phi),
\lb{eq:simplelag}
\ee
and a typical potential,  the $\phi^4$ potential is 
\be
V(\phi)=m\phi^2+\alpha\phi^3+\la\phi^4.
\lb{eq:simplepot}
\ee
Metric variation gives the stress
\be
T_{ab}=\p_a\phi\p_b\phi
+\fr{1}{2}g_{ab}\left[g^{cd}\p_c\phi\p_d\phi-V(\phi)\right],
\lb{eq:stresss}
\ee
if the derivatives of the scalar field vanish for a ``vacuum state'' 
$\phi_{vac}=<0|\phi|0>$ then
\be
T^{vac}_{ab}=-\fr{1}{2}V(\phi_{vac})g_{ab}=-\rho_{vac}g_{ab},
\lb{eq:Tvac}
\ee
and then assuming that the stress is of the perfect fluid form
\be
T_{ab}=(\rho+p)U_aU_b+p~g_{ab},
\lb{eq:Tfluid}
\ee
gives the pressure
\be
-p_{vac}=\rho_{vac}=\fr{\Lambda}{8\pi G}.
\lb{eq:pressure}
\ee
This can be thought of as the vacuum expectation value of a field giving a
cosmological constant contribution to the field equations,  compare \S4.1;
this sort of contribution might produce inflation,  compare \S4.2.
There are at least {\it three} problems with this approach to the vacuum.
The {\it first} is that the assumption $\p_a\phi=0$ is odd,  especially if one 
wants to argue that inflation is due to vacuum expectation of a scalar field;
in an inherently non-static situation the assumption that $\p_t\phi=0$ is 
unlikely to hold.
The {\it second} is that $V(\phi)$ is position dependent because 
$\phi$ is position dependent so $\p_a\phi\ne0$;  
so that $\phi= a~ constant$ does not hold
as required for it to be a cosmological constant.
Of course it can be still
represented by a perfect fluid with $p=\rho$.
The {\it third} is that I do not think that scalar fields,  at least of the
above form,  are fundamental.   
Perfect fluids form a gauge system of scalar fields and
I believe that all fundamental fields are gauge
fields;  how symmetry breaking can then be achieved using perfect fluids 
is discussed in Roberts \cite{mdrvel} (1997) and \S2.10 below.

An alternative approach to vacuum energy is given in Roberts 
\cite{mdr84} (1984);  there a quantum-mechanical plausibility argument 
is introduced which suggests that matter always has a minuscule scalar 
field associated with itself,
perhaps this can be thought of as there always being a non-zero vacuum 
expectation value in the form of a scalar field;
and this is relate to quintessence, see \S4.3.
When considering quantum field theory on a curved 
spacetime it is usually assumed 
that the field equations can be used in the form
$G_{ab}=8\pi<T_{ab}>$.
Hawking \cite{hawking} (1975) has shown that this gives rise to an 
uncertainty in the local energy density of the order $B^2$,  
where $B={\rm l.u.b.}|R_{abcd}|$.
This local uncertainty would necessitate a global modification of the field
equations to the form $G_{ab}=8\pi<T_{ab}>+U$,  where $U$ is an extremely small
term which perturbs the equations from there classical value.   
Here justification
is given for relating this with the splitting of $T_{ab}$ 
into a matter term and a massless scalar term.

In Roberts \cite{mdr84} (1984) the original form of the local uncertainty 
of local energy density due to Hawking is not used,  
but instead a later variant derived by 
H\'aj'i\v{c}ek \cite{bi:hajicek77} (1977).
Briefly,  in curved space-time there is a difficulty in defining positive 
frequency,  but expanding $g_{ab}$ to second order 
$g_{ab}=\eta_{ab}-\frac{1}{3}R_{asbt}x^sx^t$,  now defining 
$R=\max_{a,b,s,t=0,1,2,3}R_{asbt}$,
$d={\rm l.u.b.}x^a(q)$,
one sees that the corrections to the flat metric $\eta_{ab}$ are of order 
$(d/R)^2$ and are small if $d<<R$,  so positive frequency can be used,
after considering what states of a field can be localized in a region with
radius $d$;  Hajicak claims the the following uncertainty relations hold:
\be
\frac{\Delta E}{E}\sim\frac{\hb c}{Ed}+\frac{d^2}{R^2},
\ee
so that
\be
\Delta E\sim\frac{\hb c}{d}+\frac{d^2E}{R^2},
\ee
and we also have
\be
\Delta G_{00}+G_{00}=8\pi(T_{00}+\Delta T_{00}).
\ee
Provided $d<<10^{-33}$cm,  where quantum gravity would be expected to dominate,
see for example Hawking \cite{hawking} (1975),  
one can assume that the geometry is given by the right-hand side,  
{\it i.e.} we can put $\Delta G_{00}=0$.   
Therefore
\be
G_{00}=8\pi(T_{00}+\Delta T_{00})
      =8\pi\left(T_{00}+\frac{\hb c}{d}+\frac{d^2T_{00}}{R^2}\right).
\ee
For regions with significant curvature $\hb c/(Ed)<d^2/R^2$ so that
\be
G_{00}=8\pi T_{00}(1+d^2/R^2).
\ee
Substituting in values for $E$ the energy from what one might typically 
expect for elementary particles suggests that the inequality would happen
with a curvature radius of about $10^{-8}$cm,  but this could clearly vary
by a few orders of magnitude;  thus my arguments work for radii of curvature
from about $10^{-33}$ to $10^{-8}$cm.   
Extrapolating,  one would expect there to be uncertainty momentum 
relationships as well - assuming they give rise to similar expressions,  then
\be
G_{ab}=8\pi T_{ab}(1+(d/R)^2).
\ee

The position of the extra term in the field equations hides the fact that it is
both massless and scalar;  massless because,  if it had mass,  there would be
an extra mass associated with any region one chose,  so that if it had mass it
would be infinite;  scalar because it does not depend explicitly on any term 
involving indices only on $\max|R_{abcd}|$ and on an arbitrary volume.
Another way of considering the masslessness of the new term is by realizing 
that it might appear to give rise to an infinite series:
\be
G_{ab}=8\pi T^1_{ab}(1+(d/R)^2)
      =8\pi T^2_{ab}(1+(d/R)^2)^2
      =8\pi T^3_{ab}\ldots
\ee
this cannot occur as we are restricting ourselves to a given volume of radius
$d$,  if such a series did occur it would diverge leaving no field equations.
The presence of the extra term has some unusual consequences such as space 
being more curved than just matter terms would lead one to believe.   This is
because of the uncertainty of local energy density arising from the curvature 
due to the matter fields.   Without the new term,  the field equations are 
observer dependent,  because the observer is part of the matter fields,  but
with the new term there is an additional dependence as the observer sets up
the volume of radius $d$ in which he makes his measurements.   It is 
interesting to note that space is curved more than one would expect 
classically,  because of the quantum effects due to the presence of an observer
and a volume in which he makes measurements.

The importance of this extra term can be seen by noting 
that it can be put in the form
\be
T_{ab}(1+(d/R)^2)={}_{(m)}T_{ab}+{}_{(a)}T_{ab}=T^{total}_{ab},
\ee
where ${}_{(a)}T_{ab}$ is a massless scalar field which has stress 
given by equation \ref{eq:stresss}.
\subsection{The QED Vacuum.}
Quantum electrodynamics (QED) is the simplest quantum field theory which is
physically realized.   There are simpler QFT's which consist of just one or two
scalar fields.   QED is the quantum theory of the Maxwell field $A_a$ and the
Dirac spinor field $\psi$.   There are four basic interactions in physics:
electromagetic,  weak,  strong,  and gravity;  and QED is the quantum field 
theory which describes the first of these.   The field theories 
corresponding to the other interactions get progressively harder 
to quantize and the methods by which their quantization is attempted 
usually follow successfull models in QED.

An example of a recent calculation of the vacuum properties of QED is that of
Kong and Ravndal \cite{KR} (1998) who construct an effective field theory
theory to calculate the properties of the QED vacuum at temperatures much
less than the electron mass $T<<m_e$.   They do this by adding quantum 
fluctuation terms to the Maxwell lagrangian.
Quantum fluctuations in the vacuum due to virtual electron 
loops can be included by extending the Maxwell lagrangian by 
additional non-renormalizable terms corresponding to the 
Uehling and Euler-Heisenberg interactions. 
By a redefinition of the electromagnetic field they show that
the Uehling term does not contribute. 
The Stefan-Boltzmann energy density is thus found by them to be modified 
by a term proportional with $T^8/m^4$ in agreement with a semi-classical
result of Barton \cite{barton90} (1990),  
compare with the Toll-Scharnhorst effect \S2.3 below. 
The speed of light in blackbody radiation is smaller than one. 
Similarly, the correction to the energy density of the vacuum between two
metallic parallel plates diverges like $1/m^4z^8$ at a distance from one 
of the plates $z \to 0$. 
While the integral of the regularized energy density is thus
divergent, the regularized integral is finite and corresponds 
to a correction to the Casimir force which varies with the 
separation $L$ between the plates as $1/m^4L^8$. 
Kong and Ravndal \cite{KR} (1998) point out that
this result is in seemingly disagreement with a 
previous result for the radiative correction to 
the Casimir force which gives a correction varying
like $1/mL^5$ in a calculation using full QED.
Marcic \cite{marcic} (1990) discusses how vacuum energy alters photon 
electron scattering.

Xue and Sheng \cite{XS} (2001)
study the fermionic vacuum energy of vacua with 
and without being applied an external magnetic field. 
The energetic difference of
two vacua leads to the vacuum decaying and the vacuum-energy releasing. 
In the context of quantum field theories, they discuss why and
how the vacuum energy can be released by spontaneous photon emissions 
and/or paramagnetically screening the external magnetic field.
In addition, they quantitatively compute the vacuum energy released, 
the paramagnetic screening effect and the rate and spectrum of
spontaneous photon emissions. They discuss the possibilities of experimentally 
detecting such a vacuum-energy releasing. 

Nikolic \cite{hnikolic} (2001) studies the possibility of electron-positron 
pair creation by an electric field by using only those methods in field theory
the predictions of which are confirmed experimentally. 
These methods include the perturbative method of quantum electrodynamics 
and the bases of classical electrodynamics. 
Such an approach includes the back reaction. 
He finds that the vacuum is always stable, in the sense that pair
creation, if occurs, cannot be interpreted as a decay of the vacuum, 
but rather as a decay of the source of the electric field or as a process
similar to bremsstrahlung. 
He also finds that there is no pair creation in a static electric field, 
because it is inconsistent with energy conservation. 
He discusses the non-perturbative aspects arising from the Borel summation 
of a divergent perturbative expansion. 
He argues that the conventional methods that predict pair creation 
in a classical background electric field cannot serve even as approximations. 
He qualtatively discusses the analogy with the possibility of particle 
creation by a gravitational field. 
\subsection{The Toll-Scharnhorst Effect.}
There is the problem of when any propagation can be considered to be light-like
or null,   I have previously discussed this in \cite{mdrnote} (1987),
\cite{bi:mdr2} (1994) and \cite{mdrnm} (1998).
If one considers the propagation of light the speed $c$ is its speed of 
propagation in a vacuum,  as discussed in \S1.3 above;  there is never an 
absolute vacuum so that light never propagates at $c$.
This might have significance for the spin states of light,
for null propagation one of the spin states becomes a helicity state.
There is also the problem of what happens to gravitational 
radiation co-moving with electromagnetic radiation.   
Even a photon has a gravitational field associated with itself,
and at first sight one might expect it to co-move.   When light moves from one
medium to another,  say air to water,  its speed and direction change.
So what happens to any co-moving gravitational radiation?   It is unlikely to
change by the same speed and direction as the photon as it interacts 
differently with matter,  there might be some scattering processes that 
allows it to co-move,  but this would be a coincidence.

A more concrete example of non-null light propagation is given by 
what is now called the Scharnhorst effect although similar studies go back to
Toll \cite{toll} (1952).
Beginning in the early 1950's, 
quantum field theoretic investigations have
led to considerable insight into the relation 
between the vacuum structure and the propagation of light. 
Scharnhorst \cite{scharnhorst} (1998) notes that recent years 
have witnessed a significant growth of activity in this field of research. 
Loosely speaking the Scharnhorst effect is that boundaries change the nature
of the quantum vacuum and this in turn changes the speed of propagation of 
light.   Scharnhorst \cite{scharn90} (1990) 
considers QED in the presence of two parallel plates in a Casimir effect, 
see \S3,  type of configuration.   
The plates impose boundary conditions on photon vacuum fluctuations.
In a physically reasonable approximation Scharnhorst calculates the two-loop
corrections to the QED effective action.   From this effective action
Scharnhorst finds a change in the velocity of light propagating between
and perpendicular to the plates.
Barton \cite{barton90} (1990) further discusses the effect which he says is 
due to the intensity of zero-point fields between parallel mirrors being 
less than in unbounded space.

Scharnhorst \cite{scharnhorst} (1998) notes that
QED vacua under the influence of external conditions 
(background fields, finite temperature, boundary conditions) 
can be considered as dispersive media whose complex
behaviour can no longer be described in terms 
of a single universal vacuum velocity of light $c$. 
After a short overview Scharnhorst \cite{scharnhorst} (1998) 
discusses {\it two} characteristic situations: 
{\it firstly} the propagation of light in a constant homogeneous 
magnetic field and {\it secondly} in a Casimir vacuum. 
The latter appears to be particularly interesting 
because the Casimir vacuum has been found to exhibit modes 
of the propagation of light with phase and group velocities larger than $c$
in the low frequency domain $\omega<<m$ where $m$ is the electron mass. 
The impact of this result on the front velocity of light 
in a Casimir vacuum is discussed by means of the
Kramers-Kronig relation. 

Winterberg \cite{winterberg} (1998) notes that if the observed superluminal
quantum correlations are disturbed by turbulent fluctuations of the zero-point
vacuum energy field,  with the perturbed energy spectrum assumed to obey the 
universal Kolmogoroff form derived above,  which the correlations 
are conjectured to break.   
A directional dependence of this length would establish a preferred 
reference system at rest with zero-point energy.
Assuming that the degree of turbulence is given by the small anisotropy
of the cosmic microwave background radiation,  a length similar to $60$ Km. 
is derived above which the correlations would break.

Cougo-Pinto {\it et al} \cite{CFST} (1999) 
consider the propagation of light in the QED vacuum 
between an unusual pair of parallel plates, namely: 
a perfectly conducting one ($\epsilon\rightarrow\infty$) 
and an infinitely permeable one ($\mu\rightarrow\infty$). 
For weak fields and in the soft photon approximation 
they show that the speed of light for propagation normal 
to the plates is smaller than its value in unbounded space 
in contrast to the original Scharnhorst \cite{scharn90} (1990) effect. 

Liberati {\it et al} \cite{LSV} (2000) consider the Scharnhorst effect 
at oblique incidence, calculating both photon
speed and polarization states as functions of angle. 
The analysis is performed in the framework of nonlinear electrodynamics 
and they show that many features of the situation can be extracted solely 
on the basis of symmetry considerations. Although birefringence is common in
nonlinear electrodynamics it is not universal; in particular they verify that 
the Casimir vacuum is not birefringent at any incidence angle.
On the other hand, group velocity is typically not equal to phase velocity, 
though the distinction vanishes for special directions or if one is
only working to second order in the fine structure constant. 
They obtain an ``effective metric'' that is subtly 
different from previous results.
The disagreement is due to the way that ``polarization sums'' 
are implemented in the extant literature, and they demonstrate that a fully
consistent polarization sum must be implemented via a bootstrap procedure 
using the effective metric one is attempting to define.
Furthermore, in the case of birefringence, they show that the polarization 
sum technique is intrinsically an approximation. 

These effects are too small by many orders of magnitude to be measured.
\subsection{The Yang-Mills Vacuum.}
Yang-Mills theory is a generalization of the Maxwell theory where instead
of one vector field $A_a$ there are several vector fields $A^i_a$ interacting
with each other subject to a group.   Some studies of its quantum vacuum are
listed below.

Casahorran and Ciria \cite{CCi} (1995) discuss 
the stability of the vacuum in the presence of fermions.

Fumita \cite{fumita} (1995) discusses the
chiral, conformal and ghost number anomalies 
from the viewpoint of the quantum vacuum 
in Hamiltonian formalism. 
After introducing the energy cut-off, he derives
known anomalies in a new way. 
He shows that the physical origin 
of the anomalies is the zero point fluctuation of bosonic or fermionic field. 
He first points out that the chiral $U(1)$
anomaly is understood as the creation of the chirality 
at the bottom of the regularized Dirac sea 
in classical electromagnetic field. 
In the study of the (1+1) dimensional quantum
vacuum of matter field coupled to the gravity, 
he gives a physically intuitive picture of the conformal anomaly. 
The central charges are evaluated from the vacuum energy. 
He clarifies that the non-Hermitian regularization factor 
of the vacuum energy is responsible for the ghost number anomaly.

Using a variational method Cea \cite{cea} (1996) 
evaluates the vacuum energy density in the one-loop approximation
for three dimensional abelian and non-abelian 
gauge theories interacting with Dirac fermions.
It turns out that the states with a constant magnetic 
condensate lie below the perturbative ground state 
only in the case of three dimensional quantum
electrodynamics with massive fermions. 

Ksenzov \cite{ksenzov97} (1997)
shows that some non-perturbative properties of the vacuum 
are described by the quantum fluctuations around the classical 
background with zero canonical momentum. 
The vacuum state that he built was in 
in the framework of the $\sigma$-models in two dimensions. 

Natale and  da Silva \cite{NS} (1997) 
show that if a gauge theory with dynamical symmetry breaking has 
non-trivial fixed points, they will correspond to extrema of the vacuum energy.
This relationship provides a different method to determine fixed points. 

Tiktopoulos \cite{tik} (1997) applies
variational (Rayleigh-Ritz) methods to local quantum field theory. 
For scalar theories the wave functional 
is parametrized in the form of a superposition of Gaussians
and the expectation value of the Hamiltonian is expressed 
in a form that can be minimized numerically. 
A scheme of successive refinements of the superposition is proposed that
might converge to the exact functional. 
As an illustration, he works out a simple numerical approximation 
for the effective potential is worked out based on minimization 
with respect to five variational parameters. 
A variational principle is formulated for the fermion vacuum energy 
as a functional of the scalar fields to which the fermions are coupled. 
The discussion in this paper is given for scalar and fermion interactions 
in 1+1 dimensions. 
The extension to higher dimensions encounters 
a more involved structure of ultraviolet divergences and
he defers it to future work.

Kosyakov \cite{kosyakov} (1998)
discusses in a systematical way exact retarded solutions 
to the classical SU(N) Yang-Mills equations with the source composed
of several colored point particles.
He reviews anew method of finding such solutions.
Relying on features of the solutions, he suggests a toy model of quark binding.
According to this model, quarks forming a hadron are influenced 
by no confining force in spite of the presence
of a linearly rising term of the potential. 
The large-$N$ dynamics of quarks conforms well with Witten's phenomenology. 
On the semiclassical level, hadrons are color neutral in the Gauss law sense. 
Nevertheless, a specific multiplet structure is observable in the form
of the Regge sequences related to infinite-dimensional unitary representations
of $SL(4,R)$ which is shown to be the color gauge group of the
background field generated by any hadron. 
The simultaneous consideration of $SU(N)$, $SO(N)$, and $Sp(N)$ 
as gauge groups offers a plausible
explanation of the fact that clusters containing two or three quarks 
are more stable than multiquark clusters. 

Gogohia and Kluge \cite{GK} (2000) 
using the effective potential approach for composite operators, 
they formulate a general method of calculation 
of the nonperturbative Yang-Mills vacuum energy
density in the covariant gauge QCD ground state quantum models. 
The Yang-Mills vacuum energy density is defined as an integration 
of the truly nonperturbative part 
of the full gluon propagator over the deep infrared
region (soft momentum region). 
A nontrivial minimization procedure 
makes it possible to determine the value 
of the soft cutoff in terms of the corresponding nonperturbative scale
parameter, which is inevitably present in any nonperturbative model 
for the full gluon propagator. 
They show for specific models 
of the full gluon propagator 
that explicitly use of the infrared enhanced and finite gluon propagators 
lead to the vacuum energy density which is finite, 
always negative and it has no imaginary part (stable vacuum), while
the infrared vanishing propagators lead to unstable vacuum 
and therefore they are physically unacceptable. 

Guendelman and Portnoy \cite{GPort} (2000) review and analyze
the stability of the vacuum of several models. 
1) In the standard Glashow-Weinberg-Salam (GWS) model they review the
instability towards the formation of a bubble of lower energy density 
and how the rate of such bubble formation process compares 
with the age of the Universe for the known values of the GWS model. 
2)They also review the recent work by Guendelman 
concerning the vacuum instability question in the context 
of a model that solves the
cosmological constant problem. 
Gundelman and Portnoy \cite{GPort} claim that it turns out that in such model 
the same physics that solves the cosmological constant problem 
makes the vacuum stable. 
3)They review their recent work concerning the instability 
of elementary particle embedded in our vacuum, 
towards the formation of an infinite Universe. 
they say that such process is not catastrophic and 
it leads to a "bifurcation type" instability 
in which our Universe is not eaten by a bubble 
(instead a baby universe is born). 
This universe does not replace our Universe rather it disconnects
from it (via a wormhole) after formation. 

Polychronakos \cite{polychronakos} (2000) points out
that the space noncommutativity parameters $\theta^{\mu \nu}$ 
in noncommutative gauge theory can be considered as a
set of superselection parameters, 
in analogy with the $\theta$-angle in ordinary gauge theories. 
As such, they do not need to enter explicitly
into the action. 
He suggests a simple generic formula to reproduce the Chern-Simons action 
in noncommutative gauge theory,
which reduces to the standard action in the commutative 
limit but in general implies a cascade of lower-dimensional Chern-Simons
terms. The presence of these terms in general alters the vacuum structure 
of the theory and nonstandard gauge theories can emerge
around the new vacua. 
\subsection{The QCD Vacuum.}
Quantum chromdynamics (QCD) is a particular Yang-Mills theory in which
the group structure is taken to be of a specific form which can describe
the strong interactions of particle physics.   The coupling constants
in the theory are large,  which means that,  unlike QED,  perturbations
in these coupling constants do not converge.   Thus the question arises
as to what the nonperturbative vacuum could be, 
as discussed in Goghia and Kluge
\cite{GK} (2000) \S2.4 above and the papers discussed below.
Grundberg and Hanson \cite{GH} (1994) use arguments taken from 
the electrodynamics of media to deduce the QCD trace anomaly from 
the expression for vacuum energy in the presence of an external colour 
magnetic field.

One way of approaching nonperturbative vacuum energy is via instanton effects,
see \S2.12 below and
Shuryak and Sch\"afer \cite{SSch} (1997) who review recent progress 
in understanding the importance of instanton effects in QCD.
Instantons explain the appearance of a non-perturbative vacuum energy density,
as calculated from correlation functions as a bridge between vacuum 
and hadronic structures.

Goghia {\it et al} \cite{GTSK} (1998)
use the effective potential approach for composite operators to 
formulate the quantum model of the QCD vacuum. 
It is based on the existence and importance of the nonperturbative 
$q^{-4}$-type dynamical, topologically nontrivial excitations of the 
gluon field configuration. The QCD vacuum is found to be
stable since the vacuum energy density has no imaginary part. 
Moreover, they discover a possible stationary ground state of the 
nonperturbative Yang-Mills (quenched QCD) vacuum. 
The vacuum energy density 
at stationary state depends on a scale at which nonperturbative effects 
become important. 
The quark part of the vacuum energy density depends in addition on the 
constant of integration of the corresponding Schwinger-Dyson equation. 
The value of the above mentioned scale is determined from the bounds 
for the pion decay constant in the chiral limit. 
Their value for the chiral QCD vacuum energy density is one order
of magnitude bigger than the instanton based models can provide 
while a fair agreement with recent phenomenological and lattice 
results for the chiral condensate is obtained. 

Kosyakov \cite{kosyakov} (1998) approaches QCD through exact solutions
of the Yang-Mills-Wong equations,  see the previous section \S2.4.

Schmidt and Yang \cite{SY} (1999)
discuss QCD condensate contributions to the gluon propagator both 
in the fixed-point gauge and in covariant gauges for the external 
QCD vacuum gluon fields with the conclusion that a covariant gauge 
is essential to obtain a gauge invariant QCD vacuum energy density 
difference and to retain the unitarity of the quark scattering amplitude. 
The gauge-invariant QCD condensate contributions to the effective 
one-gluon exchange potential are evaluated by using the
effective gluon propagator which produces a gauge-independent 
quark scattering amplitude. 

Luo \cite{luo} (1998) investigates
the vacuum properties of lattice QCD with staggered quarks
by an efficient simulation method. 
He presents data for the quark condensate with
flavour number $N_f=0, ~ 1, ~ 2, ~ 3, ~ 4$ and many quark masses, 
including the vacuum energy in the chiral limit. 
Obvious sea quark effects are observed in some parameter space. 
He also describes a mechanism to understand this and a formula 
relating the chiral condensate and zero modes. 

Velkovsky and Shuryak \cite{VS} (1998)
calculate the contribution of the instanton -- anti-instanton ($I\bar I$) 
pairs to the vacuum energy of QCD-like theories with $N_f$ light fermions 
using the saddle point method. 
They find a qualitative change of the behavior: 
for $N_f \ge 6$ it starts to oscillate with $N_f$. 
Similar behaviour was known for
quantum mechanical systems interacting with fermions. 
They discuss the possible consequences of this phenomenon, 
and its relation to the mechanism of chiral
symmetry breaking in these theories. 
They also discuss the asymptotics of the 
perturbative series associated with the $I\bar I$ contribution, 
comparing their results with those in literature. 

Ksenzov \cite{ksenzov00} (2000) describes
the non-perturbative part of the vacuum energy density 
for static configuration in pure SU(2) Yang-Mills theory, 
and also he constructs a vacuum state. 

Montero {\it et al} \cite{MNS} (1997) claim to find that
nonperturbative infrared finite solutions for 
the gluon polarization tensor. 
The possibility that gluons might have a dynamically 
generated mass is supported by recent Monte Carlo simulation on the lattice. 
These solutions differ among themselves, 
due to different approximations performed 
when solving the Schwinger-Dyson equations
for the gluon polarization tensor. 
Only approximations that minimize energy are meaningful, and, 
according to this, they compute an effective potential 
for composite operators as a function of these solutions 
in order to distinguish which one is selected by the vacuum. 

Paniak {\it et al} \cite{PSZ} (1996) 
address the issue of topological angles in the context 
of two dimensional SU(N) Yang-Mills theory coupled 
to massive fermions in the adjoint representation. 
Classification of the resulting multiplicity of vacua 
is carried out in terms of asymptotic fundamental Wilson loops, 
or equivalently, charges at the boundary of the world. 
They explicitly demonstrate that the multiplicity of 
vacuum states is equal to N for SU(N) gauge group. 
Different worlds of the theory are classified by the 
integer number k=0,1,...N-1 (superselection rules) 
which plays an analogous role to the $\theta$ parameter in QCD. 
Via two completely independent approaches they study the physical 
properties of these unconnected worlds as a function of k. 
{\it Firstly} they apply the well known machinery of the loop calculus 
in order to calculate the effective string tensions in the theory 
as function of $k$. 
The {\it second} way of doing the same physics is the standard particle/field 
theoretic calculation for the binding potential of a pair of infinitely 
massive fermions. 
They also calculate the vacuum energy as function of k. 

An equivariant BRST-construction is used by
Schaden \cite{schaden} (1998) to define 
the continuum SU(3) gauge theory on a finite torus. 
Schaden corroborate previous results using renormalization
group techniques by explicitly computing the measure 
on the moduli-space of the model with 3 quark flavours to two loops. 
Schaden finds that the correction to the maximum of the one-loop effective 
action is indeed of order $g^2$ in the critical covariant gauge. 
The leading logarithmic corrections from higher loops are also
shown to be suppressed by at least one order of $g^2$. 
Schaden therefore is able to relate the expectation value 
of the moduli to the asymptotic scale parameter of
the modified minimal subtraction scheme. 
An immediate consequence is the determination 
of the non-perturbative proportionality constant in the relation
between the vacuum expectation value of the trace of 
the energy momentum tensor and $\Lambda_{QCD}$ 
for the modified minimal subtraction scheme with three quark flavours. 
The result compares favorably with phenomenological estimates 
of the gluon condensate from QCD sum rules for the charmonium system
and $\Lambda_{QCD}$ from $\tau$-decay. 

Gabadadze and Shifman \cite{GShi} (2000) show that
large N gluodynamics to have a set of metastable vacua with 
the gluonic domain walls interpolating between them. The walls
may separate the genuine vacuum from an excited one, 
or two excited vacua which are unstable at finite N 
(here N is the number of colours). 
One might attempt to stabilize them by switching on the axion field. 
They study how the light quarks and the axion affect the
structure of the domain walls. In pure gluodynamics (with the axion field) 
the axion walls acquire a very hard gluonic core. Thus, they deal
with a wall "sandwich" which is stable at finite N. 
In the case of the minimal axion, the wall "sandwich" 
is in fact a $"2-\pi"$ wall, i.e., the
corresponding field configuration interpolates 
between identical hadronic vacua. 
The same properties hold in QCD with three light quarks
and very large number of colours. 
However, in the realistic case of three-colour QCD the phase 
corresponding to the axion field profile in the
axion wall is screened by a dynamical phase associated 
with the $\eta'$, so that the gluon component of the wall is not excited. 
They propose a toy Lagrangian which models these properties and allows 
one to get exact solutions for the domain walls. 
\subsection{The SUSY Vacuum.}
From the point of view of inquiring what the quantum vacuum is,  supersymmetry
is important.   In supersymmetric theories the overall infinite
vacuum energies of bosons and fermions cancel out term by term.
Of course if supersymmetry is realized in nature then it is badly broken,
whether the above type of term by term cancellation occurs in broken 
supersymmetry is an open question,  as the correct way to break supersymmetry 
to establish contact with the physical world has not been found.
A description of the term by term cancellations is given in
Peskin and Schroeder \cite{PesS} page 796 (1995) who say:
\begin{quote}
We have noted already that bosonic fields give positive contributions to the
vacuum energy through their zero-point energy,  
and fermionic fields give negative contributions.
We now see that,  in a supersymmetric model,  
these contributions cancel exactly,
not only at the leading order but to all orders in perturbation theory.
\end{quote}

Following Carroll \cite{carroll} (2000)
SUSY is a spacetime symmetry relating fermions and bosons to each other.
Just as ordinary symmetries are associated with conserved charges,
supersymmetry is associated with ``supercharges'' $Q_\al$,  where $\al$
is a spinor index.  As with ordinary symmetries,  a theory might be
supersymmetric even though a given state is not supersymmetric;
a state which is annihileted by the supercharges, $Q_\al|\psi>=0$,
presevres supersymmetry,  while states with $Q-\al|\psi>\ne0$
are said to ``spontaneously break'' SUSY.

Consider ``globally supersymmetric' theories,  which are
defined in flat spacetime.   Unlike most nongravitating field theories,  
in supersymmetry the total energy of a state has an absolute meaning:  
the Hamiltonian is related to the supercharges in a straightforward way:
\be
{\cal{H}}=\sum_\al\{Q_\al,Q_\al^\dagger\},
\lb{eq55}
\ee
where brackets represent the anticommutator.   Thus,  in a completely 
supersymmetric state (in which $Q-\al|\psi>=0 \forall\al$),
the energy vanishes automatically, $<\psi|H|\psi)=0$.
More concretely,  in a given supersymmetric theory one can explicitly calculate
the contributions to the vacuum energy from vacuum fluctuations and the scalar
potential $V$.  In the case of vacuum fluctuations,  contributions from bosons
are exactly canceled by equal and opposite contributions when supersymmetry
is unbroken.   Meanwhile,  the scalar-field potential in supersymmetric 
theories takes on a special form;  scalar fields $\psi^i$ must be complex
(to match the degrees of freedom of the fermions),  
and the potential is derived from a function called 
the superpotential $W(\psi^i)$ which is necessarily holomorphic 
(written in terms of $\psi^i$ and not its complex conjugate $\bar{\psi}^i$).
In the simple Wess-Zumino models of spin-0 and spin-1/2 fields,  for example,
the scalar potential is given by
\be
V(\psi^i,\bar{\psi}^j)=\sum_1|\p_iW|^2,
\lb{eq56}
\ee
where $\p_iW=\p W/\p\psi^i$.   In such a theory,  one can show that SUSY will
be unbroken only for values of $\psi^i$ such that $\p_iW=0$,
implying that $V(\psi^i,\bar{\psi}^j)=0$.

So the vacuum energy of a supersymmetric state in a globally supersymmetric
theory will vanish.  This represents rather less progress than 
it might appear at first sight,  since {\it firstly} supersymmetric states
manifest a degeneracy in the mass spectrum of bosons and fermions,
a feature not apparent in the observed world;  
and {\it secondly} the above results imply that non-supersymmetric states 
have a positive definite vacuum energy.
Indeed,  in a state where SUSY was broken at an energy scale $M_{SUSY}$
it would be expected that the corresponding vacuum energy 
$\rho_\Lambda\sim M^4_{SUSY}$.   In the real world,  the fact that 
the accelerator experiments have not discovered superpartners 
for the known particles of the Standard Model implies that $M_{SUSY}$ 
is of order $10^3$ GeV or higher.
Thus a discrepancy
\be
\fr{M_{SUSYU}}{M_{VAC}}\ge10^{15}
\lb{eq57}
\ee
is left.
Comparison of this discrepancy with the naive discrepancy of Carroll 
\cite{carroll} eq.54 (2000), 
is the source of the claim that SUSY can solve the cosmological constant 
problem halfway,  at least on a log scale.

This analysis is strictly valid only in flat space.
In curved spacetime,  the global transformations of ordinary supergravity 
are promoted to the position dependent,  or gauge, 
transformations of supergravity.
In this context the Hamiltonian and the supersymmetry generators play 
different roles than that in flat spacetime,  
but it is still possible to express the vacuum energy,
here meaning cosmological constant,
in terms of the scalar field potential $V(\phi^i,\bar{\phi}^j)$.
In supergravity $V$ depends not only on the superpotentioal $W(\phi^i)$,  
but also on a ``K\"ahler potential'' $K(\phi^i,\bar{\phi}^j)$,
and the K\"ahler metric $K_{i\bar{j}}$ constructed from the K\"alher
potential by $K_{i\bar{j}}=\p^2K/\p\phi^i\p\bar{\phi}^j$.
(The basic role of the K\"alher metric is to define the kinetic term 
for the scalars,  which takes the form 
$g^{\mu\nu}K_{i\bar{j}}\p_\mu\phi^i\p_\nu\bar{\phi}^j$.)   
The scalar potential is
\be
V(\phi^i,\bar{\phi}^j)=\exp\left(\fr{K}{M^2_{Pl}}\right)
\left[K^{i\bar{j}}(D_iW)(D_{\bar{j}}\bar{W})-3M^{-2}_{PL}|W|^2\right],
\lb{eq58}
\ee
where $D_iW$ is the K\"ahler derivative,
\be
D_iW=\p_iW+M^{-2}_{Pl}(\p_iK)W.
\lb{eq59}
\ee
Note that,  if we take the canonical K\"ahler metric 
$K_{i\bar{j}}=\de_{i\bar{j}}$,  
in the limit $M_{Pl}\rightarrow\infty(G\rightarrow0)$
the first term in square brackets reduces to the flat result equation 
\ref{eq56}.
But with gravity,  in addition to the non-negative first term there 
is a second term providing a non-positive contribution.
Supersymmetry is unbroken when $D-IW=0$;  the effective cosmological constant
is thus non-positive.  
One is,  in equation \ref{eq58} therefore free to imagine a scenario 
in which supersymmetry is broken in exactly the right way,  
such that the two terms in parentheses cancel to fantastic accuracy,  
but only at the cost of an unexplained fine-tuning.
At the same time,  supergravity is not by itself a 
renormalizable quantum theory,  and therefore it might not be reasonable 
to hope that a solution can be found purely within this context.
Some recent work on SUSY vacua includes the below.

Brignole {\it et al} \cite{BFZ} (1995) 
construct $N=1$ supergravity models where the gauge symmetry 
and supersymmetry are both spontaneously broken, 
with naturally vanishing classical vacuum energy,
here again meaning the cosmological constant,  and
unsuppressed Goldstino components along gauge non-singlet directions. 
They discuss some physically interesting situations where such a mechanism 
could play a role, and identify the breaking of a grand-unified gauge group 
as the most likely possibility. 
They show that, even when the gravitino mass is much smaller 
than the scale $m_X$ of gauge
symmetry breaking, important features can be missed 
if they first naively integrate out the degrees of freedom 
of mass ${\cal O} (m_X)$, in the limit of unbroken supersymmetry,
and then describe the super-Higgs effect in the resulting effective theory. 
They also comment on possible connections with extended supergravities 
and realistic four-dimensional
string constructions.

Anselm and Berezhiani \cite{ABe} (1996) produce 
an analogy to the case of axion, which converts the $\Theta$-angle 
into a dynamical degree of freedom, they try to imagine 
a situation where the quark mixing angles turn
out to be dynamical degrees of freedom (pseudo-Goldstone bosons), 
and their vacuum expectation values are obtained from the minimization 
of the vacuum energy. 
They present an explicit supersymmetric model with horizontal symmetry, 
where such a mechanism can be realized. 
It implies one relation between the quark masses and the CKM mixing
angles: $s_{13}s_{23}/s_{12}=(m_s/m_b)^2$, which is fulfilled within 
present experimental accuracy. 
They believe, however, that the idea might be more general than this
concrete model, and it can be implemented in more profound frameworks. 

Lythe and Stewart \cite{LS} (1996) claim that supersymmetric theories can 
develop a vacuum expectation value of $M>>10^3$ GeV,  when the temperature
of the early universe falls below some number related to this there is
thermal inflation,  see also \S4.2.

Das and Pernice \cite{DP} (1997) propose a new mechanism 
for symmetry breaking which naturally 
avoids the constraints following from the usual theorems 
of symmetry breaking. 
In the context of supersymmetry, for example, 
the breaking might be consistent with a vanishing vacuum energy. 
A 2+1 dimensional super-symmetric gauge field theory is explicitly shown to
break super-symmetry through this mechanism while maintaining 
a zero vacuum energy.
They claim that 
this mechanism might provide a solution to two long standing problems 
{\it firstly} dynamical super-symmetry breaking and 
{\it secondly} the cosmological constant problem. 

Matsuda \cite{matsuda} (1996) examines 
the phase structures of the supersymmetric $O(N)$ $\sigma$-model 
in two and three dimensions by using the tadpole method. 
Using this simple method, the tadpole calculation is
largely simplified and the characteristics of this theory become clear. 
He also examines the problem of the fictitious negative energy state. 

Oda \cite{oda} (1997) studies the
vacuum structures of supersymmetric (SUSY) Yang-Mills theories 
in $1+1$ dimensions with the spatial direction compactified. 
SUSY allows only periodic boundary conditions for both fermions and bosons. 
By using the Born-Oppenheimer approximation for the weak coupling limit, 
Oda finds that the vacuum energy vanishes, and hence the SUSY is unbroken. 
He studies other boundary conditions, 
especially the antiperiodic 
boundary condition for fermions which is related to the system 
in finite temperatures. 
In that case Oda finds for gaugino bilinears 
a nonvanishing vacuum condensation which indicates instanton contributions. 

Chernyak (1999) \cite{chernyak} 
shows that there is no chirally symmetric vacuum state in the ${\cal N}=1$ 
supersymmetric Yang-Mills theory. The values of the gluino condensate and the
vacuum energy density are found out through a direct instanton calculation. 
A qualitative picture of domain wall properties is presented, 
and a new explanation
of the phenomenon of strings ending on the wall is proposed.

Riotto \cite{riotto} (1998) says that 
the scale at which supersymmetry is broken and the mechanism 
by which supersymmetry breaking is fed down to the observable 
sector has rich implications on the way nature might have chosen 
to accomplish inflation,  see also \S4.4. 
Riotto discusses a simple model for slow rollover inflation 
which is minimal in the sense that the inflaton might be 
identified with the field responsible for the generation of the $\mu$-term. 
Inflation takes place at very late times and is characterized by a very
low reheating temperature. 
This property is crucial to solve the gravitino problem 
and might help to ameliorate the cosmological moduli problem. 
The COBE normalized value of the vacuum energy driving inflation 
is naturally of the order of $10^{11}$ GeV. 
This favours the N=1 supergravity scenario where
supersymmetry breaking is mediated by gravitational interactions. 
Nonetheless, smaller values of the vacuum energy are not excluded 
by present data on the
temperature anisotropy and the inflationary scenario may be implemented 
in the context of new recent ideas about gauge mediation where the 
standard model gauge interactions can serve as the messengers of 
supersymmetry breaking. 
In this class of models supersymmetry breaking masses are 
usually proportional to the F-term of a gauge singlet superfield. 
The same F-term might provide the vacuum energy density necessary to 
drive inflation. 
The spectrum of density perturbations is characterized 
by a spectral index which is significantly displaced from one. 
The measurements of the temperature anisotropies in the cosmic
microwave background radiation at the accuracy expected to result 
from the planned missions will be able to confirm or disprove this 
prediction and to help in
getting some deeper insight into the nature of supersymmetry breaking. 
\subsection{The Supergravity Vacuum.}
Supergravity was first done in four dimensions with extra asymmetric fields.
Then it was noted that it could be more conveniently expressed as 
an eleven dimensional theory,  in which the extra seven dimensions 
were simply a notational device which simplified its presentation.
Freund and Rubin \cite{bi:FR} (1980),  
see also Roberts \cite{pram} (1991),
took the extra dimensions to be there 
and assumed a non-zero asymmetric field $F$,
this can be thought of as assuming a non-zero vacuum.   
The non-zero field causes the four external and seven internal dimensions
to ``curl up'' into maximally symmetric spaces with the constant of
curvature related to the asymmetric field $F$.
For the four external dimensions spacetime is anti-deSitter spacetime
with the constant of curvature being the cosmological constant with a value
outside that allowed for by observations.
Recent work on the Freund-Rubin ansatz includes
L\"utken and Ordonez \cite{LO1,LO2} (1987) who say that
for the Freund-Rubin solution of $d=11$ supergravity to have small extra 
dimensions it is necessary for the corresponding spacetime 
to be anti-deSitter spacetime
with an unphysically large cosmological constant.   The aim of their two papers
is to investigate whether there are solutions which are a product of Minkowski
spacetime and have small extra dimensions,  when quantum mechanical vacuum
energy is taken into account.   The authors' conclusion is that,  for the case
of the vanishing Freund-Rubin parameter,  there are none.
In their {\it first} paper the effective potential for gravitinos,  in $d=11$ 
supergravity on the background of a product of Minkowski spacetime 
and the seven-sphere,  is calculated by a background field method to one loop.
The effective potential is lower than that for lower spin fields,  
but smaller than that for gravitons.   In the {\it second} paper the effective 
potential for the bosonic sector is calculated,  for the case of vanishing 
Freund-Rubin parameter;   the case of nonvanishing Freund-Rubin parameter
is intractable by this method.   It is found that the three-index antisymmetric
gauge field $F$ has a lower effective potential than the gravitino,  
and that the graviton has a higher effective potential 
than that of the gravitino.
Fr\'e {\it et al} \cite{FGPT} (1996) exhibit
generic partial supersymmetry breaking of $N=2$ supergravity 
with zero vacuum energy and with surviving unbroken arbitrary 
gauge groups, and they give specific examples. 

Ellwanger \cite{ellwanger} (1995) constructs
supergravity models are constructed in which the 
effective low energy theory contains only ``super-soft'' 
explicit supersymmetry breaking: masses of the scalars and
pseudoscalars within a multiplet are split in opposite directions. 
With this form of supersymmetry breaking the radiative corrections 
of the matter sector to the vacuum energy are
bounded by ${\cal O} (M^4_{Susy})$ to all orders in perturbation theory, 
and he requires $Str M^2 = 0$ including the hidden sector. 
The models are based on K\"ahler potentials
obtained in recent orbifold compactifications, 
and he describes the construction of realistic theories. 
\subsection{String and Brane Vacua.}
The string vacuum is similar to the supergravity vacuum \S2.7 above,
in that the nature of the vacuum is related to the splitting of space
into four spacetime dimensions and internal dimensions.
For example Candelas {\it et al} \cite{CHSW} (1985) note that
the vacuum state should be of the form $M_4\times K$,
where $M_4$ is the spactime manifold and $K$ is some six dimensional compact
manifold.   
To determine $K$ they say one needs geometry of the above form with
$M_4$ maximally symmetric.  
They also say one needs unbroken $N=1$ supersymmetry 
in $d=4$ as this might resolve the gauge hierarchy and Dirac large number 
problem.   
Furthermore the gauge group and fermionic spectrum should be realistic.
The manifold forced on Candelas {\it et al}
by phenomenological considerations in the field 
theory limit are precisely those manifolds on which it is possible to 
formulate a consistent string theory.   
They say that indications are that the candidate vacuum 
configurations singled out by phenomenological 
considerations obey the equations of motion.

Largely following Carroll \cite{carroll} \S4.2 (2000) note that
unlike supergravity,  string theory appears to be a consistent 
and well-defined theory of quantum gravity,  
and therefore calculating the value of the cosmological constant 
should at least in principle be possible.
Cosmological constant here means something related to potentials $V$,
and thus is subject to the caveats at the end of \S2.1\P2.
On the other hand,  the number odd vacuum states seems to be quite large,
and none of them (to the best of our current knowledge) 
features three large spatial dimensions,  broken supersymmetry,  
and a small cosmological constant.
At the same time,  there are reasons to believe that any realistic vacuum of
string theory must be strongly coupled (although it is not clear what to,
perhaps self-coupling is meant);  
therefore,  our inability to find an appropriate solution 
might simply be due to the technical difficulty of the problem.

String theory is naturally formulated in more than four spacetime dimensions.
Studies of duality symmetries have revealed that what used to be thought of as
five distinct ten-dimensional superstring theories - Type I,  
Types IIA and IIB,  and heterotic theories based on gauge groups 
$E(8)\times E(8)$ and $SO(32)$ - 
are along with eleven-dimensional supergravity,
different low-energy weak-coupling limited of a single underlying theory,  
sometimes known as M-theory.
In each of these six cases,  the solution with the maximum number 
of uncompactified flat flat spacetime dimensions is a stable vacuum 
preserving all of the supersymmetry.
To bring the theory closer to the world that is observed,  the extra dimensions
can be compactified on a manifold whose Ricci tensor vanishes.   
there are a large number of possible compactifications,
many who preserve some but not all of the original supersymmetry.
If enough SUSY is preserved,  the vacuum energy will remain zero;
generically there will be a manifold of such states,  
known as the moduli space.

Of course,  to describe our world we want to break all supersymmetry
in order for there to be correspondence with the observed world.
Investigations in contexts where this can be done in a controlled way have 
found that the induced cosmological constant vanishes at the classical level,
but a substantial vacuum energy is typically induced by quantum corrections.
Moore \cite{moore} (1987) has suggested that Atkin-Lehner symmetry,  
which relates strong and weak coupling on the string world sheet,  
can enforce the vanishing of the one-loop quantum contribution 
in certain models;  generically however there would still be an 
appreciable contribution at two loops.

Thus,  the search is still on for a four-dimensional string theory vacuum with 
broken supersymmetry and vanishing or very small cosmological constant,
see Dine \cite{dine} (1999) below for a general discussion of the vacuum 
problem in string theory.   
Carroll \cite{carroll} (2000) notes that
the difficulty of achieving this in conventional 
models has inspired a number of more speculative models:
\begin{enumerate}
\item
In three spacetime dimensions supersymmetry can remain unbroken,  maintaining
a zero cosmological constant,  in such a way as to break the mass degeneracy
between bosons and fermions.
This mechanism relies crucially on special properties of spacetime in (2+1)
dimensions,  but in string theory it sometimes happens that the strong-coupling
limit of one theory is another theory in a higher dimension.
\item
More generally,  it is now understood that (at least in some circumstances)
string theory obeys the ``holographic principle'',  the idea that a theory with
gravity in $D$ dimensions is equivalent to a theory without gravity in $D-1$
dimensions.   In a holographic theory,  the number of degrees of of freedom
in a region grows as the area of its boundary,  rather than its volume.
Therefore,  the conventional computation of the cosmological constant due
to vacuum fluctuations conceivably involves a vast overcounting of degrees
of freedom.   It might be imagined that a more correct counting would yield
a much smaller estimate of the vacuum energy,  although no reliable calculation
has been done yet.   See also Thomas \cite{thomas} (2000).
\item
The absence of manifest SUSY in our world leads us to ask whether the 
beneficial aspect of canceling contributions to the vacuum energy
could be achieved even without a truly supersymmetric theory.
Kachru {\it et al} (1998), see also this section below,  
have constructed such a string theory,  
and argue that the perturbative contributions to the cosmological 
constant should vanish (although the actual calculations are somewhat 
delicate,  and not everyone agrees.)
If such a model could be made to work,  it is possible that the small 
non-perturbative effects could generate a cosmological constant 
of an astrophysically plausible magnitude.
\item
A novel approach to compactification starts by imagining that 
the fields of the standard model are confined to a (3+1)-dimensional 
manifold (or ``brane'',  in string theory parlance) embedded in a larger space.
While gravity is harder to confine to a brane,  
phenomenologically acceptable scenarios can be constructed 
if either the extra dimensions are any size less than a millimeter,  
or if there is significant spacetime curvature 
in a non-compact extra dimension.
Although these scenarios do not offer a simple solution 
to the cosmological constant problem,  the relationship between 
the vacuum energy and the expansion rate can differ from our 
conventional expectation,  and one is free to imagine that further study 
might lead to a solution in this context.
\end{enumerate}

Buchbinder {\it et al} \cite{BOS} (1992)
discuss some techniques used in string theory calculations.

Chang and Dowker \cite{CD} (1992) calculate vacuum energy 
for a free, conformally-coupled 
scalar field on the orbifold space-time R$\times S2/\Gamma$ 
where $\Gamma$ is a finite subgroup of O(3)
acting with fixed points. 
The energy vanishes when $\Gamma$ is composed of pure rotations 
but not otherwise. 
It is shown on general grounds that the same conclusion holds for
all even-dimensional factored spheres and the vacuum energies are given 
as generalized Bernoulli functions (i.e. Todd polynomials). 
The relevant $\zeta$- functions are analyzed
in some detail and several identities are incidentally derived. 
Chang and Dowker give a general discussion
in terms of finite reflection groups. 

Cornwall and Yan \cite{CY} (1995) 
give a quantitative 
(if model-dependent) estimate of the relation between
the string tension and a gauge-invariant measure of the 
Chern-Simons susceptibility, due to vortex linkages, 
in the absence of a Chern-Simons term in the action,
based on a model of the d=3 SU(2) pure gauge theory vacuum 
as a monopole-vortex condensate, they 
give a quantitative 
(if model-dependent) estimate of the relation between
the string tension and a gauge-invariant measure of the 
Chern-Simons susceptibility, due to vortex linkages, 
in the absence of a Chern-Simons term in the action. 
They also give relations among these quantities 
and the vacuum energy 
and gauge-boson mass. Both the susceptibility and the string tension 
come from the same physics:  namely the
topology of linking, twisting, and writhing of closed vortex strings. 
The closed-vortex string is described via a complex scalar field theory 
whose action has a precisely-specified functional form, 
inferred from previous work giving the exact form of a 
gauge theory effective potential at low energy.

Lopez and Nanopoulos \cite{LN} (1995) 
show that the presence of an anomalous $\rm U_A(1)$ 
factor in the gauge group of string derived models 
might have the new and important phenomenological consequence of
allowing the vanishing of ${\rm Str}\,{\cal M}^2$ in the ``shifted" vacuum, 
that results in the process of canceling the anomalous $\rm U_A(1)$. 
The feasibility of this effect
seems to be enhanced by a vanishing vacuum energy, 
and by a ``small" value of ${\rm Str}\,{\cal M}^2$ in the original vacuum. 
In the class of free-fermionic models with
vanishing vacuum energy that they focus on, 
a necessary condition for this mechanism 
to be effective is that ${\rm Str}\,{\cal M}^2>0$ in the original vacuum. 
They say that a vanishing ${\rm Str}\,{\cal M}^2$ ameliorates 
the cosmological constant problem and is a necessary element 
in the stability of the no-scale mechanism,
again they take the cosmological constant 
is taken to be related to vacuum energy.

Naculich \cite{naculich} (1995) notes that
Z-strings in the Weinberg-Salam model including fermions 
are unstable for all values of the parameters. 
The cause of this instability is the fermion vacuum energy in the Z-string
background. Z-strings with non-zero fermion densities, 
however, might still be stable. 

Leontaris and Tracas \cite{LT} (1996) 
analyse the constraints from duality invariance 
on effective supergravity models with an intermediate gauge symmetry. 
Requiring vanishing vacuum energy and invariance 
of the superpotential couplings, they find that modular weights 
of the matter and Higgs fields are subject to various constraints.
In addition, the vacuum expectation values of the Higgs fields 
breaking the intermediate gauge group, are determined in terms 
of their modular weights and the moduli. 
They also examine the possibility of breaking 
the intermediate gauge symmetry radiatively. 

Berntsssen {\it et al} \cite{BBO} (1997)
study the Casimir effect for a string.

Kim and Rey \cite{KRe} (1997) study
matrix theory on an orbifold and classical two-branes 
therein with particular emphasis to heterotic 
M(atrix) theory on $S_1/Z_2$ relevant to strongly coupled
heterotic and dual Type IA string theories. 
By analyzing orbifold condition on Chan-Paton factors, 
they show that three choice of gauge group are possible for heterotic M(atrix)
theory: SO(2N), SO(2N+1) or USp(2N). 
By examining area preserving diffeomorphism 
that underlies the M(atrix) theory, 
they find that each choices of gauge group restricts
possible topologies of two-branes. 
The result suggests that only the choice of SO(2N) or SO(2N+1) 
groups allows open two-branes, hence, relevant to heterotic M(atrix) theory. 
They show that requirement of both local vacuum energy cancellation 
and of worldsheet anomaly cancellation of resulting heterotic string 
identifies supersymmetric twisted
sector spectra with sixteen fundamental representation 
spinors from each of the two fixed points. 
Twisted open and closed two-brane configurations are obtained in the large N
limit. 

Aldazabal {\it et al} \cite{AFIUV} (1998) consider $D=6, N=1, Z_M$ orbifold 
compactifications of heterotic strings 
in which the usual modular invariance constraints are violated. 
They argue that in the presence of non-perturbative effects many 
of these vacua are nevertheless consistent. 
The perturbative massless sector can be computed explicitly from
the perturbative mass formula subject to an extra shift in the vacuum energy. 
They say that this shift is associated to a non-trivial antisymmetric 
B-field flux at the orbifold fixed points. 
The non-perturbative piece is given by five-branes either moving in the bulk 
or stuck at the fixed points, giving rise to Coulomb phases with
tensor multiplets. 
The heterotic duals of some Type IIB orientifolds belong to this class 
of orbifold models. 
Aldazabal {\it et al} also discuss how to carry out this type of
construction to the $D=4, N=1$ case and specific $Z_M\times Z_M$ 
examples are presented in which non-perturbative transitions 
changing the number of chiral generations do occur. 

Buonanno {\it et al} \cite{BDV} (1998) discusses how string vacuum energy might
give rise to prebig bang bubbles.

Kachru {\it et al} \cite{KKS} (1998) 
present a nonsupersymmetric orbifold of type II string theory 
and show that it has vanishing cosmological constant at the one 
and two loop level,  here the cosmological constant is taken 
to be related to the vacuum energy.
They argue heuristically that the cancellation persists at higher loops. 

Kachru and Silverstein \cite{KS} (1998)
propose and test correspondences between 4D QFT's
with N=2,1,0 conformal invariance and type IIB string theory on 
various orbifolds of $AdS_5\times S^5$.
This allows them to translate the problem of finding stable nontrivial 
nonsuper string background into the problem of realizing nontrivial
renormalization group fixed point QFT's on branes.
Renormalization group fixed lines in this context correspond 
to string theories in which no 
vacuum energy is generated quantum mechanically.

Kawamura \cite{kaw} (1998) studies the magnitudes of soft masses 
in heterotic string models with 
anomalous U(1) symmetry model-independently. 
In most cases, D-term contribution
to soft scalar masses is expected to be comparable to or dominant over other 
contributions provided that supersymmetry breaking is mediated by the
gravitational interaction and/or an anomalous U(1) symmetry and the magnitude 
of vacuum energy is not more than of order $m_{3/2}^2 M^2$. 

Angelantonj {\it et al} \cite{AAF} (1999) study open descendants 
of non-supersymmetric type IIB asymmetric 
(freely acting) orbifolds with zero cosmological constant,  here again the 
cosmological constant is taken to be related to the vacuum energy.
A generic feature of these models is that supersymmetry remains 
unbroken on the brane at all mass levels, 
while it is broken in the bulk in a way that preserves Fermi-Bose degeneracy
in both the massless and massive (closed string) spectrum. 
This property remains valid in the heterotic dual of the type II 
model but only for the massless excitations. 
A possible application of these constructions concerns scenarios 
of low-energy supersymmetry breaking with large dimensions. 

Bianchi {\it et al} \cite{BGMN} (1999) 
determine the spectrum of D-string bound states 
in various classes of generalized type I vacuum configurations 
with sixteen and eight supercharges. 
They say that the precise matching of the BPS spectra 
confirms the duality between 
unconventional type IIB orientifolds with quantized NS-NS antisymmetric 
tensor and heterotic CHL models in D=8. A similar
analysis puts the duality between type II (4,0) 
models and type I strings {\it without open strings} on a firmer ground. 
The analysis can be extended to type II (2,0) asymmetric
orbifolds and their type I duals that correspond to 
unconventional K3 compactifications. 
Finally they discuss BPS-saturated threshold corrections 
to the corresponding low-energy effective lagrangians. 
In particular they show how the exact moduli dependence 
of some $F^4$ terms in the eight-dimensional type II (4,0) 
orbifold is reproduced by the infinite sum
of D-instanton contributions in the dual type I theory. 

Dine \cite{dine} (1999) says that recently, 
a number of authors have challenged the conventional assumption 
that the string scale, Planck mass, and unification scale are
roughly comparable. It has been suggested that the string scale 
could be as low as a TeV. The greatest obstacle to developing a string 
phenomenology is our lack of understanding of the ground state. 
He explains why the dynamics which determines this state is not likely to
be accessible to any systematic approximation. 
He notes that the racetrack scheme, often cited as a counterexample, 
suffers from similar difficulties. 
He stresses that the weakness of the gauge couplings, 
the gauge hierarchy, and coupling unification suggest that it might be
possible to extract some information in a systematic approximation. 
He reviews the ideas of K\"ahler stabilization, an attempt to reconcile
these facts. 
He considers whether the system is likely to sit at extremes 
of the moduli space, as in recent proposals for a low string scale.
Finally he discusses the idea of Maximally Enhanced Symmetry, 
a hypothesis which is technically natural, 
and hoped to be compatible with basic facts about
cosmology, and potentially predictive. 

Dvali and Tye \cite{DT} (1999) present a novel inflationary scenario 
in theories with low scale (TeV) 
quantum gravity, in which the standard model particles are localized 
on the branes whereas gravity propagates in the bulk of large extra 
dimensions.   They say that
this inflationary scenario is natural in the brane world picture. 
In the lowest energy state, a number of branes sit on top of each 
other (or at an orientifold plane), so the vacuum energy cancels out. 
In a cosmological setting, some of the branes "start out" relatively 
displaced in the extra dimensions and the resulting vacuum energy 
triggers the exponential growth of the 3 non-compact dimensions. 
They say that the number of e-foldings can be very large due to the very weak 
brane-brane interaction at large distances. 
In the effective four-dimensional field theory, the
brane motion is described by a slowly rolling scalar 
field with an extremely flat plateau potential. 
They say that when branes approach each other to a critical distance, the
potential becomes steep and inflation ends rapidly. 
Then the branes "collide" and oscillate about the equilibrium point, 
releasing energy mostly into radiation on the branes. 
See also \S4.2.

King and Riotto \cite{KR} (1999) note that dilaton stabilization 
is usually considered to pose a serious obstacle to 
successful $D$-term inflation in superstring theories. 
They argue that the physics of gaugino condensation is likely to be modified 
during the inflationary phase in such a way as to enhance the gaugino 
condensation scale. 
This enables dilaton stabilization during inflation with the $D$-term still 
dominating the vacuum energy at the stable minimum.
See also \S4.2.

Blumenhagen {\it et al} \cite{BGKL} (2000) 
investigate the D-brane contents of asymmetric orbifolds. 
Using T-duality they find that the consistent description 
of open strings in asymmetric orbifolds requires to turn on
background gauge fields on the D-branes. 
They derive the corresponding noncommutative 
geometry arising on such D-branes with mixed 
Neumann-Dirichlet boundary conditions
directly by applying an asymmetric rotation 
to open strings with pure Dirichlet or Neumann boundary conditions. 
As a concrete application of their results they construct asymmetric
type I vacua requiring open strings with mixed boundary conditions 
for tadpole cancellation.

Donets {\it et al} \cite{DIST} (2000) discuss the brane vacuum as a chain
of rotators by using the noncomuatative U(1) sigma model.
 
Ellis {\it et al} \cite{EMN} (2000) note that
classical superstring vacua have zero vacuum energy and are 
supersymmetric and Lorentz-invariant. 
They argue that all these properties may be destroyed when quantum aspects
of the interactions between particles and non-perturbative 
vacuum fluctuations are considered. 
A toy calculation of string/D-brane interactions using a world-sheet approach
indicates that quantum recoil effects - reflecting the gravitational 
back-reaction on spacetime foam due to the propagation of 
energetic particles - induce non-zero vacuum
energy that is linked to supersymmetry breaking 
and breaks Lorentz invariance. 
This model of space-time foam also suggests 
the appearance of microscopic event horizons. 
Ellis {\it et al} \cite{ELPT} (2000)
have not identified a vacuum with broken supersymmetry and zero vacuum energy.

Hata and Shinohara \cite{HShin} (2000) note that
tachyon condensation on a bosonic D-brane 
was recently demonstrated numerically in Witten's 
open string field theory with level truncation approximation. 
This non-perturbative vacuum, which is obtained by solving 
the equation of motion, has to satisfy furthermore
the requirement of BRST invariance. 
This is indispensable in order for the theory around 
the non-perturbative vacuum to be consistent.
They carry out the numerical analysis of the BRST invariance 
of the solution and find that it holds to a good accuracy. 
They also mention the zero-norm property of the solution. 
The observations in this paper are expected to give clues 
to the analytic expression of the vacuum solution. 

Nudelman \cite{nudelman} (2000)
considers certain linear objects, termed physical lines;
and introduces initial assumptions concerning their properties. 
He investigates a closed physical line in the form of a circle 
called a J-string.
He showns that this curve consists of indivisible line
segments of length $\ell_\Delta$. 
It is assumed that a J-string has an angular momentum whose value is $\hbar$.
It is then established that a J-string of radius $R$ possesses a mass $m_J$, 
equal to $h/2\pi c R$, a corresponding energy, as well as a charge $q_J$, 
where $q_J= (hc/2\pi)^{1/2}$. 
He also establishes that $\ell_\Delta = 2\pi(hG/c^3)^{1/2}$, 
where $c$ is the speed of light and $G$ is the gravitational constant. 
Based upon investigation of the properties and characteristics of J-strings, 
he develops a method for the computation of the
Planck length and mass $(\ell^*_P, m^*_P)$. 
Using the methods developed in the paper the values of $\ell^*_P$ 
and $m^*_P$ are computed these values differ from the currently accepted ones.
  
Toms \cite{toms} (2000) considers 
the quantization of a scalar field in the five-dimensional model 
suggested by Randall and Sundrum. 
Using the Kaluza-Klein reduction of the scalar field, discussed
by Goldberger and Wise, he sums the infinite tower of modes 
to find the vacuum energy density. 
Dimensional regularisation is used to compute the pole term needed for
renormalisation, as well as the finite part of the energy density. 
He makes some comments concerning the possible self-consistent 
determination of the radius. 
\subsection{Lattice Models.}
Unlike QFT's which are defined at each point of spacetime,  lattice models
are only defined at,  a usually finite number of,  discrete points.
A finite number of points might result in less of the infinite objects
which occur in QFT's.   
Luo \cite{luo} (1998) investigate the properties of lattice QCD,  see \S2.5.
Some recent work on vacua occurring in lattice models is listed below.

Hollenberg \cite{hollenberg} (1994) computes the vacuum energy density 
for a SU(2) lattice gauge theory and gets results which might apply to 
beyond the reach of the strong to weak transition point $g^2_c\approx2.0$.

Bock {\it et al} \cite{BHS} (1995) note that
lattice proposals for a nonperturbative formulation of the 
Standard Model easily lead to a global U(1) symmetry corresponding 
to exactly conserved fermion number. 
The absence of an anomaly in the fermion current would then appear 
to inhibit anomalous processes, such as electroweak baryogenesis 
in the early universe. 
One way to circumvent this
problem is to formulate the theory such that this U(1) symmetry 
is explicitly broken. 

Adam \cite{adam} (1997) gives a detailed discussion 
of the mass perturbation theory of the massive Schwinger model. 
After discussing some general features and briefly reviewing the
exact solution of the massless case, he computes the vacuum energy 
density of the massive model and some related quantities. 
He derives the Feynman rules of mass perturbation
theory and discuss the exact $n$-point functions 
with the help of the Dyson-Schwinger equations. 
Furthermore he identifies the stable and unstable bound states of the 
theory and computes some bound-state masses and decay widths. 
Finally he discuss scattering processes, where he claims that the resonances 
and particle production thresholds of the model are
properly taken into account by his methods.

Aroca \cite{aroca} (1999) studies the 
Schwinger model in a finite lattice 
by means of the P-representation. 
The evaluate the vacuum energy, mass gap and chiral condensate 
showing good agreement with the expected values in the continuum limit. 

Cea and Cosmai \cite{CCo} (1999) study 
the vacuum dynamics of SU(2) lattice gauge theory is studied by means 
of a gauge-invariant effective action defined using the lattice 
Schr\"odinger functional. 
they perform numerical simulations both at zero and finite temperature. 
They probe the vacuum using an external constant Abelian 
chromomagnetic field. 
Their results suggest that at zero temperature the external field is screened 
in the continuum limit. On the other hand at finite temperature 
they say that it seems that 
confinement is restored by increasing the strength of the applied field. 

Maniadas {\it et al} \cite{MTBZ} (1999) 
use a collective coordinate approach to investigate 
corpuscular properties of breathers in nonlinear lattice systems.
They calculate the breather internal energy and inertial mass and 
use them to analyze the reaction pathways of breathers with kinks 
that are performed in the lattice.
They find that there is an effective kink breather interaction potential,  
that,  under some circumstances,  is attractive and has a double well shape.
Furthermore,  they find that in some cases the internal energy of a moving 
breather can be realized during the reaction with the kink and subsequently
transformed to kink translational energy.   
These breather properties seem to be model independently.
See also \S2.14.

Actor {\it et al} \cite{ABR} (2000) 
present a Hamiltonian lattice formulation of static Casimir 
systems at a level of generality appropriate 
for an introductory investigation.
Background structure - represented by a lattice potential V(x) - is 
introduced along one spatial direction with translation invariance in all 
other spatial directions.
After some general analysis they analyze two
specific finite one dimensional lattice QFT systems. 
See also \S3.
\subsection{Symmetry Breaking.}
One of the outstanding problems of particle physics is whether the Higgs 
field exists;  this hypothetical field is conjectured to make
non-Abelian gauge fields describe massive particles as needed by
the standard particle physics model,  see \S2.1. 
Mass comes from the vacuum expectation value of the field differing
from zero,  in this manner symmetry breaking can be thought of as a 
vacuum energy effect.
Baum \cite{baum} (1994) discusses a positive definite action
which leads to no cosmological constant and breaks symmetry. 
The way that symmetry is broken invokes properties of the vacuum, 
see equation \ref{eq:2.14}.
Vacuum energy essentially ``shifts'' the value of the scalar field:
and this shift corresponds to mass.
For purposes of illustration symmetry breaking in scalar electrodynamics (SEL)
is presented below;  although because electromagnetism is massless and the 
only theory with just one vector $A_a$ symmetry breaking is not physically
realize here.
The scalar electrodynamic Lagrangian \cite{bi:higgs},  
\cite{bi:IZ}p.68,  \cite{bi:HE}p.699 is
\be
{\cal{L}}_{sel}=-D_a\psi D^a\bar{\psi}-V(\psi\bar{\psi})-\fr{1}{4}F^2,
\label{eq:2.1}
\ee
where the covariant derivative is
\be
D_a\psi=\p_a\psi+ieA_a,
\label{eq:2.2}
\ee
and $D_a\bar{\psi}=\bar{D_a\psi}$;
and so is related to the lagrangian in \ref{eq:simpleaction} \S2.1 by 
changing the partial derivatives to covariant derivatives involving the 
vector potential $A_a$.
The variation of the corresponding action with respect to
$A_a,  \psi$, and $\bar{\psi}$ are given by
\ber
\fr{\de I}{\de A_c}&=&F^{ab}_{..;b}
                     +ie(\psi D^a\bar{\psi}-\bar{\psi}D^a\psi)\nonumber\\
\fr{\de I}{\de \psi}&=&(D_aD^a-V')\bar{\psi},~~~V'=\fr{dV}{d(\psi\bar{\psi})},
\label{eq:2.3}
\ear
and its complex conjugate.   Variations of the metric give the stress
\be
T_{ab}=2D_{(a}\psi D_{b)}\bar{\psi}+F_{ac}F^{~c}_{b.}+g_{ab}{\cal{L}}.
\label{eq:2.4}
\ee
The complex scalar field can be put in "polar" form by defining
\be
\psi= \rho\exp(i\nu),
\label{eq:2.5}
\ee
giving the Lagrangian
\be
{\cal{L}}=\rh^2_a+({\mathcal D}_a\nu)^2-V(\rh^2)-\fr{1}{4}F^2,
\label{eq:2.6}
\ee
where
\be
{\mathcal D}_a\nu=\rh(\nu_a+eA_a).
\label{eq:2.7}
\ee
The variations of the corresponding action with respect to 
$A_a   ,\rh$,  and $\nu$ are given by
\ber
\fr{\de I}{\de A_a}&=&F^{ab}_{..;b}+2e\rh{\mathcal D}^a\nu,\nonumber\\
\fr{\de I}{\de \rh}&=&2\left(\Box+(\nu_a+eA_a)^2+V'\right),\nonumber\\
\fr{\de I}{\de \nu}&=&2\left(\Box\nu+eA^a_{.a;a}\right).
\label{eq:2.8}
\ear
Variation of the metric gives the stress
\be
T_{ab}=2\rh_a\rh_b
      +2{\mathcal D}_a\nu{\mathcal D}_b\nu+F_{ac}F^{~c}_{b.} 
      +g_{ab}{\cal{L}}.
\label{eq:2.9}
\ee
Defining
\be
B_a=A_a+\nu_a/e,
\label{eq:2.10}
\ee
$\nu$ is absorbed to give Lagrangian
\be
{\cal{L}}=\rh^2_a+\rh^2e^2B_a^2-V(\rh^2)-\fr{1}{4}F^2,
\label{eq:2.11}
\ee
which does not contain $\nu$;  equation \ref{eq:2.10} 
is a gauge transformation when there are no discontinuities 
in $\nu$,  i.e. $\nu_{;[ab]}=0$.

The requirement that the corresponding quantum theory is renormizable
restricts the potential to the form
\be
V(\rh^2)=m^2\rh^2+\la\rh^4.
\label{eq:2.12}
\ee
The ground state is when there is a minimum,  for $m^2,\la>0$ this is $\rh=0$,
but for $m^2<0,\la>0$ this is
\be
\rh^2=\fr{-m^2}{2\la}=a^2,
\label{eq:2.13}
\ee
thus the vacuum energy is
\be
<0|\rh|0>=a.
\label{eq:2.14}
\ee
To transform the Lagrangian \ref{eq:2.11} to take this into account substitute
\be
\rh\rightarrow\rh'=\rh+a,
\label{eq:2.15}
\ee
to give
\be
{\cal{L}}=\rh_a^2+(\rh+a)^2e^2B_a^2-V\left((\rh+a)^2\right)-\fr{1}{4}F^2.
\label{eq:2.16}
\ee
Now apparently the vector field has a mass $m$ from the $a^2e^2B_a^2$ 
term it is given by $m=ae$.   The cross term $2a\rh e^2B_a^2$ is ignored.

There are well known techniques by which stresses involving scalar 
fields can be rewritten as fluids, so that it is possible to re-write 
the standard model with fluids instead of Higgs scalars.   
It is preferable to use fluids rather than Higgs 
scalar fields for {\it two} reasons.
The {\it first} is
because Higgs scalar fields are {\it ad hoc} whereas 
fluids might arise from the statistical properties of the non-Abelian gauge 
fields:  furthermore taking an extreme view of the principle of equivalence 
Roberts \cite{mdrsym} (1989) 
suggests that Higgs scalars cannot be fundamental.   
The {\it second} is that perfect fluids are gauge systems and thus all matter
under consideration is part of a gauge system.
My {\em first} attempt at 
using fluids for symmetry breaking Roberts \cite{mdrsym} (1989) 
was essentially to deploy 
rewriting procedures to convert the scalar fields to fluids;  the drawback of 
this approach is the fluids that result are somewhat unphysical,  
however an 
advantage is that symmetry breaking occurs with a change of state of the 
fluid.
My {\em second} attempt Roberts \cite{mdrvel} (1997) 
used the decomposition of a perfect fluid vector 
into several scalar parts called vector potentials and identifying one of 
these with the radial Higgs scalar $\rho$,  see equation \ref{eq:2.15}.
The gauge fields are then introduced 
by using the usual covariant substitutions for the partial derivatives of the
scalars,  for example 
$\phi_a=\partial_a\phi\rightarrow\nabla_a\phi=\partial_a\phi+ie\phi A_a$,  
and calling the resulting fluid the {\sc covariantly interacting fluid}.   
This results in an elegant extension of standard Higgs symmetry breaking,
but with some additional parameters present.   
Previous work on the vector potentials shows 
that some of these have a thermodynamic interpretation,  
it is hoped that this is inherited
in the fluid symmetry breaking models.   The additional parameters can be 
partially studied with the help of an explicit Lagrangian 
Roberts \cite{mdrgh} (1999). 
Both of my approaches have been restricted to Abelian gauge fields. 

Some recent papers on symmetry breaking include those below.
Coleman and Weinberg \cite{ColW} \S2 (1973) show that functional methods allow
the definition of an effective potential,  the minimum of which,  without
any approximation,  gives the true vacuum states of a theory.
Kujat and Scherrer \cite{KSch} discuss the cosmological implications of a
time dependent Higgs field,  see \S4.5 above.

Kounnas {\it et al} \cite{KPZ} (1994) note that 
in the minimal supersymmetric standard model (MSSM), 
the scale $M_{SUSY}$ of soft supersymmetry breaking 
is usually {\em assumed} to be of the order of the electroweak scale. 
They reconsider here the possibility of treating $M_{SUSY}$ 
as a dynamical variable. Its expectation value should be 
determined by minimizing the vacuum energy, after
including MSSM quantum corrections. 
They point out the crucial role of the cosmological term 
for a dynamical generation of the desired hierarchies $m_Z, M_{SUSY} << M_P$,
here yet again the cosmological term is related to the vacuum energy.
Inspired by four-dimensional superstring models, 
they also consider the Yukawa couplings as dynamical variables. 
They find that the top Yukawa coupling is attracted close to its
effective infrared fixed point, corresponding to a top-quark mass 
in the experimentally allowed range. 
As an illustrative example, they present the results 
of explicit calculations for a special case of the MSSM. 

Ferrara {\it et al} \cite{FGP} (1995) 
show that the minimal Higgs sector of a generic N=2 supergravity 
theory with unbroken N=1 supersymmetry must contain a Higgs hypermultiplet 
and a vector multiplet.
When the multiplets parameterize the quaternionic manifold SO(4,1)/SO(4), 
and the special K\"ahler manifold SU(1,1)/U(1), respectively, 
a vanishing vacuum energy with a sliding
massive spin 3/2 multiplet is obtained. 
Potential applications to N=2 low energy effective actions 
of superstrings are briefly discussed. 

Boyanovsky {\it et al} \cite{BVHS} (1996) 
analyze the phenomenon of preheating,i.e. explosive particle production 
due to parametric amplification of quantum fluctuations in the unbroken case, 
or spinodal instabilities in the broken phase, using the Minkowski space 
$O(N)$ vector model in the large $N$ limit to study the non-perturbative 
issues involved. 
They give analytic results for weak couplings and times short compared 
to the time at which the fluctuations become of the same order 
as the tree level,  as well as numerical results including 
the full backreaction.
In the case where the symmetry is unbroken, the analytic results agree 
spectacularly well with the numerical ones 
in their common domain of validity. 
In the broken symmetry case, slow roll initial conditions from the unstable 
minimum at the origin, give rise to a new and unexpected phenomenon: 
the dynamical relaxation of the vacuum energy.
That is,  particles are abundantly produced at the expense 
of the quantum vacuum energy while the zero mode comes back 
to almost its initial value.
In both cases they obtain analytically
and numerically the equation of state 
which turns to be written in terms of an effective 
polytropic index that interpolates between vacuum 
and radiation-like domination. They find that simplified analysis 
based on harmonic behavior of the zero mode, 
giving rise to a Mathieu equation for the non-zero modes 
misses important physics.
Furthermore, they claim that analysis that do not include 
the full backreaction do not conserve energy, 
resulting in unbound particle production. 
Their results do not support the recent claim of symmetry restoration by
non-equilibrium fluctuations.
Finally estimates of the reheating temperature are given,
as well as a discussion of the inconsistency of a kinetic 
approach to thermalization when a
non-perturbatively large number of particles is created. 

Das and Pernice \cite{DP} (1996) propose
a new mechanism for symmetry breaking which naturally 
avoids the constraints following from the usual theorems 
of symmetry breaking. 
In the context of super-symmetry, for example, 
the breaking may be consistent with a vanishing vacuum energy. 
A 2+1 dimensional super-symmetric gauge field theory is explicitly shown to
break super-symmetry through this mechanism while maintaining 
a zero vacuum energy. 
This mechanism may provide a solution to two long standing problems, namely,
dynamical super-symmetry breaking and the cosmological constant problem,
see \S4.1.

Axenides and Perivolaropoulos \cite{AP} (1997) 
demonstrate that field theories involving 
explicit breaking of continuous symmetries, 
incorporate two generic classes of topological 
defects each of which is stable for a particular range of parameters. 
The first class includes defects of the usual type where the symmetry 
gets restored in the core and vacuum energy gets trapped there. 
However they show that these defect solutions become unstable for 
certain ranges of parameters and decay not to the vacuum but to 
another type of stable defect where the symmetry in not restored in the core. 
In the wall case, initially spherical, bubble-like configurations are 
simulated by them numerically and shown to evolve generically towards 
a planar collapse. 
In the string case, the decay of the symmetric core vortex resembles 
the decay of a semilocal string to a skyrmion with the important 
difference that while the skyrmion is unstable and
decays to the vacuum, the resulting non-symmetric 
vortex is topologically stable. 

Natale and da Silva \cite{NS} (1997) discuss how to dynamically break symmetry.

Lepora and Kibble \cite{LK} (1999) 
analyze symmetry breaking in the Weinberg-Salam model 
paying particular attention to the underlying geometry of the theory. 
In this context they find two natural metrics upon the vacuum manifold: 
an isotropic metric associated with the scalar sector, 
and a squashed metric associated with the gauge sector.
Physically, the interplay between these metrics gives 
rise to many of the non-perturbative features of Weinberg-Salam theory. 

Foot {\it et al} \cite{FLV} (2000) note that 
if the Lagrangian of nature respects parity invariance 
then there are two distinct possibilities: 
either parity is unbroken by the vacuum or it is spontaneously broken. 
They examine the two simplest phenomenologically consistent gauge models 
which have unbroken and spontaneously broken parity symmetries, respectively. 
These two models have a
Lagrangian of the same form, but a different parameter range 
is chosen in the Higgs potential. 
They both predict the existence of dark matter 
and can explain the MACHO events.
However, the models predict quite different neutrino physics. 
Although both have light mirror (effectively sterile) neutrinos, 
the ordinary-mirror neutrino mixing angles are
unobservably tiny in the broken parity case. 
The minimal broken parity model therefore 
cannot simultaneously explain the solar, atmospheric and LSND data. 
By contrast, the unbroken parity version can explain 
all of the neutrino anomalies. 
Furthermore, they argue that the unbroken case provides 
the most natural explanation of the neutrino physics
anomalies (irrespective of whether evidence from the 
LSND experiment is included) because of its characteristic 
maximal mixing prediction.
\subsection{$\Phi^4$ Theory and the Renormalization Group.}
If one starts with the view that a quantum description of reality is correct
then one must assume that a quantum field theory (QFT) description 
of reality will be correct,  as opposed to a classical 
(no Planck's constant $\hbar$) field theory description of reality.
There are only a limited number of classical field theories that can be 
quantized to produce a QFT,  the quantization schemes do not 
necessarily give a unique QFT.   
The reasons that classical field theories cannot be quantized are
either technical or that they give infinite results or both.
There might be classical theories which for a variety of reasons there 
is no corresponding quantum theory,   
starting with the view that only a quantum description of reality 
is fundamentally correct suggests that such theories 
are of no fundamental importance.
The simplest classical field theories involve only scalar fields 
and those that give finite QFT's have an action \ref{eq:simpleaction} 
which has lagrangian \ref{eq:simplelag} and potential \ref{eq:simplepot}.
In the potential
the $\phi^3$ term does not seem to correspond to anything,
thus $\phi^4$ theories are picked out as well defined
scalar QFT's.   Lagrangians of this form are used in symmetry breaking,
see the previous section \S2.10.
Langfeld {\it et al} \cite{LSR} (1995) discuss $\Phi^4$ theory 
and the Casimir effect.
Two recent papers on $\Phi^4$ theory are those below.

Borsanyi {\it et al} \cite{BVHS} (2000) study
thermalisation of configurations with initial white noise power spectrum 
in numerical simulations of a one-component $\Phi^4$ 
theory in 2+1 dimensions, coupled to a
small amplitude homogeneous external field. 
The study is performed for energy densities 
corresponding to the broken symmetry phase of the system in equilibrium. 
The effective
equation of the order parameter motion is reconstructed 
from its trajectory which starts from an initial value 
near the metastable point and ends in the stable ground state. 
They say that this phenomenological theory quantitatively 
accounts for the decay of the false vacuum. 
The large amplitude transition of the order parameter between 
the two minima displays characteristics reflecting the dynamical effect 
of the Maxwell construction. 

Sch\"utzhold {\it et al} \cite{SKMPS} (2000) show 
that the massless neutral $\lambda\Phi^4$-theory 
does not possess a unique vacuum. 
Based on the Wightman axioms the nonexistence of a state which preserves
Poincar{\'e} and scale invariance is demonstrated non-perturbatively 
for a non-vanishing self-interaction. 
They conclude that it is necessary to break the scale invariance in order
to define a vacuum state. 
They derive the renormalized vacuum expectation value 
of the energy-momentum tensor 
as well as the $\phi$-onic and scalar condensate from the
two-point Wightman function employing the point-splitting technique. 
They point out possible implications to other self-interacting field theories 
and to different approaches in quantum field theory. 

Kastening \cite{kast3} (1992) studies the renormalization
group running of the effective potential which includes a renormalization 
group running of the vacuum energy.
This is one of the
motivations for the computations in the four- and
five-loop papers of his \cite{kast2,kastening} discussed below.

{\bf Renormalization} is a process by which the infinities of QFT's 
can sometimes be removed,  
often such a process has some freedom which forms a ``group''.
Some recent work on the vacuum and this includes
Christie {\it et al} \cite{CES} (1999) who
use asymptotic Pade-approximant methods to estimate from prior orders of
perturbation theory the five-loop contributions to the coupling-constant 
$\beta$-function $\beta_g$, the anomalous mass dimension $\gamma_m$, the
vacuum-energy $\beta$-function $\beta_v$, and the anomalous dimension 
$\gamma_2$ of the scalar field propagator, 
within the context of massive N-component $\phi^4$ scalar field theory.
They compare these estimates with
explicit calculations of the five-loop contributions to $\beta_g$, $\gamma_m$,
$\beta_v$, and are seen to be respectively within 
5\%, 18\%, and 27\% of their true
values for $N$ between 1 and 5. 
They then extend asymptotic Pade-approximant methods to predict 
the presently unknown six-loop contributions to $\beta_g$,
$\gamma_m$, and $\beta_v$. These predictions, as well as the six-loop 
prediction for $\gamma_2$, provide a test of asymptotic Pade-approximant 
methods against future calculations. 

Kastening \cite{kastening} (1997) 
uses dimensional regularization in conjunction
with the MSbar scheme to analytically compute
the $\beta$-function of the vacuum energy density   
at the five-loop level in O(N)-symmetric 
$\phi^4$ theory, see \S2.11.
The result for the case of a cubic anisotropy is also given. 
It is pointed out how to also obtain the beta function 
of the coupling and the gamma function of
the mass from vacuum graphs. 
They say that this method may be easier than traditional approaches.
The four loop equivalent of this,  where the conventions are set is,
Kastening \cite{kast2}.
\subsection{Instantons and $\Theta$-Vacua.}
Instantons are defined by Kaku \cite{kaku} \S16.6 (1993) 
as finite action solutions to equations of motion with positive 
definite metric,  they allow tunneling between different vacua 
because they connect vacua at $x_4\rightarrow\pm\infty$ thus
the naive vacuum is unstable.   
The instanton allows tunneling between all possible vacua labeled by winding 
number $n$.   Thus the true vacuum must be a superposition of the various
vacua $|n>$ belonging to some different homotopy class.
The effect of a gauge transformation $\Om_1$ 
is to shift the winding number $n$ by one:
\be
\Om_1:|n>\rightarrow|n+1>.
\lb{k1}
\ee
Since the effect of $\Om_1$ on the true vacuum can change 
it only by an overall phase factor,  
this fixes the coefficients of the various vacua $|n>$ within
the true vacuum.   This fixes the coefficients of $|n>$ as follows:
\be
|vac>_\th=\sum^\infty_{n=-\infty}e^{in\th}|n>.
\lb{k2}
\ee

Some recent work on $\theta$-vacua includes
Yi \cite{yi} (1994) finds that the sign of the vacuum energy density 
effects the geometry of dilatonic extremal black holes.

Etesi \cite{etesi} (2000) studies
the existence of $\theta$-vacuum states in Yang-Mills theories 
defined over asymptotically flat, stationary spacetimes taking into 
account not only the topology but the complicated causal structure 
of these spacetimes. By a result of Chrusciel and Wald, 
apparently causality makes all vacuum states, 
seen by a distant observer, homotopically equivalent making the
introduction of $\theta$-terms unnecessary in causally 
effective Lagrangians. 
He claims that a more careful study shows that certain twisted classical 
vacuum states survive even in this case eventually leading to the conclusion
that the concept of ``$\theta$-vacua'' is meaningful in the case of general 
Yang-Mills theories. 
He gives a classification of these vacuum states
based on Isham's results showing that the Yang-Mills vacuum 
has the same complexity as in the flat Minkowskian case hence the
general CP-problem is not more complicated than the well-known flat one. 

Zhitnitsky \cite{zhitnitsky} (2000)
notes that it has been recently argued that an arbitrary induced 
$\theta$-vacuum state could be created in the heavy ion collisions, 
similar to the creation 
of the disoriented chiral condensate with an arbitrary isospin direction. 
It should be a large domain with a wrong $\theta$ 
orientation which will mimic 
the physics of the world when the fundamental $\theta$ is non-zero. 
He suggest a few simple observables which can hopefully be measured 
on an event by event basis at RHIC, 
and which uniquely determine whether 
the induced $\theta$-vacuum state is created. 

Halpern and Zhitnitsky \cite{HZ} (1998) 
suggest that the topological susceptibility in gluodynamics 
can be found in terms of the gluon condensate using renormalizability 
and heavy fermion representation of the anomaly. 
Analogous relations can be also obtained for other 
zero momentum correlation functions involving 
the topological density operator.
Using these relations, they find the $\theta$ dependence of the 
condensates $<GG>, <G \tilde{G}>$ and of the partition function 
for small $\theta$ and an arbitrary number of colours. 

Some recent work on instantons includes
Kochelev \cite{kochelev} (1995) who shows 
that specific properties of the instanton induced 
interaction between quarks leads to the anomalous violation 
of the OZI-rule in the $N\bar N\rightarrow \Phi\Phi$,
$N\bar N\rightarrow \Phi\gamma$ reactions. 
In the framework of instanton model of the QCD vacuum, 
the energy dependence of the cross sections of these reactions is
calculated. 

Yung \cite{yung} (1995) considers instanton dynamics in the broken phase of
the topological $\sigma$ model with the black hole metric 
of the target space. 
It has been shown before that this model is in the phase with BRST-symmetry
broken. 
In particular he says that vacuum energy is non-\-zero and correlation 
functions of observables show the coordinate dependence. 
However he says that these quantities turned out to be infrared (IR) 
divergent. 
Yung shows that IR divergences disappear after the sum 
over an arbitrary number of additional instanton-\-anti-\-instanton
pairs is performed. 
The model appears to be equivalent to Coulomb gas/Sine Gordon system.

Zhitnitsky \cite{zhitnitsky} (2000)
notes that it has been recently argued that an arbitrary induced 
$\theta$-vacuum state could be created in the heavy ion collisions, 
similar to the creation 
of the disoriented chiral condensate with an arbitrary isospin direction. 
It should be a large domain with a wrong $\theta$ 
orientation which will mimic 
the physics of the world when the fundamental $\theta$ is non-zero. 
He suggest a few simple observables which can hopefully be measured 
on an event by event basis at RHIC, 
and which uniquely determine whether 
the induced $\theta$-vacuum state is created. 
\subsection{Solitons,  Integrable Models and Magnetic Vortices.}
Solitons are solutions to differential equations which have well defined 
properties,  such as their energy does not disperse,  
see Roberts \cite{mdrsol} (1985).
Roughly speaking,  a soliton is a stable,  localized,  finite energy solution
to a classical equation.   The idea of a soliton has been extended to curved
spacetimes and reasonable necessary conditions for a 
classical solution to be a soliton are:
\begin{enumerate}
\item the solution is asymptotically flat and admits a 
one-dimensional symmetry group whose trajectories are time-like;\\
\item the energy density is localized and the total energy is finite;\\
\item the solution is classically stable;\\
\item the solution is quantum mechanically stable.
\end{enumerate}

In gravitational theory 
usually just Kerr-Newman solutions are considered as possible candidates for
solitons.   However,  generic Kerr-Newmann solutions have horizons and so are 
considered quantum mechanically unstable because of the Hawking effect.
Naked Kerr-Newmann solutions have negative energy and extreme Kerr-Newmann 
solutions are destroyed by the fermionic vacuum;   the static
spherically symmetric Einstein scalar fields are discussed in
Roberts \cite{mdrsol} (1985) do not possess event horizons.   
All the solutions discussed are 
asymptotically flat and static and so satisfy 1.   The qualitative features
of these solutions,  such as no event horizon,  remain,  no matter how small
the scalar field is and so they can be made to have a total energy arbitrary 
similar to the Schwarzschild solution,  and in any case their Tolman energy
is positive and they obey reasonable energy conditions
and so 2 is satisfied.   The solutions are static solutions
of the field equations for which perturbations do not change the qualitative
features and so 3 is satisfied.   It has been suggested that curvature 
always produces particles Roberts \cite{bi:mdr86} (1986).   In this case there
would be no quantum mechanical stability.   However,  it is usually thought
that only spacetimes with event horizons produce particles,  but Einstein 
scalar solutions do not have event horizons and so satisfy 4.   Whether
there is a mechanism by which the Einstein scalar solutions can excite the
vacuum remains to be investigated.

Some recent work on solitons and the vacuum include
M\"uller-Kirsten {\it et al} \cite{MTTZ} (1998) who
suggest the the shift in vacuum energy in the O(3) Skyrme
model is due to instantons and Moss \cite{moss} (1999) who relates the
the quantum properties of solitons at one loop 
to phase shifts of waves on the soliton background. 
These can be combined with heat kernel methods to 
calculate various parameters. 
The vacuum energy of a CP(1) soliton in 2+1 dimensions 
is calculated as an example. 

Some two dimensional theories can be solved completely and are called integral
models.   Some recent work on the vacuum in integrable models include 
the paper of Delfino {\it et al} \cite{DMS} (1996) who 
approaches the study of non--integrable models 
of two--dimensional quantum field theory as perturbations 
of the integrable ones. By exploiting the knowledge of the exact
$S$-matrix and form factors of the integrable field theories 
they obtain the first order corrections to the mass ratios, 
the vacuum energy density and the $S$-matrix of the
non-integrable theories. 
As interesting applications of the formalism Delfino {\it et al} \cite{DMS} 
study the scaling region of the Ising model 
in an external magnetic field at $T \sim T_c$ and the scaling
region around the minimal model $M_{2,7}$. 
For these models they observe a remarkable agreement 
between the theoretical predictions 
and the data extracted by a numerical
diagonalization of their Hamiltonian. 

Babansky,A.Yu and  Sitenko,Ya,A. \cite{BSit} (1999) discuss vacuum energy 
induced by a singular magnetic vortex.
Langfeld {\it et al} \cite{LRT} (1998) note that 
the magnetic vortices which arise in SU(2) lattice gauge theory 
in center projection are visualized for a given time slice. 
They establish that the number of vortices piercing a given
2-dimensional sheet is a renormalization group invariant 
and therefore physical quantity. 
They find that roughly 2 vortices pierce an area of 1 $fm^2$. 
\subsection{Topological Quantum Field Theory.}
There are a variety of objects which can be ``quantized'',  i.e. starting with
a classical (no $\hbar$) system one can produce a quantum system which 
corresponds to it.   Topology is the study of shapes and one can 
ask whether shapes can be quantized;  in particular
Witten \cite{witten} (1988) notes that
topological quantum field theory has the formal 
structure of a quantum theory
(e.g. dealing with probabilities),
but that the information they produce is purely topological
(e.g. information about the nature of a knot).
Topological QFT's have been recently reviewed by Schwarz \cite{schwarz} (2000).
That Casimir energy might be topological is discussed in 
Williams \cite{williams} (1997).
\subsection{SAZ Approach.}
A way of approaching
the quantum vacuum is the Sakarov \cite{sakharov}(1967)
-Adler \cite{adler} (1983) -Zee \cite{zee} (1983) 
approach (explained for example in Misner,  Thorne and Wheeler 
\cite{MTW} page 426 (1970)) where the quantum vacuum again induces
a cosmological constant.
Their approaches are covariant and so are compatible with general
relativity,  no doubt there are approaches that are not covariant.
Because of covariance the quantum vacuum problem and the inertia 
problem become separate issues.   
Pollock and Dahder \cite{PD} (1989) discuss how the SAZ 
approach fits in with inflation.

Belgiorno and Liberati \cite{BL} (1997) show 
an analogy between the subtraction procedure in the Gibbons-Hawking 
Euclidean path integral approach to horizon's thermodynamics 
and the Casimir effect. 
Then they conjecture about a possible Casimir nature of the 
Gibbons-Hawking subtraction is made in the framework of 
Sakharov's induced gravity. 
In this framework it appears that the
degrees of freedom involved in the Bekenstein-Hawking 
entropy are naturally identified with zero--point modes of the matter fields. 
They sketch some consequences of this view. 

Consoli \cite{consoli} (2000) notes that 
the basic idea that gravity can be a long-wavelength effect {\it induced} 
by the peculiar ground state of an underlying quantum field theory leads 
to consider the implications of
spontaneous symmetry breaking through an elementary scalar field. 
He point out that Bose-Einstein condensation implies the existence 
of long-range order and of a gap-less
mode of the (singlet) Higgs-field. 
This gives rise to a $1/r$ potential and couples 
with infinitesimal strength to the inertial mass of known particles. 
If this is interpreted as the
origin of Newtonian gravity one finds 
a natural solution of the hierarchy problem. 
As in any theory incorporating the equivalence principle, 
the classical tests in weak gravitational
fields are fulfilled as in general relativity. 
On the other hand, our picture suggests that 
Einstein general relativity may represent 
the weak field approximation of a theory generated
from flat space with a sequence of conformal transformations. 
This explains naturally the absence of a {\it large} 
cosmological constant from symmetry breaking. 
Finally, one also predicts new phenomena that 
have no counterpart in Einstein theory such as 
typical `fifth force' deviations below the centimeter scale 
or further modifications at distances
$10^{17}$ cm in connection with the Pioneer anomaly 
and the mass discrepancy in galactic systems. 

Guendelman and Portnoy \cite{GPort} (1999) 
consider a model of an elementary particle as a 2 + 1 
dimensional brane evolving in a 3 + 1 dimensional space. 
Introducing gauge fields that live in the brane
as well as normal surface tension can lead to 
a stable "elementary particle" configuration. 
Considering the possibility of non vanishing vacuum energy inside
the bubble leads, when gravitational effects are considered, 
to the possibility of a quantum decay of such "elementary particle" 
into an infinite universe. 
Some remarkable features of the quantum mechanics of this process 
are discussed, in particular the relation between possible boundary 
conditions and the question of instability towards Universe 
formation is analyzed. 

Guendelman \cite{guendelman} (1999) discusses
the possibility of mass in the context of scale-invariant, 
generally covariant theories. 
Scale invariance is considered in the context of a
gravitational theory where the action, in the first order formalism, 
is of the form $S = \int L_{1} \Phi d^4x$ + $\int L_{2}\sqrt{-g}d^4x$ 
where $\Phi$ is a
density built out of degrees of freedom independent of the metric. 
For global scale invariance, a "dilaton" $\phi$ has to be introduced, 
with non-trivial potentials
$V(\phi)$ = $f_{1}e^{\alpha\phi}$ in $L_1$ and 
$U(\phi)$ = $f_{2}e^{2\alpha\phi}$ in $L_2$. 
This leads to non-trivial mass generation and a potential for
$\phi$ which is interesting for inflation. 
The model can be connected to the induced gravity model 
of Zee \cite{zee} (1983), 
which is a successful model of inflation.
Models of the present universe and a natural transition from 
inflation to a slowly accelerated universe at late times are discussed. 
\subsection{Superfluids and Condensed Matter.}
Fields are not the only objects which one can think 
of as occupying spacetime,  there are also fluids;  
when there is no equation of state specified 
they are more general than fields.
Quantum fluids as usually understood are a limited class 
of object used to discuss low temperature phenomena.
There are many analogies between such ``superfluids'' and QFT's,  
see Volkovik \cite{volovik2} (2000). 
So far there has been no contact between low temperature superfluids and the
quantized fluids discussed by Roberts \cite{mdrqc} (1999).
Some recent work on vacua and superfluids includes
Fischer {\it et al} \cite{FSFFS} (1993) who find that all systems under their
investigation exhibit an occupied alkali-mixture state near the Fermi energy
as well as a series of unoccupied states converging towards the vacuum energy,
which are identified as image-potential states;  
thus in this case vacuum energy is associated with a state of the system.
Duan \cite{duan} (1993) uses the concept of finite compressibility of a Fermi
superfluid to reconsider the problem of inertial mass of vortex lines in both 
neutral and charge superfluids at zero temperature $T=0$.

Chapline \cite{chapline} (1998) shows 
that a simple model for 4-dimensional quantum gravity 
based on a 3-dimensional generalization of anyon superconductivity 
can be regarded as a discrete form of Polyakov's string theory. 
This suggests that there is a universal negative pressure that 
is on the order of the string tension divided by the
square of the Robertson-Walker scale factor. 
This is in accord with recent observations 
of the brightness of distant supernovae, which suggest that at the
present time there is a vacuum energy whose magnitude 
is close to the mass density of an Einstein-de Sitter universe. 

Volovik \cite{volovik2} (2000) notes that
superfluid 3He-A gives an example of how chirality, Weyl fermions, 
gauge fields and gravity appear in low energy corner together 
with corresponding symmetries, including Lorentz
symmetry and local SU(N). This supports idea that quantum field theory 
Standard Model or GUT is effective theory describing low-energy phenomena.
{\it Seven} reasons for this are:  
{\it firstly} momentum space
topology of fermionic vacuum provides topological stability 
of universality class of systems, where above properties appear. 
{\it Secondly} BCS scheme for 3He-A incorporates both
``relativistic'' infrared regime and ultraviolet ``transPlanckian'' range: 
subtle issues of cut-off in quantum field theory and anomalies 
can be resolved on physical grounds. 
This allows to separate ``renormalizable'' terms in action, 
treated by effective theory, 
from those obtained only in ``transPlanckian'' physics. 
{\it Thirdly} energy density of superfluid vacuum
within effective theory is ~ $E_{Planck}^4$. 
Stability analysis of ground state beyond effective 
theory leads to exact nullification of vacuum energy: 
equilibrium vacuum is not
gravitating. In nonequilibrium, vacuum energy is of order energy density 
of matter. 
{\it Fourthly} 3He-A provides experimental prove for anomalous nucleation 
of fermionic charge according to Adler-Bell-Jackiw. 
{\it Fifthly} helical instability in 3He-A is described by the same equations 
as formation of magnetic field by right electrons in 
Joyce-Shaposhnikov scenario. 
{\it Sixthly} macroscopic parity violating effect and angular momentum paradox 
are both described by axial gravitational Chern-Simons action. 
{\it Seventhly} high energy dispersion of quasiparticle
spectrum allows treatment of problems 
of vacuum in presence of an event horizon.

Elstov {\it et al} \cite{EKKRV} (2000) note that
spin-mass vortices have been observed to form in rotating superfluid 
3He-B following the absorption of a thermal neutron and a rapid
transition from the normal to superfluid state. 
The spin-mass vortex is a composite defect 
which consists of a planar soliton (wall) which
terminates on a linear core (string). 
This observation fits well within the framework 
of a cosmological scenario for defect formation, 
known as the Kibble-Zurek mechanism. 
It suggests that in the early Universe analogous 
cosmological defects might have formed. 

Fischer {\it et al} \cite{FSFFS} (1993) note that high-resolution 
photoemission and two-photon photoemission spectroscopy have been 
used for a comparative study of the electronic structure of a single layer 
of $Na$ and $K$ on $Cu(111)$,  $Co(0001)$ and $Fe(110)$.
All of the systems under investigation exhibit an occupied alkali-induced 
state near the Fermi energy as well as a series of unoccupied states 
converging toward the vacuum energy,  which are identified as 
image-potential states....
\section{The Casimir and Related Effects.}
\subsection{Introduction.}
Lamoreaux \cite{lam1} (1999) provides an introductory guide to the literature 
on the Casimir effect with 368 references;
he points out that there are over a $1,000$ references on this subject 
and the number doubles every five years;
in part this is due to what is called the ``Casimir effect'' changing,
contemporary usage does not necessarily require fixed plate boundaries.
The Casimir \cite{casimir} (1948) effect can be characterized
as a force that arises when some of the
normal modes of a zero rest-mass field such as the electromagnetic field 
are excluded by boundary conditions.
If one places in Minkowski space-time two parallel flat perfect conductors,
then the boundary conditions on the conductors ensure that normal modes whose
wavelength exceeds the spacing of the conductors are excluded.
If now the conductors are moved slightly apart,  new normal modes are
permitted and the zero-point energy is increased.   
Work must be done to achieve this energy increase,  
and so there must be an attractive force between the plates.
This force has been measured,  and the zero-point calculation verified.   
This agreement with experiment is important,  
since it shows that calculations with zero-point energy 
of a continuous field does have some correspondence with reality,  
although the total energy associated with modes 
of arbitrarily high frequency is infinite.
The Casimir effect shows that finite differences between
different configurations of infinite energy do have physical reality.
Other example of this principle for zero-point effects is the Lamb shift 
and the Toll-Scharnhorst effect, see \S2.3 above.
\subsection{History of the Casimir Effect.}
This subsection largely follows Fulling \cite{fulling} (1989).   
Casimir \cite{casimir} (1948) tried to calculate the van der Waals force 
between two polarizable atoms.   The charge fluctuations in one atom can
create an electric field capable of polarizing the charge in the other atom,
so that there is a net force between them.   To simplify the analysis,
Casimir \cite{casimir} (1948) considered a similar force between an atom and
a conducting plate.   From there he was led to the problem of two parallel
plates.   It was found that to get agreement with experiment in any of these
problems,  it was necessary to take into account the finiteness of the speed
of light,  which implies that the influence of charge fluctuations in one
part of the system reaches the other part only after a delay.   
This means that
the energy stored in the electromagnetic field passing between the two bodies
must be taken into account.   In fact,  it turned out that the {\it long range
limit} of the force can be associated {\it entirely} with the change in the 
energy of the field as the distance between the bodies varies;  the charge 
fluctuations faded into the background of the analysis.

Formally the field energy is a sum over the divergent modes of the field.
Casimir calculated the energy change $\Delta E$, by inserting an 
{\it ad hoc} convergence factor,  usually $e^{-\omega_n/\Lambda}$ 
with $\Lambda$ a large constant,  to make the two energies finite.   
Their difference, $\Delta E$, 
has a finite limit as $\Lambda\rightarrow\infty$.
Replacing $\omega_n$ by $n$ in the exponential gives a different answer.   
This exponential cutoff is algebraically equivalent to an analytic 
continuation of the point separation along the time axis:  
as can be seen by taking $\Lambda=i(t-t')$.

The physical interpretation offered for this procedure was this - no physical
conductor is perfect at arbitrarily high frequencies;  a very high-frequency 
wave should hardly notice the presence of the plates at all.   Therefore,
in a real experiment $\Delta E$ {\it must} be finite.   An integral or sum
defining it must have an effective cutoff depending on the detailed physics
of the materials;  $e^{-\omega_n/\Lambda}$ is a plausible model.   
The fact that the result is independent of $\Lambda$ in the limit of large
$\Lambda$ suggests that this model is roughly correct 
and a more detailed model
is unnecessary for a basic understanding.

This paragraph largely follows Lamoreaux \cite{lamoreaux} (1999). 
For conducting parallel flat plates separated by distance $r$,
this force per unit area $A$ has the magnitude Casimir \cite{casimir} (1948)
\be
\frac{F(r)}{A}=\fr{\pi^2}{240}\fr{\hb c}{r^4}
\approx \fr{1.3\times10^{-2}}{r^4}{\rm dyn(\mu m)^4.cm.^2}.
\lb{lamc}
\ee
Casimir derived this relation by considering the electromagnetic mode structure
between the two parallel plates of infinite extent,  as compared to the mode
structure when the plates are infinitely far apart,  and by assigning a 
zero-point energy of $1/2\hb\omega$ to each electromagnetic mode (photon).
The change in total energy density between the plates,  as compared to 
free space,  as a function of separation $r$,  leads to the force of 
attraction.   This result is remarkable partly because it was one of 
the first predictions of a physical consequence directly due to zero-point 
fluctuations,  and was contemporary with,  but independent of,  
Bethe's treatment of the Lamb shift.
Lifshitz \cite{lifshitz} (1956) first developed the theory
for the attractive force between two plane surfaces made of a material with
generalized susceptibility.

The attraction between smooth,  
very close surfaces was eventually demonstrated
experimentally Tabor and Winterton \cite{TW} (1969).  
For technical reasons they used dielectric materials instead of conductors.
The theory of the effect for dielectrics is more complicated 
than for conductors but leads to similar results.

The existence of pointlike charged particles has always been
problematic within the framework of classical electrodynamics.
In view of the negative vacuum energy in the configuration of conducting
parallel plates,  Casimir proposed a model of the electron as a charged
sphere with properties like those of a microscopic conductor.   
The hope was that an attractive force arising from the dependence of the vacuum
energy on the sphere's radius would balance the electrostatic self-repulsion
of the charge distribution,  thereby holding the electron together stably.
Boyer \cite{boyer} (1970) and Davies \cite{davies} (1972) succeeded in 
calculating a conducting spherical shell's vacuum energy.
Boyer's results were fatal to the Casimir electron model;
the Casimir force turned out to be {\it repulsive} in this case:
\be
F=-\frac{\p E}{\p R}>0.
\ee 
Furthermore,  he found that the energy associated with the presence of a 
spherical conducting shell is infinite;  that is,  the energy difference 
between a configuration with an inner shell and one without did not converge 
as the cutoff was removed.   In the presence of a {\it curved} conducting
boundary the electromagnetic field behaves as the scalar field with canonical
stress tensor at a flat boundary.

The rise of interest in quantum field theory in curved spacetime in the 1970's
attracted renewed attention to the Casimir effect,  as a more tractable model
of field-theoretical effects associated with the geometry of space.
It was in this new era that calculations of the energy-momentum tensor,
not just the total energy,  were made,  and the question of the geometric
covariance of the cutoff procedure was raised.

Among those who approached the subject from a general-relativistic motivation
were Deutsch and Candelas \cite{DC} (1979).   
They consider boundary conditions
({\it not} quantization effects) for general curved surfaces,  
and for several types of quantum field theory.   
They avoided eigenfunction expansions by working directly with 
the Green functions of the elliptic operators involved.

For $x\in \Omega$ and close to the boundary of $\Omega$,  let $\xi$ be the 
point on the boundary closest to $x$ and $\epsilon$ be the distance from $x$
to $\xi$.  Then $(\xi,\epsilon)$ provides a convenient coordinate system in the
vicinity of the smooth boundary.   Deutch and Candelas \cite{DC} (1979) 
{\it assume} that near 
the boundary the renormalized stress tensor has an expansion of the form
\be
T_{\mu\nu}(x)\sim\Sigma^\infty_{n=-4}A^{(n)}_{\mu\nu}(\xi)\epsilon^n.
\ee
Note that this is consistent with Fulling's \cite{fulling} (1989)
findings for the flat plate,  where $\epsilon=z$. 
Deutch and Candelas \cite{DC} also {\it assume} that $A^{(n)}(\xi)$ 
depends only on the geometry of the boundary at $\xi$;  
that is,  it can be expressed in terms of a function defining the surface 
and its derivatives,  evaluated at $\xi$ only.
Geometrical covariance and dimensional analysis 
then imply that each $A^{(n)}$ is a linear combination 
of finitely many scalar quantities built out of the second fundamental form 
of the surface at $\xi$.

Deutch and Candelas \cite{DC} (1979) 
were then able to calculate the coefficients in their 
general series by matching it against various special cases for which 
the answers were known or could be easily found.  For $\Omega\subset{\cal R}^3$
they find a hierarchy of terms.   
Higher-order terms are of the order $\epsilon^0$,  hence they are nonsingular.
The conclusion of Deutch and Candelas \cite{DC} and \cite{fulling} 
is that the finiteness of the 
Casimir force for the slab and the sphere is an accident of there geometries.
In general,  the electromagnetic force on a perfect conductor will turn out to
be infinite.  A spherical shell is unstable against wrinkling.
The perfect-conductor boundary condition must be judged to be a pathological
idealization in this context.  Good physics requires that the infinite terms 
be replaced by {\it cutoff-dependent} terms related to the detailed 
properties of realistic materials.

Rosu \cite{rosu} (1999) notes that if
stationary, the spectrum of vacuum field noise (VFN) 
is an important ingredient to get information 
about the curvature invariants of classical worldlines 
(relativistic classical trajectories). 
For scalar quantum field vacua there are six stationary cases 
as shown by Letaw some time ago, these are reviewed here. 
However, the non-stationary vacuum
noises are not out of reach and can be processed 
by a few mathematical methods which he briefly comments on. 
Since the information about the kinematical curvature invariants of
the worldlines is of radiometric origin, 
hints are given on a more useful application 
to radiation and beam radiometric standards at relativistic energies.
\subsection{Zero-Point Energy and Statistical Mechanics.}
This section largely follow Sciama \cite{sciama} (1991),  a different approach
can be found in Lima and Maia \cite{limamaia} (1995).
Usually the boundary conditions associated with a physical system
limit the range of normal modes that contribute to the ground state of
the system and so to the zero-point energy.   An example is the
harmonic oscillator,  which has a single normal mode of
frequency $\nu$ and so has a zero-point energy of $\fr{1}{2}h\nu$.   
In more complicated cases the range of normal modes may depend 
on the configuration of the system.   This would lead to a dependence
of the ground-state energy on the variables defining the configuration and so,
by the principle of virtual work,  to the presence of an associated set of
forces.   One important example of such a force is the homopolar binding
between two hydrogen atoms when their electric spins are antiparallel, 
see Hellman \cite{hellman} (1927).   When the protons are close together,  
each electron can occupy the volume around either proton.   
The resulting increase in the uncertainty of the electron's position leads 
to a decrease in the uncertainty of its momentum and so to a {\it decrease} in
its zero-point energy.   Thus,   there is a binding energy associated 
with this diatomic configuration,  and the resulting attractive force 
is responsible for the formation of the hydrogen molecule.
By contrast,  when the electron spins are parallel,  the Pauli exclusion 
principle operates to limit the volume accessible to each electron.
In this case the effective force is repulsive.

Zero-point fluctuations occur
when the associated density of states is large,  
for example in a continuous field.   
In such a case the zero-point fluctuations 
can be an important source of {\it noise} and {\it damping},
these two phenomena being related by a fluctuation-dissipation theorem.
Some examples of these zero-point effects are:
{\it firstly} X-ray Scattering by Solids.
The noise associated with zero-point fluctuations was discovered in
1914 by Debye during his study of X-ray scattering by solids.   His
main concern was to calculate the influence of the thermal vibrations
of the lattice on the X-ray scattering,  but he showed in addition that,  if
one assumes with Planck that the harmonic oscillators representing
these vibrations have a zero-point motion,  then there would be an
additional scattering which would persist at the absolute zero of
temperature.   This additional scattering is associated with the
emission of phonons by X-rays,  this emission being induced by the 
zero-point fluctuations.   Thus,  noise and damping are related together
as in Einstein's theory of Brownian motion.   Here the 
example of a 'spontaneous' radiation process,  which can be regarded
as being induced by the coupling of the radiating system to a field of
zero-point fluctuations.   
{\it Secondly} Einstein fluctuations in black body radiation.
Sciama \cite{sciama} (1991) notes in passing here that the interference 
between the zero-point 
and thermal fluctuations of the electromagnetic field gives 
a characteristic contribution to the Einstein fluctuations 
of the energy in a black body radiation field,
namely the term 'linear' in the energy density.
This term would be absent for a classical radiation 
field pictured as an assembly of waves.
{\it Thirdly} the  Lamb Shift.
An electron,  whether bound or free, is always subject 
to the stochastic forces produced by the zero-point fluctuations of the 
electromagnetic field,  and as a result executes Brownian motion.
The kinetic energy associated with this motion is infinite,
because of the infinite energy in the high-frequency components 
of the zero-point fluctuations.
This infinity in the kinetic energy can be removed by renormalizing the
mass of the electron Weisskopf \cite{weisskopf} (1949).
As with the Casimir effect,  physical significance can be given 
to this process in situations where one is dealing with different 
states of the system for which the difference in the total 
renormalized Brownian energy (kinetic plus potential) is finite.
An example of this situation is the Lamb shift between the 
energies of the $s$ and  $p$ electrons in the hydrogen atom;  
according to Dirac theory,  the energy levels should degenerate.
Welton \cite{welton} (1948) pointed out that a large part of this shift 
can be attributed to the effects of the induced Brownian motion of the 
electron,  which alters the mean Coulomb potential energy.   
This change in electron energy is itself different for 
an $s$ and $p$ electron,  
and so the Dirac degeneracy is split.   
This theoretical effect has been well verified by the observations.
One can also regard the Lamb shift as the change in zero-point energy 
arising from the dielectric effect of introducing a dilute distribution 
of hydrogen atoms into the vacuum.   
The frequency of each mode is simply modified by a refractive index factor, 
see Feynman \cite{feynman} (1961),
Power \cite{power} (1966), and
Barton \cite{barton} (1970).

The vacuum can be considered to be just a dissipative system.
A physical system containing a large number of closely spaced modes 
behaves as a dissipater of energy,  as well as possessing 
fluctuations associated with the presence of those modes.
Now the vacuum states of the electromagnetic field constitute an example
of a system with closely spaced energy levels,
there being $(8\pi/c^3)\nu^2{\rm d}\nu$ modes with frequencies lying 
between $\nu$ and $\nu+{\rm d}\nu$.
Sciama \cite{sciama} (1991) expects this state also to possess 
a dissipative character
related to the zero-point fluctuations;
Callen and Welton \cite{CW} (1951) proved this in 
their quantum mechanical derivation of the fluctuation-dissipation theorem.
In this derivation the dissipation is represented simply by the 
{\it absorption} of energy by the dissipative system.   
This can be justified by the expectation that the energy absorbed 
is divided up among so many modes that the possibility 
that it is later re-admitted into its original modes with its initial 
phase relations intact can be neglected.
The absorption rate is calculated by second-order perturbation theory and 
is found to depend quadratically on the external force acting on the system.
This enables an impedance function to be defined in the usual way.   
This function also appears in a linear relation between the external 
force acting on the system and the response of the system to this force.
A familiar example would be the relation 
between an impressed electric field,  
the resulting current flow,  and the electrical resistance of the system.   
The rate of energy absorption,  and hence the resistance, clearly depends on 
the coupling between the external disturbance and the system,  
and on the density of the states of the system.

Blasone {\it et al} \cite{BJV} (2000) note the proposal that information loss 
in certain Casimir systems might lead to an apparent quantization 
of the orbits which resemble the quantum structure seen in the real world.
They show that the dissipation term in the Hamiltonian for a couple of
classically damped-amplified oscillators manifests itself as a 
geometric phase and is actually responsible for the appearance of zero-point 
energy in the quantum spectrum of the $1D$ linear harmonic oscillator.   
They also discuss the thermodynamical features of their system.

Callen and Welton \cite{CW} calculate 
the quantum fluctuations 
of a dissipative system in its unperturbed state.   
These fluctuations also depend on the density of states.   
Apparently the procedure is quite general,  but for simplicity they restrict 
themselves to the case where the system is in thermal equilibrium 
at temperature $T$,  so that its states are occupied in accordance 
with the Boltzmann distribution $e^{-h\nu/kT}$.
By eliminating the density of states,  the following relation is obtained
between the mean square force $<V^2>$ associated with the fluctuations
and the frequency-dependent impedance function, $R(\nu)$:
\be
<V^2>=\fr{2}{\pi}\int^\infty_0R(\nu)E(\nu,T){\rm d}\nu,
\ee
where
\be
E(\nu,T)=\fr{1}{2}h\nu+\fr{h\nu}{e^{h\nu/kT}-1},
\ee
which is the mean energy of a harmonic oscillator at temperature $T$.
The presence of the zero-point $\fr{1}{2}h\nu$ shows that the zero-point 
fluctuations (as well as the thermal fluctuations) are a source 
of noise power and damping,  and therefore satisfy 
the fluctuation-dissipation theorem.

This theorem can be interpreted in three ways.
\begin{enumerate}
\item It shows that the damping rate $R$ is determined 
by the equilibrium fluctuations $<V^2>$,
Nyquist \cite{nyquist} (1928) relation.

\item It shows that the noise {\it power} $<V^2>$ is determined 
by the absorption coefficient $R$ (Kirchhoff relation).

\item It shows how the damping rate is proportional to the density of states
in the dissipative system.   In fact,  the mean square force $<V^2>$ is
proportional to the mean square electric field,  which in turn is proportional 
to the energy of the dissipative system.   This energy is given by the Planck
distribution plus the zero-point contribution.   This is the content of the
fluctuation-dissipation theorem when $R(\nu)$ is regarded as proportional to
the density of states,  that is proportional to $\nu^2{\rm d}\nu$.
When the dissipative system is the vacuum $(T=0)$,  this remains true.
Thus,  the impedance of the vacuum is proportional to $\nu^2$.
\end{enumerate}

An important illustration of these ideas was also given by Callen and Welton 
\cite{CW} (1951),  who showed that the {\sc radiation damping} 
of an accelerated charge could be interpreted in this way.   
The non-relativistic form of this damping force is
\be
\fr{2}{3}\fr{e^2}{c^3}\fr{{\rm d}^2v}{{\rm d}t^2},
\ee
and if the charge is oscillating sinusoidally with frequency $\nu$,
then the associated impedance function turns out to be
\be
\fr{2}{3}\fr{e^2}{c^3}\nu^2.
\ee
This is precisely of the expected form,  that is,  proportional to $\nu^2$,
the impedance of the vacuum.   Thus,  the dependence of the damping force on
${\rm d}^2v/{\rm d}t^2$ is simply a manifestation of the density of states
for the vacuum electromagnetic field.

Another example is Weber's \cite{weber} (1954) 
discussion of the {\sc zero-point noise of an electric circuit}.
One can quantize such a circuit and show that its zero-point fluctuations
represent a measurable source of noise.   For example,  a beam of electrons
passing near such a circuit would develop a noise component from this source.
Associated with this noise is a damping of the electron beam caused by
'spontaneous' emission by the beam into the circuit.
The zero-point noise is thus physically real,  
and is unaffected by a change in the origin of the energy 
which might be introduced to remove in a formal way 
the infinite total energy associated with the zero-point fluctuations 
of arbitrary high frequency.

The work of Callen and Welton \cite{CW} (1951) 
and of Weber \cite{weber} (1954)
was generalized by Senitzky \cite{senitzky60} (1960),
who studied the {\sc damping of a quantum harmonic oscillator} 
coupled to a loss mechanism (reservoir) 
idealized as a system whose Hamiltonian 
is unspecified but which possesses a large number 
of closely spaced energy levels.   
Senitzky assumed that the coupling was switched on at a time $t_0$,
so that for $t<t_0$ the harmonic oscillator executed its zero-point 
motion undisturbed by the reservoir.
After $t_0$ the fluctuating forces exerted by the reservoir would 
damp out the zero-point motion of the oscillator by a now familiar mechanism.
To avoid conflict with the Heisenberg uncertainty principle,  
he required that the zero-point fluctuations of the reservoir 
also introduce sufficient noise into the oscillator to restore 
its zero-point motion to the full quantum mechanical value $\fr{1}{2}h\nu$.
Senitzky (1960) \cite{senitzky60} showed that after a few damping times 
the zero-point motion of the oscillator would be effectively driven 
by the zero-point motions of the reservoir.   
This is a 
quantum example of the relation between noise,  damping,  and equilibrium 
which Einstein discovered in 1905.

{\sc Response theory} was initiated by Onsager \cite{onsager} (1931),  
who extended Einstein's 
theory of Brownian motion by taking into account the perturbation 
of the dissipative system by the system being dissipated.
He related this perturbation to the equilibrium fluctuations of the 
unperturbed dissipative system by an ansatz which was very much 
in Einstein's spirit.
He assumed that if,  as a result of these fluctuations,  
the system at one instant deviated appreciably from its mean configuration,
then on average it would regress back to the mean at the same rate
as if the deviation had been produced by an external perturbation.
This average regression would represent an irreversible approach to 
equilibrium and so would determine the generalized friction coefficient 
associated with the response of the system to the external perturbation.
In the approximation he envisaged,  
the perturbation would have a linear effect,
and as an example consider an electric current as the response 
to an impressed voltage.
The linear coefficient of this response,  the resistance,  
not only would govern the rate at which the current would die 
irreversibly away after the voltage is removed,  
but also would govern the rate of dissipation 
associated with Joule heating.   
With Onsager's ansatz it is expected that this resistance is determined by 
the equilibrium fluctuations of the unperturbed system,  
giving a further reason for obtaining a 
fluctuation-dissipation relation.

Of course, the friction coefficient determines only the 
out-of-phase response of the system.
However,  one would also expect the in-phase response,  
that is the reactance of the system,  
to be determined by the equilibrium fluctuation spectrum.
This follows from the dispersion relations which are a direct consequence
of the causality requirement that the response of the system 
should not precede the disturbance of it.
The reactance at any frequency can then be determined by an integral 
of the friction overall frequencies;  
the friction is,  of course,  itself determined by the fluctuation spectrum.

One can also calculate the total response directly:  
this was first done by Kubo \cite{kubo} (1957).
Kubo used the fact that an averaging of an observable over a system 
in thermal equilibrium involves multiplying the observable by $E^{-H/kT}$,
where $H$ is the Hamiltonian of the system and $T$ is the temperature.
This is very similar to multiplying the observable by a quantum mechanical 
time-evolution operator,  with the temperature acting as the reciprocal 
of an imaginary time.
By exploiting this analogy,  and using complex variable methods,
Kubo arrived at the following fluctuation-response relation:
\be
\sigma_{ab}(\nu)=P(\nu,T)\int^\infty_{-\infty}
                    e^{-i\nu t}<j_a(t)j_b(0)>{\rm d}t,
\ee
where the complex conductivity $\sigma_{ab}$ is given in terms of an external 
field $E_b(\nu)$ by
\be
j_a=\sigma_{ab}B_b,
\ee
$P(\nu,T)$ is the Plankian function (with zero-point contribution),
\be
P(\nu,T)=\fr{1-e^{-h\nu/kT}}{2\nu},
\ee
and $<j_a(t)j_b(0)>=<j_a(0)j_b(t)>$ represents the quantum thermal correlation
function of the unperturbed current fluctuations in the system.
Comparing with equation \ref{defB}
\be
2B(\nu,T)P(\nu,T)=h\exp{-\fr{h\nu}{kT}}.
\ee
Note,  in particular,  that by microscopic reversibility,
\be
<j_a(t)j_b(0)=<j_a(0)j_b(t)>,
\ee
so that
\be
\sigma_{ab}=\sigma_{ba}~~~;
\ee
these are just Onsager's \cite{onsager} (1931) reciprocal relations.

Some recent work on statistical mechanics and the Casimir effect includes 
Power and Thirunamachandran \cite{PT} (1994) who
find in a systematic way the fully retarded dispersion interaction potentials,
including many-body interactions,  
among neutral molecules.
The method used relates the total zero-point energy of all the 
electromagnetic modes with the spectral sum of a linear operator.
The difference between the zero-point energies with and without 
molecules present is given as a contour integral.
From the value of this integral it is possible to extract 
the N-body dispersion energy by locating those terms which 
depend on the product of the polarizabilities of those N molecules.
The Casimir-Polder pairwise energy is the two-body result.
General formulas are found,  and the special cases for N=3 and 4 are
discussed in detail.
The non-retarded interaction potentials are found as their asymptotic limits 
for small intermolecular separations,
and the London and Axilrod-Teller results are the N=2 and N=3 special cases.
The N=4 near zero limit is presented in its explicit form. 
It is of interest to note that,  for the one-body case,
the energy shift given by this method is the nonrelativistic
Lamb shift.

Herzog and Bergou \cite{HB} (1997)
investigate specific nonclassical maximum-entropy states of a simple
harmonic-oscillator mode that arise when only number states (Fock states) 
differing by a multiple of a certain integer $k(k\ge1)$ 
are allowed to be occupied.
For $k=2$ the number-probability distribution of the even-number maximum-entropy
state has a close resemblance to that of a squeezed vacuum state.
These maximum entropy states can be obtained as the stationary solutions
of a master equation which takes into account $k$-quantum absorption as
well as $k$-quantum emission processes only.   The steady-state
solution of this master equation depends on their initial conditions.
For the vibrational motion of a trapped ion such nonclassical
maximum-entropy states could be produced with the help of the 
recently proposed method of laser-assisted quantum reservoir engineering.

Rosu \cite{rosu} (2000) discusses vacuum field noise,
see the end of \S3.2 above.

Some recent work on the Casimir effect 
and the liquid-crystal ground state includes
Kivelson {\it et al} \cite{KFE} (1998) who notes that
the character of the ground state of an antiferromagnetic insulator is
fundamentally altered following addition of even a small amount of charge.
The added charge is concentrated into domain walls across which a $\pi$ phase
shift in the spin correlations of the host material is induced.
In two dimensions,  these domain walls are 'stripes' which can be insulating or
conducting - that is,  metallic 'rivers' with their own low -energy degrees
of freedom.   However,  in arrays of one-dimensional metals,  which occur
in materials such as organic conductors,  interactions between strings
typically drive a transition to an insulating ordered charge-density-wave
(CDW) state at low temperatures.   
Here they show that such a transition is
eliminated if the zero-point energy of transverse stripe fluctuations is
sufficiently large compared to the CDW coupling stripes.   As a consequence,
there should exist electronic quantum liquid-crystal phases,  which 
constitute new states of matter,  and which can be either high-temperature
superconductors or two-dimensional anisotropic 'metallic' non-Fermi liquids.
Neutron scattering and other experiments in the copper oxide superconductors
$La_{1.6-x}Nd_{0.4}Sr_xCuO_4$ already provide evidence for the existence of 
these phases in at least one class of materials.

Sentizky \cite{senitzky98} (1998) studies
the effects of zero-point energy, in comparison with those 
of excitation energy,  are investigated by studying the behaviour 
of two coupled harmonic oscillators for four types of coupling:
rotating-wave coupling,  counter-rotating wave coupling,  
dipole-dipole (or dipole field) coupling and elastic-attraction coupling.
It is found that in all but the rotating-wave coupling case,  which has only 
slowly varying terms in their interaction Hamiltonian,  the interaction
coupling between the ground-state oscillators is followed 
by an oscillating energy increase in both systems has a lower 
zero-point energy that the energy of the state in which 
the individual oscillators are in their ground states,  
thus leading to the presence of a van der Waals attraction.
This result motivates the conclusion that only the rapidly 
varying terms in the interaction Hamiltonian 
are responsible for the attraction.
The energies of the individual oscillators while the coupled system 
is in its ground state are calculated.  
In the van der Waals cases, these energies tern out 
to be greater than the zero-point energies of the individual oscillators.
he offers a physical explanation for the increase.
he discusses the feasibility of the utilization or absorption of 
zero-point energy.
\subsection{Casimir Calculations.}
There are a very large number of papers calculating the Casimir effect,
some recent ones are mentioned below.
The technique of $zeta$-function regularization which is used for many of these
calculations is discussed in the book of 
Elizade {\it et al} (1994) \cite{EORBZ}.

Duru and Tomak \cite{DT} (1993) discuss 
the magnitude of the Casimir energies for the electrons 
and the electromagnetic
fields resulting from confinement of the corresponding particles in 
dis-ordered materials.   
They estimate that the vacuum energy of the confined
light is of the order of $eV$,  while for electrons it is negligible small.

Fendley {\it et al} \cite{FSZ} (1993) 
study the spectrum, the massless S-matrices and the ground-state energy 
of the flows between successive minimal models of conformal field theory, 
and within the sine-Gordon model with imaginary coefficient 
of the cosine term (related to the minimal models by ``truncation''). 
For the minimal models, they find exact S-matrices which
describe the scattering of massless kinks, 
and show using the thermodynamic Bethe ansatz that 
the resulting non-perturbative c-function (defined by the Casimir energy on a
cylinder) flows appropriately between the two theories, 
as conjectured earlier. 
For the non-unitary sine-Gordon model, they find unusual behavior. 
For the range of couplings they
can study analytically, the natural S-matrix deduced from the minimal 
one by ``undoing'' the quantum-group truncation does not reproduce 
the proper c-function with the TBA. 
They say that it does describe the correct properties of the model 
in a magnetic field. 

Vanzo \cite{vanzo} (1993) discusses
the functional integral for a scalar field confined in a cavity and subjected
to linear boundary conditions.   
He shows how the functional measure can be conveniently 
dealt with by modifying the classical action with boundary corrections.
The nonuniqueness of the boundary actions is described with a three-parameter 
family of them giving initial boundary conditions.
In some cases,  the corresponding Greens' function 
will define a kind of generalized Gaussian measure on function space.
He discusses vacuum energy,  paying due attention to its anomalous scale
dependence,  and the physical issues involved are considered.

Bender and Kimball \cite{BK} (1994) compute
the Casimir force on a $D$-dimensional sphere due to the confinement 
of a massless scalar field as a function of $D$, 
where $D$ is a continuous variable that
ranges from $-\infty$ to $\infty$. 
The dependence of the force on the dimension 
is obtained using a simple and straightforward Green's function technique. 
They find that the
Casimir force vanishes as $D\to +\infty$ ($D$ non-even integer) 
and also vanishes when $D$ is a negative even integer. 
The force has simple poles at positive even integer values of $D$. 

Villarreal \cite{villarreal} (1995) 
investigates the effective potential up to the two-loop level 
for a scalar field with a quartic self-interaction defined on 
a Minkowskian spacetime endowed with a nontrivial topology 
$T^4=S^1\times S^1\times S^1\times S^1$. 
The periodicity in the time dimension allows one to get 
simultaneously expressions for the effective potential 
and the self-energy valid for arbitrary temperature. 
The spatial
periodicity generates Casimir forces that compactify 
the associated spatial dimensions. 
He finds that for a critical value of the coupling constant 
the self-interaction of the field balances the
Casimir interaction, and a ``phase transition'' occurs, 
creating a stable vacuum structure with the topology of a two-torus. 
He says that thermal fluctuations might destroy this structure in another phase
transition occurring at high temperature.
Closely related to the work of Villarreal \cite{villarreal} is the work 
Elizalde and Kirsten \cite{EKir} (1994) on topological symmetry breaking in
self-interacting theories on toroidal spacetime.

Garattini \cite{garattini} (1999) 
calculates by variational methods, the Schwarzschild and flat space 
energy difference 
following the subtraction procedure for manifolds with boundaries, 
calculate by variational methods, the Schwarzschild and flat space 
energy difference. 
He considers the one loop approximation for TT tensors. 
An analogy between the computed energy difference in momentum space 
and the Casimir effect is illustrated. 
He finds a singular behaviour in the ultraviolet-limit, 
due to the presence of the horizon when $r=2m.$ 
He finds $r>2m$ this singular behaviour disappears, 
which is in agreement with various other models
previously presented. 

Leseduarte and Romeo \cite{LR} (1996) 
study the simultaneous influence of boundary conditions 
and external fields on quantum fluctuations by considering 
vacuum zero-point energies for quantum fields in the
presence of a magnetic fluxon confined by a bag, 
circular and spherical for bosons and circular for fermions. 
They calculate Casimir effect in a generalized cut-off regularization
after applying $\zeta$-function techniques to eigenmode sums 
and using recent techniques involving Bessel 
$\zeta$-functions at negative arguments. 

Krech and Landau \cite{KL} (1996) note that 
if a critical system is confined to a finite geometry critical fluctuations
of the order parameter generate long-range forces between the system 
boundaries.   These Casimir forces,  are
characterized by universal amplitudes and scaling functions. 
They derive a hybrid Monte Carlo algorithm and used it to measure the
Casimir amplitudes directly and accurately.   
They apply the algorithm to a
critical {\it q}-state Potts model confined to a rectangular $MXL$ geometry
in d=2 dimensions and to a critical Ising model confined to a $M^2\times L$
geometry in d=3 dimensions.   
They find good agreement with rigorous results in 
d=2 and compare their results with field-theoretical estimates of the Casimir 
amplitude in d=3.

Weingert \cite{weigert} (1996) discusses free quantum electromagnetic 
radiation is enclosed in a one dimensional cavity.
The contribution of the {\it k}th mode of the field to the energy,  
contain in a region $\mathcal{R}$ of the cavity,  is minimized.   
For the resulting {\it squeezed} state,  the energy expect ion in $\mathcal{R}$
is {\it below} its vacuum value.   Pressure zero-point energy out of the 
spatial region can be used to temporarily {\it increase} the Casimir force.

Bayin and Mustafa \cite{bayin} (1997) 
calculate the Casimir energy of the massless conformal scalar field on the
surface (S-2) of a 3-dimensional Reimann sphere by using the point-splitting,
mode sum and the $\zeta$-function renormalization methods.   
They also consider the half space case with both Dirichlet and the 
Neuwman boundary conditions.   This problem is interesting since 
the Casimir energy could be calculated analytically by various methods,  
thus allowing them to compare different regularization schemes.

Williams \cite{williams} (1997) discusses how topology and Casimir 
might be related,  see also \S2.14

Bukina and Shtykov \cite{BS} (1997) evaluate
the effective one-loop potential on $R^{m+1}\times S^N$ 
spaces for massless tensor fields is evaluated. 
The Casimir energy is given as a value of $\zeta-$ function by means
of which regularization is made. 
In even-dimensional spaces the vacuum energy contains divergent terms 
coming from poles of $\zeta(s,q)$ at $s=1$, whereas in
odd-dimensional spaces it becomes finite. 

Gosdzinsky and Romeo \cite{GR} (1998) find 
values for the vacuum energy of scalar fields under Dirichlet and Neuman 
boundary conditions on an infinite cylindrical surface are found, 
and they happen to be of opposite signs. 
In contrast with classical works, 
they apply a complete $\zeta$-function regularization scheme. 
These fields are regarded as interesting
both by themselves and as the key to describing the electromagnetic case. 
They claim that the electromagnetic Casimir effect in the presence of this
surface, found by De Raad and Milton, is now confirmed. 

Lu and Huang (1998) \cite{LH} find that ``unlike a closed string the 
Casmir energy of an open string can be either positive or negative''.

Cognola {\it et al} \cite{CEK} (1999) calculate Casimir energies, 
in an arbitrary number of dimensions,
for massless quantized fields in spherically 
symmetric cavities is carried out. 
All the most common situations, including scalar and spinor fields, 
the electromagnetic field, and various boundary conditions 
are treated with care. 
The final results are given as analytical
closed expressions in terms of Barnes $\zeta$-functions. 
They perform a direct, straightforward numerical evaluation of the formulas 
is then performed, which yields highly accurate numbers of,
in principle, arbitrarily good precision.

Katzgraber {\it et al} \cite{KBB} (1999) calculate 
the attractive long-range vortex-vortex interactions 
of the van der Waals type presented in anisotropic 
and layered superconductors.   
This allows them to define a 2D Casimir problem 
and determine the attractive force between two half planes.

Kirsten \cite{Khab} (1999) among other things
discusses the use of spectral functions in calculation of Casimir energy.
He says that
several functions of the spectrum of second order elliptic differential 
operators play a central role in the analysis of properties of physical 
systems. E.g. in statistical mechanics relevant spectral functions 
comprise of various partition sums for the evaluation of
thermodynamical quantities as critical temperatures or fluctuations 
of the ground state occupation. 
In quantum field theory under external conditions relevant quantities 
are effective actions (closely related to functional determinants) 
and ground state or vacuum energies, which describe e.g. the influence of 
external fields or of boundaries on the properties of the vacuum. 
In this context, results are generically divergent and need 
a renormalization to give a physical meaning to them. 
The renormalization procedure at one-loop is completely determined by the 
heat-kernel coefficients, central objects of spectral geometry. 
All mentioned spectral functions can be related
to an associated $\zeta$-function. 
In recent years an analysis of $\zeta$-functions in spherically symmetric 
situations has become available. 
These results (and techniques involved) allow the analysis of vacuum 
properties in the presence of spherically symmetric boundaries or background 
fields as well as the determination of thermodynamical properties of ideal 
gases in magnetic traps. 
Furthermore, when combined with other methods, it provides an effective 
scheme for the calculation of heat-kernel coefficients on arbitrary smooth
Riemannian manifolds with smooth boundaries. 

Lambiase {\it et al} \cite{LNB} (1999) propose a simple method is proposed 
to construct the spectral $\zeta$-functions required 
for calculating the electromagnetic vacuum energy with boundary conditions
given on a sphere or on an infinite cylinder. 
When calculating the Casimir energy in this approach no exact divergences 
appear and no renormalization is needed. 
The starting point of the consideration is the representation of the 
$\zeta$-functions in terms of contour integral, 
further the uniform asymptotic 
expansion of the Bessel function is essentially used by them.
After the analytic continuation, needed for calculating the Casimir energy, 
the $\zeta$-functions are presented as infinite series containing the Riemann 
$\zeta$-function with rapidly falling down terms. 
The spectral $\zeta$-functions are constructed exactly for a material ball and
infinite cylinder placed in an uniform endless medium under the condition 
that the velocity of light does not change when crossing the interface. 
As a special case, perfectly conducting spherical and cylindrical shells are 
also considered in the same line. 
In this approach they claim to succeed in justifying in
mathematically rigorous way, the appearance of the contribution 
to the Casimir energy for cylinder which is proportional to $\ln (2\pi)$. 
This method is related to that of 
Leseduarte and Romeo \cite{LRom} (1996)
and Bordag {\it et al} \cite{BEK} (1996).

Matloob \cite{matloob} (1999) uses
the Maxwell stress tensor to introduce the radiation pressure force
of the electron field on a conducting surface.   
This expression is related to the imaginary part of the vector potential 
Green function for the fluctuating fields of the vacuum via the fluctuation 
dissipation theorem and Kubo's formula see \S3.2.
The formalism allows him to evaluate the vacuum radiation pressure on a 
conducting surface without resorting to the process of field quantization.
The later formula is used to calculate the attractive and repulsive 
Casimir force between two conducting plates.   What is more,
in this formalism,  there is no need to apply any regularization
proceedure to recover the final result.

Bordag and Kirsten \cite{BorK} (1996) and
Bordag {\it et al} \cite{BHK} (2000) 
consider the vacuum energy of a scalar field in a spherically symmetric 
background field.
The numerical procedure is refined from previous work 
and applied to several examples. 
They provide numerical evidence that repulsive potentials lead to 
positive contributions to the vacuum energy, 
and show the crucial role played by bound-states. 

Bordag and Vassilevich \cite{BVas} (2000)
calculate the Casimir force between two parallel plates if 
the boundary conditions for the photons are modified due to presence of the
Chern-Simons term. They show that this effect should be measurable 
within present experimental technique. 

Brevik {\it et al} \cite{BMOO} (2000)
apply the background field method and the effective action formalism 
to describe the four-dimensional dynamical Casimir effect. 
Their picture corresponds to the consideration
of quantum cosmology for an expanding FRW universe 
(the boundary conditions act as a moving mirror) 
filled by a quantum massless GUT which is conformally invariant. 
They consider cases in which the static Casimir energy is repulsive 
and attractive. 
Inserting the simplest possible inertial term, they find, 
in the adiabatic (and semiclassical) approximation, 
the dynamical evolution of the scale factor and the dynamical Casimir stress 
analytically and numerically (for SU(2) super Yang-Mills theory). 
Alternative kinetic energy terms are also explored.

Garattini \cite{remogar} (2000)
calculates by variational methods the subtraction procedure for manifolds 
with boundaries the K\"ottler (Schwarzschild-deSitter) and
the deSitter space energy difference. 
By computing the one loop approximation for TT tensors he discovers 
the existence of an unstable mode even for the non-degenerate case. 
This result seems to be in agreement with the sub-maximal black hole pair 
creation of Bousso-Hawking. 
The instability can be eliminated by the boundary reduction method. 
He discusses the implications for a foam-like space. 

Herdegen \cite{herdegen} (2000) notes that
two thin conducting, electrically neutral, parallel plates forming an isolated
system in vacuum exert attracting force on each other,
whose origin is the quantum electrodynamical interaction. 
He says that this theoretical hypothesis, known as Casimir effect, 
has been also confirmed experimentally. 
Despite long history of the subject, no completely convincing 
theoretical analysis of this effect appears in the literature.
Here he discusses the effect (for the scalar field) anew, 
on a revised physical and mathematical basis. 
He uses standard, but advanced methods of
relativistic quantum theory. 
No anomalous features of the conventional approaches appear. 
The Casimir quantitative prediction for the force is shown to constitute 
the leading asymptotic term, for large separation of the plates, 
of the full, model-dependent expression. 

Marachevsky \cite{marachevsky} (2000) applies a
general formalism of quantum field theory 
and addition theorem for Bessel functions 
to derive formula for Casimir-Polder
energy of interaction between a polarizable particle 
and a dilute dielectric ball. 
He shows The equivalence of dipole-dipole interaction and Casimir
energy for dilute homogeneous dielectrics. 
He uses a novel method to derive Casimir energy 
of a dilute dielectric ball without
divergences in calculations. 
Physically realistic model of a dilute ball is discussed. 
He reviews different approaches to the calculation of Casimir
energy of a dielectric ball. 

Scandurra \cite{scandurra3} (2000) computes 
the ground state energy of a massive scalar field 
in the background of a cylindrical shell whose potential 
is given by a $\delta$-function. 
The zero-point energy is
expressed in terms of the Jost function of the related scattering problem, 
the renormalization is performed with the help of the heat kernel expansion. 
The energy is found to be
negative for attractive and for repulsive backgrounds as well.

Hagen \cite{hagen} (2001)
considers the Casimir force between two conducting planes in both 
the electromagnetic and scalar field cases. This is done by the usual
summation over energy eigenmodes of the system as well as by 
a calculation of the stress tensor in the region between the planes. The
latter case requires that careful attention be given to singular operator 
products, an issue which is accommodated here by invoking the
point separation method in conjunction with a scalar cutoff. 
This is shown to yield cutoff dependent and divergent contributions to the
Casimir pressure which are dependent on the separation parameters, 
but entirely consistent with Lorentz covariance. Averaging over the
point splitting parameters allows finite results to be obtained, 
but fails to yield a unique Casimir force. 
\subsection{The Dynamical Casimir Effect.}
In the past,  the Casimir effect has been considered as a {\it static} effect.
Growing interest in recent years has been drawn to the {\it dynamical} variety
of the effect,  meaning,  in essence,  that not only the geometric 
configurations of the extremal boundaries (such as plates) 
but also their velocities play a physical role,
see Brevik {\it et al} \cite{BMOO} (2000) and \S4.7.
One can also consider the Universe to form a sort of boundary,
in which case the ``Casimir Effect'' is just QFT in some cosmological model,
see \S3.9.   Padmanabhan and Coudhury \cite{PC} (2000) use the fact that
the Universe forms some sort of boundary to move from some sort of quantum
gravity theory to QFT in curved spaces.   Having the Universe forming some
sort of boundary can be thought of as being an aspect of Mach's principle,
compare Roberts \cite{mdrse} (1998).   
There is a method is based on the requirement that the zero-point 
fluctuations must vanish on the boundary.
This requirement leads to a finite change in the zero-point energy,
which can be extracted from the infinite total energy by special techniques.
The two-dimensional problem can be solved exactly by de Witt's point
splitting method discussed in de Witt \cite{dewitt} (1975) and  
Fulling and Davies \cite{FD} (1975), and one again finds 
a radiation damping force which in the non-relativistic 
limit is proportional to ${\rm d}^2v/{\rm d}t^2$.
Some recent dynamical Casimir calculations are mentioned below.

Langfeld {\it et al} \cite{LSR} (1993),  compare \S2.11 study
non-trivial $\phi ^{4}$-theory is studied in a renormalisation group invariant
approach inside a box consisting of rectangular plates and where the scalar 
modes satisfy periodic
boundary conditions at the plates. 
They find that the Casimir energy exponentially approaches 
the infinite volume limit, the decay rate given by the scalar condensate. 
It therefore
essentially differs from the power law of a free theory. 
This might provide experimental access to properties 
of the non-trivial vacuum.
At small interplate distances the system
can no longer tolerate a scalar condensate, 
and a first order phase transition to the perturbative phase occurs. 
The dependence of the vacuum energy density and the scalar
condensate on the box dimensions are presented. 

Pontual and Moraes \cite{PM} (1996) show that the existence of 
a non-zero vacuum energy density 
(the Casimir energy) for a scalar field appears in a continuous 
elastic solid due to the presence of a topological
defect, the screw dislocation. 
An exact expression is obtained for this energy density 
in terms of the Burgers vector describing the defect, 
for zero and finite temperature. 

Carlson {\it et al} \cite{CMPV} (1997) note that 
in a recent series of papers, Schwinger discussed a process 
that he calls the dynamical Casimir effect. 
The key essence of this effect is the change in zero-point energy
associated with any change in a dielectric medium. 
In particular, if the change in the dielectric medium 
is taken to be the growth or collapse of a bubble, this effect may have
relevance to sonoluminescence, see \S4.5.
The kernel of Schwinger's result is that the change 
in Casimir energy is proportional to the change in volume of the dielectric, 
plus finite-volume corrections. 
They say that other papers have called into question this result, 
claiming that the volume term should actually be discarded, 
and that the dominant term remaining is proportional to
the surface area of the dielectric. 
They claim to present a careful and critical review of the relevant
analyses. 
They find that the Casimir energy, defined as the change in 
zero-point energy due to a change in the medium, 
has at leading order a bulk volume dependence. 
This is in full
agreement with Schwinger's result, 
once the correct physical question is asked. 
They say they have nothing new to say about sonoluminescence itself. 

Li {\it et al} \cite{LCLZ} (1997) re-examine
the Casimir effect giving rise to an attractive
or repulsive force between the configuration boundaries 
that confine the massless scalar field for a 
$(D-1)$-dimensional rectangular
cavity with unequal finite $p$ edges and different spacetime dimensions $D$.
That say that 
with periodic or Neumann boundary conditions, the energy is always negative. 
The case of Dirichlet boundary conditions is more complicated. 
The sign of the Casimir energy satisfying Dirichlet conditions 
on the surface of a hypercube (a cavity with equal finite $p$ edges) 
depends on whether $p$ is even or odd. 
In the general case (a cavity with unequal $p$ edges), 
however, we show that the sign of the Casimir energy 
does not only depend on whether $p$ is odd or even. 
Furthermore, they find that the
Casimir force is always attractive if the edges are chosen appropriately. 
It is interesting that the Casimir force may be repulsive for odd $p$ cavity 
with unequal edges, in contrast with the same
problem in a hypercube case. 

Sitenko and Babansky \cite{SBa} (1997) study
the combined effect of the magnetic field background 
in the form of a singular vortex and the Dirichlet boundary 
condition at the location of the vortex on the vacuum of quantized
scalar field. 
They find the induced vacuum energy density and current 
to be periodic functions of the vortex flux and holomorphic functions 
of the space dimension. 

Dalvit and Neto \cite{DN} (2000)
derive a master equation for a mirror interacting 
with the vacuum field via radiation pressure. 
They say that the dynamical Casimir effect leads 
to decoherence of a 'Schr\"odinger cat' 
state in a time scale that depends on the degree of 'macroscopicity' 
of the state components, and which may be much shorter 
than the relaxation time scale. 
Furthermore they say that coherent states 
are selected by the interaction as pointer states.

Hofmann {\it et al} (2000)
study the stabilization of one spatial dimension in $p+1+1$-dimensional 
spacetime in the presence of $p$-dimensional brane(s), a bulk
cosmological constant and the Casimir force generated by 
a conformally coupled scalar field. 
They find general static solutions to the metric
which require two fine tunings: setting the size of the extra dimension 
to its extremal value and the effective cosmological constant to zero.
Taking these solutions as a background configuration, 
they perform a dimensional reduction and study the effective theory 
in the case of one- and two-brane configurations. 
They show that the radion field can have a positive mass squared, 
which corresponds to a stabilization
of the extra dimension, only for a repulsive nature of the Casimir force. 
This type of solution requires the presence of a negative tension brane. 
The solutions with one or two positive tension branes arising in this theory 
turn out to have negative radion mass squared, and therefore are not stable. 

Milton \cite{milton} (2000) notes that  
zero-point fluctuations in quantum fields give rise 
to observable forces between material bodies, 
which he notes are the so-called ``Casimir forces''. 
In this paper he presents some results of the theory of the Casimir effect, 
primarily formulated in terms of Green's functions. There is an intimate
relation between the Casimir effect and van der Waals forces. 
Applications to conductors and dielectric bodies of various shapes are
given for the cases of scalar, electromagnetic, and fermionic fields. 
The dimensional dependence of the effect is described. Finally, he
asks the question: Is there a connection between the Casimir effect 
and the phenomenon of sonoluminescence? 
Sonoluminescence is discussed in \S4.5.
\subsection{Mechanical Analogs of the Casimir Effect.}
The Casimir effect has various analogs some of which are mentioned below.

Sokolov \cite{sokolov} (1994) considers
energy production due to the Casimir effect for the case 
of a superdense state of matter, which can appear 
in such cosmological objects as white dwarfs,
neutron stars, quasars and so on. 
The energy output produced by the Casimir effect 
during the creation of a neutron star turns out 
to be sufficient to explain nova and supernova explosions. 
He claims to show that the Casimir effect might be a possible source 
of the huge energy output of quasars. 

Van Enk \cite{enk} (1995) shows
that the vacuum can induce a torque between two uniaxial 
birefringent dielectric plates.
He discuss the differences and analogies between this torque 
and the Casimir force in the same configuration.
He show that the torque can be interpreted as arising from 
angular-momentum transfer from the vacuum to the plates,  
as well as from the orientation dependence of the zero-point energy.

Boersma \cite{boersma} (1996) notes that 
at sea,  on a windless day,  in a strong swell,  free floating ships will roll
heavily.   It was believed in the days of the clipper ships that under those 
circumstances two vessels at close distances will attract each other,
Boersma \cite{boersma} asks if they do.
The ships are harmonic oscillators in a wave field and as such analogous to
two atoms in the sea of vacuum fluctuations.   These atoms do attract by
the van der Waals force,  suggesting that the two vessels attract.

Larraza and Denardo \cite{LD} (1998) present
theoretical and experimental results for the force law 
between two rigid,  parallel plates due to the radiation pressure of 
band-limited acoustic noise.   
They claim excellent agreement is shown between 
theory and experiment.   While these results constitute an acoustic 
analog for the Casimir effect,  an important difference is that 
the band-limited noise can cause the force to be {\it attractive} 
or {\it repulsive} as a function of the distance of separation of the plates.
Applications of the acoustic Casimir effect to background noise transduction 
and non-resonant acoustic levitation are suggested.

Widom {\it et al} \cite{WSSS} (1998) note that
one loop field theory calculations of free energies quite often yield 
violations of the stability conditions associated with the thermodynamic 
second law. 
Perhaps the best known example involves the equation of state of black holes. 
They point out that the Casimir force between two parallel 
conducting plates also violates a thermodynamic stability
condition normally associated with the second law of thermodynamics.

Mulhopaclhgay and Law \cite{ML} (1999)
study the critical Casimir force per unit area which determines the film
thickness of critical binary liquid wetting films for the specific case of
opposite boundary conditions within the film.
A universal scaling function is derived in the one-phase region.
At criticality this scaling function reduces to a universal amplitude 
with value $\Delta_{+,-}\sim0.0053$
\subsection{Applications of the Casimir Effect.}
Iacopini \cite{iacopini} (1993) notes that 
the possibility of observing the Casimir force at macroscopic distances
(a few centimeters) using a conformal optical resonator is discussed 
and a possible experimental apparatus is also suggested.
Long range effects are also discussed in Spruch \cite{spruch} (1996),
Yam \cite{yam} (1997)
and Winterberg \cite{winterberg} (1998).
Kiers and Tytgat \cite{KT} (1997) note that
it has recently been argued that long range forces 
due to the exchange of massless neutrinos give rise 
to a very large self-energy in a dense, finite-ranged, weakly-charged medium. 
Such an effect, if real, would destabilize a neutron star. 
To address this issue they have studied the related problem 
of a massless neutrino field in the presence of an
external, static electroweak potential of finite range. 
To be precise, they have computed to one loop the exact 
vacuum energy for the case of a spherical square well potential of
depth $\alpha$ and radius $R$. 
For small wells, the vacuum energy is reliably 
determined by a perturbative expansion in the external potential. 
For large wells, however, the perturbative expansion breaks down. 
A manifestation of this breakdown is that the vacuum carries 
a non-zero neutrino charge. 
The energy and neutrino charge of the ground state are, to a
good approximation for large wells, those of a neutrino condensate 
with chemical potential $\mu=\alpha$. 
Our results demonstrate explicitly that long-range forces due to the
exchange of massless neutrinos do not threaten the stability of neutron stars. 
\subsection{Experimental Testing of the Casimir Effect.}
An early experimental verification of the Casimir effect is that of
Tabor and Winton \cite{TW} (1969),  see also \S3.3\P5 above.
Some more recent measurement are mentioned below.

Storry {\it et al} \cite{SRH} (1995)
make measurements which they compare
with high-precision variational calculations.  They find that the expected
long-range Casimir effect is not present.

Grado {\it et al} \cite{GCF} (1999) notes that
the accurate measurement of the Casimir force and the search for hypothetical 
long-range interactions are subjects of growing interest. 
They propose to use a suspended
interferometric device to measure the Casimir effect between metallic flat 
surfaces. 
This allows them to perform the measurement at distances much larger 
than the ones used in previous measurements with spherical surfaces. 
The use of the proposed dynamic detection scheme can also help in the 
discrimination of the different contributions to the measured force. 
Some considerations about the physical information which can be obtained 
from this type of measurement are also made.

Lamoreaux \cite{lamoreaux} (1999) 
reviews the recent experimental verifications 
of the Casimir force between extended bodies. 
He says that with modern techniques, it now appears feasible to test the
force law with 1\% precision; he addresses the issues relating 
to the interpretation of experiments at this level of accuracy 
\subsection{Quantum Field Theory on Curved Spaces.}
The Casimir effect as originally understood depends crucially on the two
plate boundaries.   One can relax this requirement to that where the boundaries
are moving ,
and then this effect is very close to
the sort of situations which are studied in QFT on curved spacetimes,
where the role of the boundary is now taken over by the metric tensor.
Padmanabhan and Choudhury \cite{PC} (2000),  see \S4.7 above,
try to justify QFT on curved spacetimes as a well defined limit
from so as yet unknown theory of quantum gravity.
Meissner and Veneziano \cite{MV} (1991) discuss how to produce 
spacetime invariant vacua.
There are many papers on this topic;  
some recent calculations,  which specifically mention vacuum energy,
are mentioned below.

Svaiter and Svaiter \cite{SS} (1993) 
discuss analytic regularization methods used to obtain the 
renormalized vacuum energy of quantum fields in an arbitrary 
ultrastatic spacetime. 
After proving that the $\zeta$-function method is
equivalent to the cutoff method with the subtraction of the polar terms, 
they present two examples where the analytic extension method gives 
a finite result, but in disagreement with the cutoff and
$\zeta$-function methods.

Linet \cite{linet} (1994) determines generally the spinor Green's function 
and the twisted spinor Green's function in an Euclidean space 
with a conical-type line singularity. In particular, in
the neighbourhood of the point source, 
he expresses them as a sum of the usual Euclidean spinor 
Green's function and a regular term. 
In four dimensions, he uses these determinations
to calculate the vacuum energy density 
and the twisted one for a massless spinor field 
in the spacetime of a straight cosmic string. 
In the Minkowski spacetime, he determines
explicitly the vacuum energy density for a massive twisted spinor field. 

Soleng \cite{soleng} (1994) notes that 
quantum field theory in curved spacetime implies that the 
strong equivalence principle is violated outside a spherically symmetric,  
static star.   Here he assumes that the quantum gravity effects restore 
the strong equivalence principle.   Together with the assumption 
that the effective vacuum polarization energy-momentum tensor is traceless,  
this leads to a specific algebraic form of the energy-momentum tensor 
for which an exact solution of Einstein's field equations is found.
The solution gives the post-Newtonian parameters $\ga=1$ and $\bt=1+3\de$,
where $\de$ is a dimensionless constant which determines the energy density 
of the anisotropic vacuum.   The vacuum energy changes the perihelion 
precession by  a factor of $1-\de$.

Allen {\it et al} \cite{AKO} (1995) 
combine and further develop their ideas and techniques for 
calculating the long range effects of cosmic string cores on classical 
and quantum field quantities far from an (infinitely long, straight) 
cosmic string. 
They find analytical
approximations for (a) the gravity-induced ground state 
renormalized expectation values of $\hat\varphi^2$ and $\hat T_\mu{}^\nu$ 
for a non-minimally coupled quantum scalar
field far from a cosmic string (b) the classical electrostatic 
self force on a test charge far from a superconducting cosmic string. 
Surprisingly -- even at cosmologically large
distances -- all these quantities would be 
very badly approximated by idealizing the string 
as having zero thickness and imposing regular boundary conditions; 
instead they are
well approximated by suitably fitted strengths of logarithmic divergence 
at the string core. 
Their formula for ${\langle {\hat \varphi}^2 \rangle}$ 
reproduces,  with much less effort
and much more generality,  
the earlier numerical results of Allen and Ottewill. 
Both ${\langle {\hat \varphi}^2 \rangle}$ 
and ${\langle {\hat T}_{\mu}{}^{\nu} \rangle}$ turn out to
be ``weak field topological invariants'' depending on the details 
of the string core only through the minimal coupling parameter ``$\xi$'' 
(and the deficit angle). 
Their formula for the
self-force (leaving aside relatively tiny gravitational corrections) 
turns out to be attractive: 
They obtain, for the self-potential of a test charge $Q$ a distance $r$ 
from a (GUT scale) superconducting string, the formula 
$- Q^2/(16\epsilon_0r\ln(qr))$ where $q$ is an (in principle, computable) 
constant of the order of the inverse string radius. 

Bytsenko {\it et al} \cite{BCVZ} (1995) review
the heat-kernel expansion and $\zeta$-regularization techniques 
for quantum field theory and extended objects on curved space-times. 
In particular they discuss ultrastatic
space-times with spatial section consisting in manifold 
with constant curvature in detail. 
They present several mathematical results, relevant to physical applications,  
including exact solutions of the heat-kernel equation, 
a simple exposition of hyperbolic geometry and an elementary derivation 
of the Selberg trace formula. 
They consider with regards to the physical applications, 
the vacuum energy for scalar fields, 
the one-loop renormalization of a self-interacting scalar field theory 
on a hyperbolic space-time, with a
discussion on the topological symmetry breaking, 
the finite temperature effects and the Bose-Einstein condensation. 
They also present some attempts to generalize the results to
extended objects, 
including some remarks on path integral quantization, 
asymptotic properties of extended objects 
and a novel representation for the one-loop
superstring free energy. 

Davies {\it et al} (1996)
derive conditions for rotating particle detectors to respond in a variety
of bounded spacetimes and compare the results with the folklore that
particle detectors do not respond in the vacuum state appropriately to
their motions.   The breifly address applications involoving possible 
violations of the second law of thermodynamics.

Chan \cite{chan} (1997) discusses vacuum energy for static backgrounds.

Ford and Svaiter \cite{FS} (1998) note that
the imposition of boundary conditions upon 
a quantized field can lead to singular energy densities on the boundary. 
They treat the boundaries as quantum
mechanical objects with a nonzero position uncertainty, 
and show that the singular energy density is removed. 
They say that this treatment also resolves a long standing paradox concerning 
the total energy of the minimally coupled 
and conformally coupled scalar fields. 

Vilkovisky \cite{vilkovisky} (1999) presents a solution 
to the simplest problem 
about the vacuum backreaction on a pair creating source. 
The backreaction effect is nonanalytic in the coupling constant and
restores completely the energy conservation law. 
The vacuum changes the kinematics of motion like 
relativity theory does and imposes a new upper bound on the velocity of the
source.

Hu and Phillips \cite{HP} (2000) note that
from calculations of the variance of fluctuations 
and of the mean of the energy density of a massless scalar field 
in the Minkowski vacuum as a function of an intrinsic scale
defined by the world function between two nearby points 
(as used in point separation regularization) they claim that, 
contrary to prior claims, the ratio of variance to mean-squared
being of the order unity does not imply a failure of semiclassical gravity. 
It is more a consequence of the quantum nature of the state of matter field 
than any inadequacy of the
theory of spacetime with quantum matter as source. 
\section{Vacuum Energy on Large Scales.}
\subsection{The Cosmological Constant.}
The standard picture of how vacuum energy gives rise to as cosmological
constant is outlined in \S2.1 at the end and my caveats to this 
are given in \S2.1 end of \P2.  
To reiterate {\it two} of the problems.
{\it Firstly} that the cosmological constant,  
found by the standard method,  is time independent,
and a cosmological constant can be modeled by a perfect fluid with pressure
and density of opposite signs,
such a choice allows time dependence as pressure and density 
can be time dependent.
{\it Secondly} a cosmological constant implies that either the equivalent
pressure or density must be negative.
I have discussed,  Roberts \cite{mdrpluto} (1987), 
how the cosmological constant might alter the orbit of Pluto,
and I have discussed,  Roberts \cite{mdrnm} (1998),
how it might be possible to re-interpret the cosmological constant
as the object on non-metricity.
Papers on the cosmological constant are written 
are the rate of about one a day,
here I restrict myself to about 30 recent papers.

Four recent reviews are:
Carroll \cite{carroll} (2000), Weinberg \cite{weinberg} (2000)
Garriga and Vilenkin \cite{GV} (2000),   Rugh and Zinkernagel \cite{RZ} (2000).
Focusing on recent developments, 
Carroll presents a pedagogical overview of cosmology in the presence of a
cosmological constant, observational constraints on its magnitude, 
and the physics of a small (and potentially nonzero) vacuum energy,
focusing on recent developments.
Weinberg's approach is given in more detail in \S4.2

Sciama \cite{sciama} (1991) approach to the cosmological constant 
is as follows.
If the vacuum energy density is a local quantity,
one would expect that,  for reasons of symmetry,  it would have to have the
form $\lambda g_{\nu\mu}$,  and so correspond to the 'cosmological'
term in Einstein's field equations.
This is redolent of the related problem that,  
in grand unified and supergravity type theories,  
one would expect to have a cosmological term of order 
$10^{120}$ times greater than any observational cosmology.
Some very fine-tuned cancellations seem to be required to achieve 
agreement with observation.
Sciama \cite{sciama} (1991) says that:  ``One sees occasional claims 
that some varieties of superstring theory 
lead naturally to the vanishing of the cosmological constant,  
but no agreement has yet been reached on this point.
My view is in \S2.6.  Some recent work includes the papers mentioned below.

Baum \cite{baum} (1984) imposes a minimum action on a scalar field coupled to
classical gravity and find zero effective cosmological constant without 
fine tuning,  as well as a new mechanism for symmetry breaking,  see \S2.10.

Carvahlo {\it et al} \cite{CLW} (1992)
argue for the first time that the dimensional
argument of  Chen and Wu  \cite{CWu}  is  not  enough to fix
the time dependence of the Cosmological Constant. They claim to
show that a similar
argument leads to $\Lambda$ proportional to $H^{2}$. The cosmological
consequences of this new possibility are discussed in detail. In
particular, they claim to show that as opposed to the Chen and Wu model, this
new dependence solve the age problem.

Moffat \cite{moffat} (1994) presents a 
dynamical model of the decaying vacuum energy, 
which is based on Jordan-Brans-Dicke theory with a scalar field $\phi$. 
The solution of an evolutionary
differential equation for the scalar field $\phi$ 
drives the vacuum energy towards a cosmological constant 
at the present epoch that can give for the age of the universe, 
$t_0\sim 13.5$ Gyr for $\Omega_0=1$, 
which is consistent with the age of globular clusters. 

Barrow and D\c{a}browski \cite{BD} (1995) note that
if a positive cosmological constant exists then these oscillations 
will eventually cease and be replaced by an era of expansion which will 
continue unless the cosmological constant is associated with a form 
of vacuum energy that ultimately decays away. 

Krauss and Turner \cite{KrT} (1995) claim that
a diverse set of observations now compellingly 
suggest that Universe possesses a nonzero cosmological constant. 
They say that in the context of quantum-field theory a cosmological
constant corresponds to the energy density of the vacuum, 
and the wanted value for the cosmological constant corresponds 
to a very tiny vacuum energy density. 
They discuss future observational tests for a cosmological constant 
as well as the fundamental theoretical challenges---and opportunities---that 
this poses for particle physics and for
extending our understanding of the evolution 
of the Universe back to the earliest moments. 

Singh \cite{singh} (1995) notes that
increasing improvements in the independent determinations 
of the Hubble constant and the age of the universe now seem 
to indicate that a small non-vanishing
cosmological constant is needed to make the two independent 
observations consistent with each other. 
The cosmological constant can be physically interpreted as due to the vacuum
energy of quantized fields. 
To make the cosmological observations consistent 
with each other he suggests that a vacuum energy density, 
$ \rho_v \sim (10^{-3} eV)^4 $ is needed today 
(in the units $ \hbar=c=k=1 $ sometimes called cosmological units). 
It is argued in his article that such a vacuum energy density 
is natural in the context of phase transitions linked to massive neutrinos. 
In fact, the neutrino masses required to provide 
the right vacuum energy scale to remove the age verses Hubble 
constant discrepancy are consistent with those required to solve the
solar neutrino problem by the MSW mechanism. 

Lima \cite{jasl} (1996)
studies the thermodynamic behaviour of a decaying
$\Lambda-term$ coupled to a relativistic simple fluid. Using the
covariant approach for nonequilibrium thermodynamics, he shows that if
the specific entropy per particle is conserved (adiabatic decay),
some equilibrium relations are preserved. In particular, if the vacuum
decay adiabatically in photons, the Planckian form of the {\bf CBR} spectrum
is also preserved in the course of the evolution. The photon spectrum
is deduced in the Appendix.

Adler {\it et al} \cite{adler97} (1997) notes that
recent work has shown that complex quantum field theory 
emerges as a statistical mechanical approximation to an underlying 
noncommutative operator dynamics based on a total trace action. 
In this dynamics, scale invariance of the trace action 
becomes the statement $0=\Re Tr T_{\mu}^{\mu}$, 
with $T_{\mu \nu}$ the operator stress energy tensor,
and with $Tr$ the trace over the underlying Hilbert space. 
They show that this condition implies the vanishing of the cosmological 
constant and vacuum energy in the emergent quantum field theory. 
However, they say that since the scale invariance condition does not require 
the operator $T_{\mu}^{\mu}$ to vanish, the spontaneous breakdown 
of scale invariance is still permitted. 

Coble {\it et al} \cite{CDF} (1997) note that
models of structure formation with a cosmological constant $\Lambda$ 
provide a good fit to the observed power spectrum of galaxy clustering. 
However, these models suffer from several problems. 
Theoretically, it is difficult to understand 
why the cosmological constant is so small in Planck units. 
Observationally, while the power spectra of cold dark matter plus
$\Lambda$ models have approximately the right shape, 
the COBE-normalized amplitude for a scale invariant spectrum is too high, 
requiring galaxies to be anti-biased relative to the mass distribution. 
Attempts to address the first problem have led to models in which a 
dynamical field supplies the vacuum energy, which is thereby determined by
fundamental physics scales. 
They explore the implications of such dynamical $\Lambda$ 
models for the formation of large-scale structure. 
They find that there are dynamical models
for which the amplitude of the COBE-normalized 
spectrum matches the observations. 
They also calculate the cosmic microwave 
background anisotropies in these models and
show that the angular power spectra are distinguishable 
from those of standard cosmological constant models. 

Dolgov \cite{dolgov} (1997) analyses
the cosmological evolution of free massless vector 
or tensor (but not gauge) fields minimally coupled to gravity. 
He shows that there are some unstable solutions for
these fields in De Sitter background. 
The back reaction of the energy-momentum tensor 
of such solutions to the original cosmological 
constant exactly cancels the latter and the
expansion regime changes from the exponential to the power law one. 
In contrast to the adjustment mechanism realized by a scalar field 
the gravitational coupling constant in this
model is time-independent and the resulting 
cosmology may resemble the realistic one.   

Guendelman and Kaganovich \cite{GK97} (1997) 
claim to have shown that the principle of nongravitating vacuum energy, 
when formulated in the first order formalism, 
solves the cosmological constant problem. 
The most appealing
formulation of the theory displays a local symmetry associated 
with the arbitrariness of the measure of integration. 
This can be motivated by thinking of this theory as a direct
coupling of physical degrees of freedom with a "space - filling brane" 
and in this case such local symmetry is related to space-filling brane 
gauge invariance. 
The model is formulated in the first order formalism 
using the metric and the connection as independent dynamical variables. 
An additional symmetry (Einstein - Kaufman symmetry) allows to
elimination of the torsion which appears due to the introduction of 
the new measure of integration. 
The most successful model that implements these ideas 
is realized in a six or higher dimensional spacetime. 
The compactification of extra dimensions into a sphere 
gives the possibility of generating scalar masses and potentials, 
gauge fields and fermionic masses. 
It turns out that remaining four dimensional spacetime 
must have effective zero cosmological constant. 

Alvarenga and Lemos \cite{ALe} (1998) 
regard the vacuum as a perfect fluid with equation of state $p=-\rho$, 
de Sitter's cosmological model is quantized. 
Their treatment differs 
from previous ones in that it endows
the vacuum with dynamical degrees of freedom. 
Instead of being postulated from the start, 
the cosmological constant arises from the degrees of freedom 
of the vacuum regarded as a dynamical entity, 
and a time variable can be naturally introduced. 
Taking the scale factor as the sole degree of freedom 
of the gravitational field, stationary and wave packet
solutions to the Wheeler-DeWitt equation are found. 
It turns out that states of the Universe with a definite 
value of the cosmological constant do not exist. For the wave packets
investigated, quantum effects are noticeable only for small values 
of the scale factor, a classical regime being attained 
at asymptotically large times. 

Ries {\it et al} \cite{R+19} (1998) 
present observations of 10 type Ia supernovae (SNe Ia) between 
$0.16 < z < 0.62$. With previous data from their high-Z supernova search team, 
this expanded set of 16 high-redshift supernovae and 34 nearby supernovae 
are used to place constraints on the Hubble constant ($H_0$), 
the mass density ($\Omega_M$), the cosmological constant
($\Omega_\Lambda$), the deceleration parameter ($q_0$), 
and the dynamical age of the Universe ($t_0$). 
The distances of the high-redshift SNe Ia are, on average, 10\% to 15\% farther
than expected in a low mass density ($\Omega_M=0.2$) Universe without 
a cosmological constant. 
Different light curve fitting methods, SN Ia subsamples, and prior constraints
unanimously favor eternally expanding models with positive cosmological 
constant (i.e., $\Omega_\Lambda > 0$) and a current acceleration 
of the expansion (i.e., $q_0 < 0$). 
With no prior constraint on mass density other than $\Omega_M > 0$, 
the spectroscopically confirmed SNe Ia are consistent with $q_0 <0$ 
at the $2.8 \sigma$ and $3.9 \sigma$ confidence levels,
and with $\Omega_\Lambda >0$ at the $3.0 \sigma$ and $4.0 \sigma$ 
confidence levels, for two fitting methods respectively. 
Fixing a ``minimal'' mass density, $\Omega_M=0.2$, results in
the weakest detection, $\Omega_\Lambda>0$ at the $3.0 \sigma$
 confidence level. 
For a flat-Universe prior ($\Omega_M+\Omega_\Lambda=1$), 
the spectroscopically confirmed SNe Ia
require $\Omega_\Lambda >0$ at $7 \sigma$ and $9 \sigma$ 
level for the two fitting methods. 
A Universe closed by ordinary matter (i.e., $\Omega_M=1$) 
is ruled out at the $7 \sigma$ to $8 \sigma$ level. 
They estimate the size of systematic errors, including evolution, 
extinction, sample selection bias, local flows, gravitational lensing, 
and sample contamination.
Presently, none of these effects reconciles the data with 
$\Omega_\Lambda=0$ and $q_0 > 0$. 

Tegmark \cite{tegmark} (1998)
describes constraints on a ``standard'' 8 parameter open cold dark 
matter (CDM) model from the most recent CMB and SN1a data. 
His parameters are the densities of CDM, baryons, 
vacuum energy and curvature, the reionization optical depth, 
and the normalization and tilt for both scalar and tensor fluctuations.
He finds that although the possibility of reionization and gravity waves 
substantially weakens the constraints on CDM and baryon density, tilt, 
Hubble constant and curvature, allowing e.g. a closed Universe, 
open models with vanishing cosmological constant 
are still strongly disfavored. 

Adams {\it et al} \cite{AML} (1999)
explore possible effects of vacuum energy on the evolution of black holes. 
If the universe contains a cosmological constant, and if black holes can absorb
energy from the vacuum, then black hole evaporation could be greatly 
suppressed. For the magnitude of the cosmological constant suggested by current
observations, black holes larger than $\sim 4 \times 10^{24}$ g would accrete 
energy rather than evaporate. In this scenario, all stellar and supermassive 
black holes would grow with time until they reach a maximum mass scale of 
$\sim 6 \times 10^{55}$ g, comparable to the mass contained within the 
present day cosmological horizon.

Alcaniz and Lima \cite{AL} (1999) notes that 
the ages of two old galaxies (53W091, 53W069) at high redshifts 
are used to constrain the value of the cosmological constant 
in a flat universe ($\Lambda$CDM) and the density parameter 
$\Omega_M$ in Friedmann-Robertson-Walker (FRW) models with no $\Lambda$-term. 
In the case of $\Lambda$CDM models, the quoted galaxies yield 
two lower limits for the vacuum energy density parameter, 
$\Omega_\Lambda \geq 0.42$ and $\Omega_\Lambda \geq 0.5$, respectively. 
Although compatible with the limits from statistics of gravitational 
lensing (SGL) and cosmic microwave background (CMB), 
these lower bounds are more stringent than the ones 
recently determined using SNe Ia as standard candles. 
For matter dominated universes ($\Omega_\Lambda=0$), 
the existence of these galaxies imply that the universe 
is open with the matter density parameter constrained by
$\Omega_M \leq 0.45$ and $\Omega_M \leq 0.37$, respectively. 
In particular, these results disagree completely with the 
analysis of field galaxies which gives
a lower limit $\Omega_M \geq 0.40$. 

Bond and Jaffe \cite{BJ} (1999)
use Bayesian analysis methods to determine what current $CMB$ 
and $CMB+LSS$ data imply for inflation-based Gaussian fluctuations 
in tilted $\La CDM$, $\La hCDM$ and $oCDM$ model sequences with cosmological
age 11-15 Gyears,  consisting of mixtures of bayrons,  cold 'c' (and possibly
hot 'h') dark matter,  vacuum energy '$\La$',  and curvature energy 'o'
in open cosmologies.

Burdyuzha {\it et al} \cite{BLPV} (1999) discuss the
problem of the physical nature of the cosmological constant genesis. 
They say that this problem can't be solved in terms of current 
quantum field theory which operates
with Higgs and nonperturbative vacuum condensates 
and takes into account the changes of these condensates 
during relativistic phase transitions. 
They also say that the problem can't be completely solved also in terms 
of the conventional global quantum theory: 
Wheeler-DeWitt quantum geometrodynamics 
does not describe the evolution of the Universe in time
(RPT in particular). 
They have investigated this problem in the context of energies density 
of different vacuum subsystems characteristic scales of which pervade 
all energetic scale of the Universe. 
At first the phemenological solution of the cosmological constant problem 
and then the hypothesis about the possible structure of a new global quantum
theory are proposed. 
The main feature of this theory is the irreversible evolution of geometry 
and vacuum condensates in time in the regime of their self organization. 
The transformation of the cosmological constant in 
dynamical variable is inevitably. 

Jackson \cite{jackson} (1999) notes that
recent observations suggest that Hubble's constant is large, 
to the extent that the oldest stars appear to have ages which 
are greater than the Hubble time, and that the
Hubble expansion is slowing down, 
so that according to conventional cosmology 
the age of the Universe is less than the Hubble time. 
He introduces the concepts of weak and strong age
crises (respectively $t_0 < 1/H_0$ but longer than the age 
inferred from some lower limit on $q_0$, and $t_0 > 1/H_0$ and $q_0 > 0$) 
are introduced. 
These observations are reconciled in
models which are dynamically dominated by a homogeneous scalar field, 
corresponding to an ultra-light boson whose Compton wavelength is of 
the same order as the Hubble radius. 
Two such models are considered, an open one with vacuum energy comprising 
a conventional cosmological term and a scalar field component, 
and a flat one with a scalar component only, 
aimed respectively at weak and strong age crises. 
Both models suggest that anti-gravity plays a significant 
role in the evolution of the Universe. 

Krauss \cite{krauss} (1999)
says that there are two contenders to explain changes in the expansion rate 
of the Universe,  one is the cosmological constant;  he seems reluctant to 
say what the other is,  but it appears to be some form of dark matter.

Pavsic \cite{pavsic} (1999) studies 
the harmonic oscillator in pseudo euclidean space. 
A straightforward procedure reveals that although such a system 
may have negative energy, it is stable. 
In the quantized theory the vacuum state has to be suitably defined 
and then the zero-point energy corresponding to a positive-signature 
component is canceled by the one corresponding to a negative-signature 
component. 
This principle is then applied to a system of scalar fields. 
The metric in the space of fields is assumed to have signature 
(+ + + ... - - -) and it is shown that the vacuum energy, 
and consequently the cosmological constant, are then exactly zero. 
The theory also predicts the existence of stable, 
negative energy field excitations (the so called "exotic matter") 
which are sources of repulsive gravitational fields, 
necessary for construction of the time machines 
and Alcubierre's hyperfast warp drive. 

Sahni \cite{sahni} (1999) notes that 
the close relationship between the cosmological constant and the vacuum has
been emphasized in the past by Zeldovich amongst others.  Sahni briefly
discusses different approaches to the cosmological constant issue 
including the possibility that it could be generated by vacuum 
polarization in a static universe.   Fresh possibilities occur in an
expanding universe.   An inflationary universe generically leads to particle
creation from the vacuum,  the nature and extent of particle production 
depending upon the mass of the field and its coupling to gravity.
For ultra-light,  non-minimally coupled scalar fields,  particle production 
can be large and the resulting vacuum energy-momentum tensor will have the
form of a cosmological constant.   The inflationary scenario therefore,
could give rise to a universe that is both flat and $\Lambda$-dominated,
in agreement with observation.

Sivaram \cite{sivaram2} (1999) notes that
recent attempts have been made to link vacuum zero-point fields (ZPF) with a 
non-zero cosmological constant ($\Lambda$),  which is now treated as a 
cosmological free variable to be determined by observation.
he claims that in another recent paper,  
$\Lambda$ is related to a graviton mass.
He says that his is shown to be incorrect.   
Flat space propagators for both massless and
massive spin-2 particles can be written (independently of gravity) 
in the context of flat space wave equations;  however,  they do not 
correspond to full general relativity with a graviton mass.

Tkach {\it et al} \cite{TSRN} (1999) 
present a new hidden symmetry in gravity 
for the scale factor in the FRW model, for $k=0$. 
This exact symmetry vanishes the cosmological constant. 
They interpret this hidden symmetry as a dual symmetry 
in the sense that appears in the string theory. 

Vishwakarma \cite{vish} (1999) investigate some Friedmann models 
in which $\Lambda$ varies as $\rho$,
by modifying the Chen and Wu ansatz.
In order to test the consistency of the models with
observations, he studies the angular size - 
redshift relation for 256 ultracompact radio sources 
selected by Jackson and Dodgson. 
The angular sizes of these sources were
determined by using very long-baseline interferometry 
in order to avoid any evolutionary effects. 
The models fit to the data very well and demand an accelerating universe with a
positive cosmological constant. 
Open, flat as well as closed models are almost equally probable, 
though the open model provides comparatively a better fit to the data. 
The models are found to have intermediate density and postulate the 
existence of dark matter, though not as high as in the canonical 
Einstein-deSitter model. 

Axenides {\it et al} \cite{AFP} (2000) note that
Newton's law gets modified in the presence of a cosmological constant 
by a small repulsive term (antigravity) that is proportional to the distance. 
Assuming a value of the cosmological constant consistent 
with the recent SnIa data ($\Lambda \sim eq 10^{-52} m^{-2}$) 
they investigate the significance of this term on various astrophysical scales.
They find that on galactic scales or smaller (less than a few tens of kpc) 
the dynamical effects of the vacuum energy are negligible 
by several orders of magnitude. 
On scales of 1Mpc or larger however they find that vacuum energy 
can significantly affect the dynamics. 
For example they shows that the velocity data in the Local Group of galaxies 
correspond to galactic masses increased by 35\% 
in the presence of vacuum energy. 
The effect is even more important on larger low density systems 
like clusters of galaxies or superclusters. 

Buosso and Polchinski \cite{BPol} (2000) note that
a four-form gauge flux makes a variable contribution 
to the cosmological constant. 
This has often been assumed to take continuous values,  but they argue 
that it has a generalized Dirac quantization condition. 
For a single flux the steps are much larger than the observational limit,
but they show that with multiple fluxes the allowed values 
can form a sufficiently dense `discretuum'. 
Multiple fluxes generally arise in M
theory compactifications on manifolds with non-trivial three-cycles. 
In theories with large extra dimensions a few four-forms suffice;
otherwise of order 100 are needed. 
Starting from generic initial conditions, 
the repeated nucleation of membranes dynamically generates
regions with a cosmological constant in the observational range. 
Entropy and density perturbations can be produced. 

Elizalde \cite{elizalde} (2000)
introduce a simple model in which the cosmological constant 
is interpreted as a true Casimir effect on a scalar field filling the universe
(e.g. $\mathbf{R} \times \mathbf{T}^p\times \mathbf{T}^q$, 
$\mathbf{R} \times \mathbf{T}^p\times \mathbf{S}^q, ...$). 
The effect is driven by compactifying boundary conditions imposed 
on some of the coordinates, associated both with large and small scales. 
The very small,  but non zero-value of the cosmological constant obtained 
from recent astrophysical observations can be perfectly matched with the
results coming from the model, by playing just with the numbers 
of actually compactified ordinary and tiny dimensions, and being the
compactification radius (for the last) in the range $(1-10^3) l_{Pl}$, 
where $l_{Pl}$ is the Planck length. This corresponds to solving, in a
way, what has been termed by Weinberg the {\it new} 
cosmological constant problem. 
Moreover, a marginally closed universe is favored by the model, 
again in coincidence with independent analysis of the observational results.

Fiziev \cite{fiziev} (2000)
notes that in the framework of a model of minimal of dilatonic gravity (MDG) 
with cosmological potential that he considers: the relations of MDG with
nonlinear gravity and string theory; natural cosmological units, 
defined by cosmological constant; the properties of cosmological factor,
derived from solar system and Earth-surface gravitational experiments; 
universal anti-gravitational interactions, induced by positive
cosmological constant and by Nordtveld effect; a new formulation 
of cosmological constant problem using the ratio of introduced
cosmological action and Planck constant $\sim 10 ^{122}$; 
inverse cosmological problem: to find cosmological potential which yields 
given evolution of the RW Universe; 
and comment other general properties of MDG. 

Freedman \cite{freedman} (2000) notes that
rapid progress has been made recently toward the measurement 
of cosmological parameters. 
Still, there are areas remaining where future progress will be
relatively slow and difficult, and where further attention is needed. 
In this review, the status of measurements of the matter density, 
the vacuum energy density
or cosmological constant, the Hubble constant, 
and ages of the oldest measured objects are summarized. 
Many recent, independent dynamical measurements
are yielding a low value for the matter density 
of about 1/3 the critical density. 
New evidence from type Ia supernovae suggests that the vacuum energy density
may be non-zero. 
Many recent Hubble constant measurements appear to be converging 
in the range of 65-75 km/sec/Mpc. 
Eliminating systematic errors lies
at the heart of accurate measurements for all of these parameters; 
as a result, a wide range of cosmological parameter space is currently 
still open.
Fortunately, the prospects for accurately measuring cosmological 
parameters continue to increase. 

Futamase and Yoshida \cite{FY} (2000) propose
a possible measurement of the variability of the vacuum energy,
perhaps here meaning some time dependent cosmological constant, 
using strong gravitational lensing. As an example they take an
Einstein cross lens HST 14176+5226 and show that the measurement 
of the velocity dispersion with the accuracy of $\pm$ 5km/sec
determines the density parameter with the accuracy of order 0.1, 
and it clarifies the existence of the vacuum energy as well as its
variability with redshift. 

Guilini and Straumann \cite{GS} (2000) note that the principles of 
general relativity allow for a non-vanishing 
cosmological constant, which can possibly be interpreted at least partially
in terms of quantum-fluctuations of matter fields. 
Depending on sign and magnitude it can cause accelerated 
or decelerated expansion at certain stages of cosmic evolution. 
Recent observations in cosmology seem to indicate that 
we presently live in an accelerated phase. 
They recall the history and fundamental issues 
connected with the cosmological constant 
and then discuss present evidences for a positive
value, which causes the accelerated expansion. 

Kakushadze \cite{kakushadze} (2000)
considers a recent proposal to solve the cosmological constant problem 
within the context of brane world scenarios 
with infinite volume extra dimensions. 
In such theories bulk
can be supersymmetric even if brane supersymmetry is completely broken. 
The bulk cosmological constant can therefore naturally be zero. 
Since the volume of the extra
dimensions is infinite, it might appear that at large distances 
one would measure the bulk cosmological constant which vanishes. 
He points out a caveat in this argument. 
In particular, he uses a concrete model, 
which is a generalization of the Dvali-Gabadadze-Porrati model, 
to argue that in the presence of non-zero brane cosmological constant at
large distances such a theory might become effectively four dimensional. 
This is due to a mass gap in the spectrum of bulk graviton modes. 
In fact, the corresponding distance
scale is set precisely by the brane cosmological constant. 
This phenomenon appears to be responsible for the fact that 
bulk supersymmetry does not actually protect the brane
cosmological constant. 

Kehagias and Tamvakis \cite{KTam} (2000) discuss
the four-dimensional cosmological constant problem 
in a five-dimensional setting. A scalar field coupled to the SM forms 
dynamically a smooth brane with four-dimensional Poincare invariance, 
independently of SM physics. 
In this respect, their solution might
be regarded as a self-tuning solution, 
free of any singularities and fine-tuning problems. 

Parker and Raval \cite{parker} (2000) propose 
a new model where nonperturbative vacuum contributions to the 
effective action of a free quantized massive scalar field lead to a 
cosmological solution in which the scalar curvature becomes constant 
after a time $t_j$ (when the redshift $z \sim 1$) that depends on the 
mass of the scalar field and its
curvature coupling. This spatially-flat solution implies an accelerating 
universe at the present time and gives a good one-parameter fit to 
high-redshift Type Ia supernovae (SNe-Ia) data, and the present age 
and energy density of the universe. 
Here they show that the imaginary part of the nonperturbative curvature term
that causes the cosmological acceleration, implies a particle production 
rate that agrees with predictions of other methods and extends them to 
non-zero mass fields. 
The particle production rate is very small after the transition and is not 
expected to alter the nature of the cosmological solution. 
They also show that the equation of state of our model undergoes a transition 
at $t_j$ from an equation of state dominated by non-relativistic pressureless 
matter (without a cosmological constant) to an effective equation of state of 
mixed radiation and cosmological constant, 
and they derive the equation of state of the vacuum.
Finally, they explain why nonperturbative vacuum effects 
of this ultra low mass 
particle do not significantly change standard early universe cosmology. 

Straumann \cite{straumann} (2000) explains why
the (effective) cosmological constant is expected to obtain contributions
from short distance physics, corresponding to an energy at least as large 
as the Fermi scale,
after a short history of the $\Lambda$-term.
The actual tiny value of the cosmological 
constant by particle physics standards represents, therefore, 
one of the deepest mysteries of present-day fundamental physics. 
Recent proposals of an approach to the cosmological constant problem 
which make use of (large) extra dimensions are briefly discussed.
Cosmological models with a dynamical $\Lambda$, 
which attempt to avoid the disturbing cosmic coincidence problem, 
are also reviewed. 

The {\it principle of holography} is that only the boundary of a
system needs to be considered,  because the surface determines
the dynamics of the system;  effectively this means that
it is only necessary to work in one less dimension.
Thomas \cite{thomas} (2000) argues that
gravitational holography renders the cosmological constant stable 
against divergent quantum corrections. 
This provides a technically natural solution to 
the cosmological constant problem. 
A natural solution can follow from a symmetry of the action. 
Evidence for quantum stability of the cosmological constant 
is illustrated in a number of examples including, 
bulk descriptions in terms of delocalized degrees of freedom, 
boundary screen descriptions on stretched horizons, 
and non-supersymmetric conformal field theories as dual
descriptions of anti-de Sitter space. 
In an expanding universe, holographic quantum contributions 
to the stress-energy tensor are argued
to be at most of order the energy density of the dominant matter component. 

Tye and Wasserman \cite{TWas} (2000) 
consider a model with two parallel (positive tension) 
3-branes separated by a distance $L$ in 5-dimensional spacetime. 
If the interbrane space is anti-deSitter, or is not
precisely anti-deSitter but contains no event horizons, 
the effective 4-dimensional cosmological constant seen 
by observers on one of the branes (chosen to be the visible brane)
becomes exponentially small as $L$ grows large. 

Chen \cite{chen} (2001) provides a new way of looking at the vacuum which
has implications for the cosmological constant,  see \S2.1 above.

Volovik \cite{volovikge} (2001) discusses condensed matter examples, 
in which the effective gravity appears in the low-energy corner 
as one of the collective modes of quantum vacuum, 
provide a possible answer to the question, why the vacuum energy is so small. 
He says that this answer comes from the fundamental ``trans-Planckian'' 
physics of quantum liquids. In the effective theory of the low energy degrees 
of freedom the vacuum energy density is proportional to the fourth power of 
the corresponding ``Planck'' energy appropriate for this effective theory. 
However,  from the exact ``Theory of Everything'' of the quantum liquid it 
follows that its vacuum energy density is exactly zero without fine tuning,
if: there are no external forces acting on the liquid; 
there are no quasiparticles which serve as matter; 
no spacetime curvature; and no boundaries which give rise to the Casimir 
effect. Each of these four factors perturbs the vacuum state 
and induces the nonzero value of the
vacuum energy density of order of energy density of the perturbation. 
This is the reason, why one must expect that in each epoch the
vacuum energy density is of order of matter density of the Universe, 
or/and of its curvature, or/and of the energy density of smooth
component -- the quintessence. 

Gurzadyan and Xue \cite{GX} (2001) present their view on the problem 
of the cosmological constant and vacuum energy. 
They point out that only relevant modes of the
vacuum fluctuation, whose wavelengths are conditioned by the size, 
homogeneity, geometry and topology of the present Universe,  
contribute to the cosmological constant. 
As a result, the cosmological constant is expressed in terms 
of the size of the Universe and the three
fundamental constants: the velocity of light, 
Planck and Newton gravitational constants. 
They say that its present value remarkably agrees with the
recent observations and its dependence on the size 
of the Universe confronts with observations. 
\subsection{The Anthropic Principle.}
Weinberg \cite{weinberg} considers the {\it anthropic principle}.
He says that in several cosmological theories the observed big bang is just
one member of an ensemble.   The ensemble might consist of different
expanding regions at different times and locations in the same spacetime,
see also Vilenkin \cite{vilenkin} (1983) and Linde \cite{linde} (1986),
or in different terms in the ``wave function of the Universe'',
see Baum \cite{baum} (1984).   
If the vacuum energy density $\rho_{\Lambda}$ varies
among the different members of this ensemble,  then the value observed by any
species of astronomers will be conditioned by the necessity that this value 
of $\rho_{\Lambda}$ should be suitable for the evolution of intelligent life.
Perhaps one can think of the anthropic principle as only allowing certain
sorts of vacuum energy.  
The anthropic principle has also been reviewed by Gale \cite{gale} (1981).
Garriga and Vilenkin \cite{GV} (2000) argue that the anthropic principle
is necessary to explain the time coincidence of a supposed epoch
of galaxy formation and an epoch dominated by the cosmological constant.
One can also think of the anthropic principle in terms of a given set of
conditions or boundaries.   Only universes (more accurately parts of the
Universe) which create conditions for observers to occur can be measured.
One can ask when else one could apply a set of conditions.
{\it Two} places occur.  {\it One} at or near the beginning of the universe,
where a set of initial conditions could be choosen.
The {\it other} at a late time,  where it is not immediately clear 
what sort of conditions one would want to apply.   
Most cosmological models are either too hot or cold at late times and one might
invoke a ``Happy Ending Principle'' that the temperature is in the region 
for life to continue to exist.
Some recent work on the anthropic principle includes the papers 
mentioned below.

Sivaram \cite{sivaram} (1999) says that
an impressive variety of recent observations which include 
luminosity evolutions
of high redshift supernovae strongly suggest that the cosmological constant 
$\Lambda$ is not zero.  Even though the $\Lambda$-term might dominate cosmic 
dynamics at the present epoch,  such a value for the vacuum energy is 
actually unnaturally small.   The difficulty finding a suitable explanation
(based upon fundamental physics) for such a small residue value for the 
cosmological term has led several authors to resort to an anthropic 
explanation of its existence.   Sivaram presents a few examples 
which invoke phase transitions in the early universe involving strong 
or electroweak interactions to show how the cosmic term of the correct 
observed magnitude can arise from fundamental physics involving gravity.

Banks {\it et al} (2000) are
motivated by recent work of Bousso and Polchinski \cite{BP} (2000), 
they study theories which explain the small value of the cosmological constant 
using the anthropic principle. 
They argue that simultaneous solution of the gauge hierarchy problem 
is a strong constraint on any such theory. 
They exhibit {\it three} classes of models which satisfy these constraints. 
The {\it first} is a version of the BP model with precisely two large
dimensions. 
The {\it second} involves 6-branes and antibranes wrapped on 
supersymmetric 3-cycles of Calabi-Yau manifolds, and the {\it third} is
a version of the irrational axion model. 
All of them have possible problems in explaining 
the size of microwave background fluctuations. 
They also find that most models of this type predict 
that all constants in the low energy Lagrangian, 
as well as the gauge groups and representation content, 
are chosen from an ensemble and cannot be uniquely determined 
from the fundamental theory. In their opinion, this
significantly reduces the appeal of this kind of solution 
of the cosmological constant problem. 
On the other hand, they argue that the vacuum
selection problem of string theory might plausibly have an anthropic, 
cosmological solution. 

Donoghue \cite{donoghue} (2000) notes that 
one way that an anthropic selection mechanism might be manifest in a 
physical theory involves multiple domains in the universe with
different values of the physical parameters. 
If this mechanism is to be relevant for understanding 
the small observed value of the cosmological constant, 
it might involve a mechanism by which some contributions 
to the cosmological constant can be fixed at a continuous
range of values in the different domains. 
He studies the properties of four possible mechanisms, 
including the possibility of the Hubble
damping of a scalar field with an extremely flat potential. 
Another interesting possibility involves fixed random values of non-dynamical
form fields, and a cosmological mechanism is suggested. 
This case in addition raises the possibility of anthropic 
selection of other parameters.  He discusses 
further requirements needed for a consistent cosmology. 

Melchiorri and Griffiths \cite{MG} (2000)
analyse boomarang data and find the Universe is very close to flat.
\subsection{Quintessence.}
Quintessence was introduced by Peebles and Ratra \cite{PRat} (1988),
see also Weinberg \cite{weinberg} \S2 (2000).
The idea is that the cosmological constant is 
small because the Universe is old.
How instead of a cosmological constant vacuum energy might be manifest 
as a scalar fields is described in \S2.1\P3.
This scalar field perhaps can be thought of 
as a time dependent cosmological constant.
Some recent work on quintessence includes the papers below.

Zlatev {\it et al} \cite{ZWS} (1998) notes that
recent observations suggest that a large fraction of the energy density 
of the universe has negative pressure. 
One explanation is vacuum energy density; another is quintessence
in the form of a scalar field slowly evolving down a potential. 
In either case, a key problem is to explain why the energy density 
nearly coincides with the matter density today. 
The densities decrease at different rates as the universe expands, 
so coincidence today appears to require that their ratio be set to a specific,
infinitesimal value in the early universe. 
In this paper, they introduce the notion of a "tracker field", 
a form of quintessence, and show how it may explain the coincidence, 
adding new motivation for the quintessence scenario. 

Bento and Bertolami \cite{BB} (1999) 
study the possibility that the vacuum energy density of scalar 
and internal-space gauge fields arising from the process of 
dimensional reduction of higher
dimensional gravity theories plays the role of quintessence. 
They show that, for the multidimensional Einstein-Yang-Mills 
system compactified on a $R \times S^3 \times S^d$ topology, 
there are classically stable solutions such that the observed 
accelerated expansion of the Universe at present can be accounted for
without upsetting structure formation scenarios or violating 
observational bounds on the vacuum energy density. 

Chiba \cite{chiba} (1999) notes that 
dynamical vacuum energy or quintessence, a slowly varying and spatially 
inhomogeneous component of the energy density with negative pressure, 
is currently consistent with the observational data. 
One potential difficulty with the idea of quintessence 
is that couplings to ordinary matter should be strongly suppressed
so as not to lead to observable time variations of the constants of nature. 
He further explores the possibility of an explicit coupling between the 
quintessence field and the curvature. 
Since such a scalar field gives rise to another gravity force 
of long range ($\sim H^{-1}_0$), the solar system experiments put a
constraint on the non-minimal coupling: $|\xi| \sim 10^{-2}$. 

Chimento {\it et al} \cite{CJP} (2000) note that the combined effect 
of a dissipative fluid and quintessence energy can simultaneously 
drive an accelerated expansion phase at the present
time and solve the coincidence problem of our current Universe. 
A solution compatible with the observed cosmic acceleration is succinctly
presented.   In Chimento {\it et al} \cite{CJP2} (2000)
they show that the combination of a fluid with a bulk dissipative 
pressure and quintessence matter can simultaneously drive an accelerated
expansion phase and solve the coincidence problem of our current Universe. 
They then study some scenarios compatible with the observed
cosmic acceleration. 

Hebecker and Wetterich \cite{HW} (2000) 
formulate conditions for the naturalness of cosmological 
quintessence scenarios. 
They take the quintessence lagrangian is taken to be the sum
of a simple exponential potential and a non-canonical kinetic term. 
This parameterization covers most variants of quintessence and
makes the naturalness conditions particularly transparent. 
Several ``natural'' scalar models lead, for the present cosmological era, 
to a large fraction of homogeneous dark energy density and an acceleration 
of the scale factor as suggested by observation. 

Amendola \cite{amendola} (1999) notes that
a new component of the cosmic medium, a light scalar field 
or ''quintessence '', has been proposed recently to explain cosmic acceleration
with a dynamical cosmological constant. 
Such a field is expected to be coupled explicitly to ordinary matter, 
unless some unknown symmetry prevents it. 
He investigates the cosmological consequences 
of such a coupled quintessence (CQ) model, assuming an exponential
potential and a linear coupling. 
This model is conformally equivalent to Brans-Dicke Lagrangians 
with power-law potential. 
He evaluates the density perturbations on the cosmic microwave background 
and on the galaxy distribution at the present and derive bounds on the 
coupling constant from the comparison with observational data. 
A novel feature of CQ is that during the matter dominated era 
the scalar field has a finite and almost constant energy density. 
This epoch,  denoted as $\phi $MDE, is responsible of several differences 
with respect to uncoupled quintessence: the multipole spectrum of the
microwave background is tilted at large angles, 
the acoustic peaks are shifted, their amplitude is changed, 
and the present 8Mpc$/h$ density variance is diminished. 
The present data constrain the dimensionless coupling constant 
to $|\beta |\leq 0.1$.

Wiltshire \cite{wiltshire} (2000) notes that
quintessence models with a dark energy generated by pseudo 
Nambu-Goldstone bosons provide a natural framework in which to test the
possibility that type Ia supernovae luminosity distance measurements 
are at least partially due to an evolution of the sources, since these
models can have parameter values for which the expansion of the Universe 
is decelerating as well as values for which it is accelerating,
while being spatially flat in all cases and allowing for a low density 
of clumped matter. The results of a recent investigation by Wiltshire 
of current observational bounds which allow for SNe Ia source evolution 
are discussed. He finds that models with
source evolution still favour cosmologies 
with an appreciable amount of acceleration in the recent past, 
but that the region of parameter
space which is most favoured shifts significantly. 
\subsection{Inflation.}
The idea here is that the scale factor $R$ of Robertson-Walker
spacetime takes an exponential form $R=\exp(at)$ 
which is supposed to correspond
to the size of the Universe increasing exponentially.
deSitter spacetime can be caste in $R=\exp(at)$ form,
and deSitter spacetime is static;  however in the inflation picture
the Universe is expanding exponentially,  the resolution of this is
apparently in Schr\"odinger \cite{schrodinger} (1956).
Pollock and Dahdev \cite{PD} (1989),  discuss how induced gravity,
see \S 2.15 fits in with inflation.
There is a lot written on inflation,  
here about 15 recent papers are mentioned.

Knox and Turner \cite{KTu} (1993) 
present a simple model for slow-rollover inflation where the 
vacuum energy that drives inflation is of the order of $G_F^{-2}$; 
unlike most models, the conversion of vacuum
energy to radiation (``reheating'') is moderately efficient. 
The scalar field responsible for inflation is a standard-model singlet, 
develops a vacuum expectation value of the order
of $4\times 10^6 GeV$, has a mass of order $1 GeV$, 
and can play a role in electroweak phenomena. 

Spokoiny \cite{spokoiny} (1993) 
shows that it is possible to realize an inflationary scenario 
even without conversion of the false vacuum energy to radiation. 
Such cosmological models have a deflationary
stage in which $Ha^2$ is decreasing and radiation 
produced by particle creation in an expanding Universe becomes dominant. 
The preceding inflationary stage ends since the
inflaton potential becomes steep. 
False vacuum energy is finally (partly) 
converted to the inflaton kinetic energy, 
the potential energy rapidly decreases and the Universe comes
to the deflationary stage with a scale factor $a(t) \propto t^{1/3}$. 
Basic features and observational consequences of this scenario are indicated.

Copeland {\it et al} \cite{CLLSW} (1994) 
investigate chaotic inflation models with two scalar fields, 
such that one field (the inflaton) rolls while the other 
is trapped in a false vacuum state. 
The false vacuum becomes unstable when the inflaton field falls 
below some critical value, and a first or second order transition 
to the true vacuum ensues. 
Particular attention is paid to Linde's
second-order `Hybrid Inflation'; with the false vacuum dominating, 
inflation differs from the usual true vacuum case both in its cosmology 
and in its relation to particle physics.
The spectral index of the adiabatic density perturbation 
can be very close to 1, or it can be around ten percent higher. 
The energy scale at the end of inflation can be anywhere
between $10^{16}$\,GeV and $10^{11}$\,GeV, though reheating is prompt 
so the reheat temperature can't be far below $10^{11}\,$GeV. 
Topological defects are almost
inevitably produced at the end of inflation, 
and if the inflationary energy scale is near 
its upper limit they can have significant effects. 
Because false vacuum inflation occurs with
the inflaton field far below the Planck scale, 
it is easier to implement in the context of supergravity 
than standard chaotic inflation. 
That the inflaton mass is small compared with
the inflationary Hubble parameter is still a problem 
for generic supergravity theories, but remarkably this can be avoided 
in a natural way for a class of supergravity models which
follow from orbifold compactification of superstrings. 
This opens up the prospect of a truly realistic, superstring.

Gaillard {\it et al} \cite{GMO} (1995) note that supersymmetry 
is generally broken by the non-vanishing vacuum energy density 
present during inflation. 
In supergravity models, such a source of supersymmetry breaking
typically makes a contribution to scalar masses of order 
${\tilde m}^2 \sim H^2$, where $H^2 \sim V/M_P^2$ 
is the Hubble parameter during inflation. 
They show that in supergravity models which possess a Heisenberg symmetry, 
supersymmetry breaking makes no contribution to scalar masses, 
leaving supersymmetric flat directions flat at tree-level. 
One-loop corrections in general lift the flat directions, 
but naturally give small negative squared masses $\sim - g^2 H^2/(4\pi)^2$ 
for all flat directions that do not involve the stop. 
No-scale supergravity of the SU(N,1) type and the untwisted sectors 
from orbifold compactifications are special cases 
of this general set of models. 
They point out the importance of the preservation 
of flat directions for baryogenesis. 

Gilbert \cite{gilbert} (1995) examines
inflationary potentials which produce power-law density perturbations. 
The models derived are dominated by a false vacuum energy at late times 
and inflate indefinitely. 
This paper
also examines the effects on the fluctuation spectrum
of reducing the potential to end inflation. 
Small reductions of the potential result in little change 
to the perturbation spectrum. 
The effect of a
large reduction, however, is to change the sign 
of the slope of the fluctuation spectrum. 

Berera \cite{berera} (1996) notes that 
in the standard picture, the inflationary universe 
is in a supercooled state which ends with a short time, 
large scale reheating period, after which the universe goes into a radiation
dominated stage. 
Here he proposes an alternative in which the radiation energy density 
smoothly decreases 
during an inflation-like stage 
and with no discontinuity enters the subsequent radiation dominated stage. 
The scale factor is calculated from standard Friedmann cosmology 
in the presence of both radiation and vacuum energy density. 
A large class of solutions confirm the above identified regime 
of non-reheating inflation-like behavior for observationally 
consistent expansion factors and not too large a drop in the
radiation energy density. One dynamical realization of such 
inflation without reheating is from warm inflation type scenarios. 
However the solutions found here are properties of
the Einstein equations with generality beyond slow-roll inflation scenarios. 
The solutions also can be continuously interpolated from the non-reheating 
type behavior to the standard supercooled limit of exponential expansion, 
thus giving all intermediate inflation-like behavior 
between these two extremes. 
The temperature of the universe and the expansion factor are calculated 
for various cases. 
Implications for baryongenesis are discussed. 
This non-reheating, inflation-like regime also appears to have some natural
features for a universe that is between nearly flat and open. 

Boyananovsky {\it et al} (1996) \cite{BVHS} analyzes 
the phenomenon of preheating,  i.e. explosive 
particle production due to parametric amplification of 
quantum fluctuations in the unbroken case, or spinodal 
instabilities in the broken phase, using the Minkowski 
space $O(N)$ vector model in the large $N$ limit to
study the non-perturbative issues involved. 
They give analytic results for weak couplings 
and times short compared to the time at which
the fluctuations become of the same order as the tree level,
as well as numerical results including the full backreaction.
In the case where the symmetry is unbroken, 
the analytic results agree spectacularly well 
with the numerical ones in their common domain of validity. 
In the broken symmetry case, slow roll initial conditions 
from the unstable minimum at the origin, give rise to a new and unexpected
phenomenon: the dynamical relaxation of the vacuum energy.
That is, particles are abundantly produced at the expense of the quantum
vacuum energy while the zero mode comes back to almost its initial value.
In both cases we obtain analytically and numerically the
equation of state which turns to be written in terms 
of an effective polytropic index that interpolates 
between vacuum and radiation-like domination.    
They find that simplified analysis based 
on harmonic behavior of the zero mode, 
giving rise to a Mathieu equation for the
non-zero modes miss important physics. 
Furthermore, analysis that do not include 
the full backreaction do not conserve energy,
resulting in unbound particle production. 
Their results do not support the recent claim 
of symmetry restoration by non-equilibrium fluctuations.
Finally estimates of the reheating temperature are given,
as well as a discussion of the inconsistency of a kinetic approach to
thermalization when a non-perturbatively large number of particles is created. 

Yamammoto {\it et al} \cite{YST} (1996) 
first develop a method to calculate a complete set 
of mode functions which describe the quantum fluctuations 
generated in one-bubble open inflation models. 
They consider two classes of models. 
One is a single scalar field model proposed 
by Bucher, Goldhaber and Turok and by them as 
an example of the open inflation scenario, and the other is a
two-field model such as the ``supernatural'' inflation 
proposed by Linde and Mezhlumian. 
In both cases they assume the difference 
in the vacuum energy density between inside
and outside the bubble is negligible. 
There are two kinds of mode functions. 
One kind has usual continuous spectrum 
and the other has discrete spectrum with characteristic
wavelengths exceeding the spatial curvature scale. 
The latter can be further divided into two classes in terms of its origin. 
One is called the de Sitter super-curvature mode,
which arises due to the global spacetime structure of de Sitter space, 
and the other is due to fluctuations of the bubble wall. 
They calculate the spectrum of quantum fluctuations in
these models and evaluate the resulting large angular scale CMB anisotropies. 
They find there are ranges of model parameters 
that are consistent with observed CMB anisotropies. 

Dine and Riotto \cite{DR} (1997) note that
inflation, as currently understood, requires the presence of fields 
with very flat potentials. 
Supersymmetric models in which supersymmetry breaking is communicated by
supergravity naturally yield such fields, but the scales are typically 
not suitable for obtaining both sufficient inflation and a suitable 
fluctuation spectrum. 
In the context of recent ideas about gauge mediation, 
there are new candidates for the inflaton. 
They present a simple model for slow-rollover 
inflation where the vacuum energy driving inflation is related
to the same F-term responsible for the spectrum of supersymmetric 
particles in gauge mediated supersymmetry breaking models. 
The inflaton is identified with field responsible
for the generation of the $\mu$-term. 
This opens the possibility of getting some knowledge 
about the low-energy supersymmetric theory from measurements of the cosmic
microwave background radiation. 
They say that gravitinos do not pose a cosmological problem, 
while the moduli problem is ameliorated. 

Guzm\'an (1997) \cite{guzman} finds that for the Bianchi types I-II-III-V
in the Brans-Dicke theory,  
the scalar field of the theory $\phi$ has the same form in the
isotropic case.  He shows that isotropization of the Universe 
occurs in a very short time when the Universe is dominated by vacuum energy,  
proving that an isotropic Robertson-Walker model is a good approximation 
to use in the extended inflation scenario.
Petry (1997) \cite{petry} finds at the beginning of the Universe radiation,  
matter and vacuum energy given by the cosmological constant are zero 
and then emerge from gravitational energy.   
In the course of time the energy of radiation and matter 
decreases whereas the vacuum energy increases for ever.

Berera {\it et al} \cite{BGR} (1998) present
a quantum field theory warm inflation model is presented 
that solves the horizon/flatness problems. 
They say that the model obtains, 
from the elementary dynamics of particle physics,
cosmological scale factor trajectories that begin in a radiation 
dominated regime, enter an inflationary regime and then smoothly 
exit back into a radiation dominated regime, with
nonnegligible radiation throughout the evolution. 

Malik and Wands \cite{MW} (1998) 
investigate the dynamics of the recently proposed 
model of assisted inflation. 
In this model an arbitrary number of scalar fields 
with exponential potentials evolve towards an
inflationary scaling solution, even if each of the 
individual potentials is too steep to support inflation on its own. 
By choosing an appropriate rotation in field space they can write
down explicitly the potential for the weighted mean field along 
the scaling solution and for fields orthogonal to it. 
This demonstrates that the potential has a global minimum along
the scaling solution. 
They show that the potential close to this attractor 
in the rotated field space is analogous to a hybrid inflation model, 
but with the vacuum energy having an
exponential dependence upon a dilaton field. 
They present analytic solutions describing homogeneous 
and inhomogeneous perturbations about the attractor solution without
resorting to slow-roll approximations. 
They discuss the curvature and isocurvature 
perturbation spectra produced from vacuum fluctuations 
during assisted inflation. 

Starkman {\it et al} \cite{STV} (1999) note that
current observations of Type Ia supernovae provide evidence 
for cosmic acceleration out to a redshift of $z \sim 1$, 
leading to the possibility that the universe is entering an
inflationary epoch. 
However, inflation can take place only if vacuum-energy 
(or other sufficiently slowly redshifting source of energy density) 
dominates the energy density of a region of physical radius 1/H. 
They argue that for the best-fit values of $\Omega_\Lambda$ and $\Omega_m$ 
inferred from the supernovae data, one must confirm cosmic
acceleration out to at least $z \simeq 1.8$ to infer that 
the universe is inflating. 

For Guendelman's \cite{guendelman} (2000) 
recent discussion of inflation,  see \S2.18 above,
and \cite{G2,G3,G4}.
\subsection{Vacuum Energy as Critical Density.}
Vacuum energy can also be considered to take what Sciama,  see \S4.1\P2
would consider to be a ``non-symmetric'' form as
just being a density.   More specifically it can be considered to be background
critical density.   See also Sokolov \cite{sokolov} (1994) and \S3.6.
Some recent papers on this include those below.

Kosowsky and Turner \cite{KoT} (1992)
introduce an approximation to calculate the gravitational radiation 
produced by the collision of true-vacuum bubbles that is simple enough 
to allow the simulation of a phase transition by the collision of hundreds 
of bubbles. This ``envelope approximation'' neglects the complicated 
``overlap'' regions of colliding bubbles and follows only the evolution
of the bubble walls. 
The approximation accurately reproduces previous results for the 
gravitational radiation from the collision of two scalar-field vacuum bubbles.
Using a bubble nucleation rate given by $\Gamma = \Gamma_0 e^{\beta t}$, 
they simulate a phase transition by colliding 20 to 200 bubbles; 
the fraction of vacuum energy released into gravity waves is 
$E_{\rm GW}/E_{\rm vac} = 0.06(H/\beta)^2$ and the peak of the spectrum 
occurs at $\omega_{\rm max}=1.6\beta$ ($H^2=8\pi G\rho /3$ 
is the Hubble constant associated with the false-vacuum phase). 
The spectrum is very similar to that in the two-bubble case, 
except that the efficiency of gravity-wave generation is about five times
higher, presumably due to the fact that a given bubble collides with many 
others. 
Finally, they consider two further ``statistical'' approximations, 
where the gravitational radiation is
computed as an incoherent sum over individual bubbles weighted 
by the distribution of bubble sizes. These approximations provide 
reasonable estimates of the gravitational-wave
spectrum with far less computation. 

John and Babu \cite{JB} (1996) discuss a modified version of the
\"Ozer-Taha \cite{OT} (1986) nonsingular cosmological model.
John and Babu assume that the universe's radius is complex if it
is regarded as empty,  but that it contains matter when the radius is real.
Their model predicts the values:
$\Omega_M\equiv\rho_M/\rho_C\approx4/3$,
$\Omega_V\equiv\rho_V/\rho_C\approx2/3$ and
$\Omega_-\equiv\rho_-/\rho_c<<1$ in the present non-relativistic era,  where
$\rho_m=$matter density,
$\rho_V=$negative energy density,  not necessarily the cosmological constant,
but perhaps a spacetime dependent cosmological constant in the style of
Chen and Wu \cite{CWu} (1990),  and
$\rho_C=$critical density.

Lima \cite{lima} (1996) derives 
a new Planckian distribution for cosmologies 
with photon creation using thermodynamics 
and semiclassical considerations. 
This spectrum is preserved during the evolution 
of the universe and compatible with the present 
spectral shape of the cosmic microwave background radiation(CMBR). 
Accordingly, the widely spread feeling that
cosmologies with continuous photon creation 
are definitely ruled out by the COBE limits 
on deviation of the CMBR spectrum from blackbody shape should be reconsidered.
It is argued that a crucial test for this kind of cosmologies 
is provided by measurements of the CMBR temperature at high redshifts. 
For a given redshift $z$ greater than zero, the
temperature is smaller than the one predicted by the standard FRW model. 

Adams and Laughlin \cite{ALau} (1997) 
outline astrophysical issues related 
to the long term fate of the universe. 
They consider the evolution of planets, 
stars, stellar populations, galaxies, and the universe
itself over time scales which greatly exceed the current 
age of the universe. 

Beane \cite{beane} (1997) notes that
general arguments suggest the existence of at least one
unobserved scalar particle with Compton wavelength bounded 
from below by one tenth of a millimeter,
if the mechanism responsible for the smallness of the vacuum energy 
is consistent with local quantum field theory.
He shows that this bound is saturated if vacuum energy is a substantial
component of the energy density of the universe. 
Therefore, the success of cosmological models 
with a significant vacuum energy component suggests the existence of new
macroscopic forces with range in the sub-millimeter region. 
There are virtually no experimental constraints on the existence 
of quanta with this range of interaction. 

Gentry \cite{gentry} (1997) notes that 
a nonhomogeneous universe with vacuum energy, 
but without spacetime expansion, is utilized 
together with gravitational and Doppler redshifts as the basis
for proposing a new interpretation of the Hubble relation 
and the 2.7K cosmic blackbody radiation. 

Arbab \cite{arbab} (1999) finds 
the universe is found to have undergone several phases 
in which the gravitational constant had a different behaviour. 
During some epoch the energy density of the universe
remained constant and the universe remained static. 
In the radiation dominated epoch the radiation field 
satisfies the Stefan's formula while the scale factor varies linearly with.
The model enhances the formation of the structure 
in the universe as observed today. 

Hogan \cite{hogan} (1999) presents 
a brief but broad survey is presented of the flows, 
forms and large-scale transformations of mass-energy in the universe, 
spanning a range of about twenty orders of magnitude
($m_{Planck}/m_{proton}$) in space, time and mass. 
Forms of energy he considers include electromagnetic radiation, 
magnetic fields, cosmic rays, gravitational energy and
gravitational radiation, baryonic matter, dark matter, 
vacuum energy, and neutrinos; 
sources considered include vacuum energy and cosmic expansion, 
fluctuations and gravitational collapse, AGN and quasars, stars, 
supernovae and gamma ray bursts. 

Kujat and Scherrer \cite{KSch} (1999) note that
a time variation in the Higgs vacuum expectation value 
alters the electron mass and thereby changes 
the ionization history of the universe. 
This change produces a measurable
imprint on the pattern of cosmic microwave background (CMB) fluctuations. 
The nuclear masses and nuclear binding energies, 
as well as the Fermi coupling constant, are also
altered, with negligible impact on the CMB. 
They calculate the changes in the spectrum of the CMB fluctuations 
as a function of the change in the electron mass. 
They find that future CMB experiments could be sensitive to 
|$\Delta m_e/m_e| \sim |\Delta G_F/G_F| \sim 10^{-2} - 10^{-3}$. 
However, they also show that a change in the electron mass is
nearly, but not exactly, degenerate with a change in the 
fine-structure constant. 
If both the electron mass and the fine-structure constant are time-varying, 
the corresponding
CMB limits are much weaker, particularly for $l<1000$. 

Pascual-S\'achez \cite{pas} (1999) 
develops an alternative explanation for the acceleration 
of the cosmic expansion, which seems to be a result of recent high redshift 
Supernova data. 
In the current interpretation, this cosmic acceleration is explained by 
including a positive cosmological constant term (or vacuum energy), 
in the standard Friedmann models. 
Instead, he considers a locally rotationally symmetric (LRS) 
and spherically symmetric (SS), but inhomogeneous spacetime, with a
barotropic perfect fluid equation of state for the cosmic matter. 
The congruence of matter has acceleration, shear and expansion. 
Within this framework the kinematical acceleration of the cosmic fluid or, 
equivalently, the inhomogeneity of matter, is just the responsible of the 
SNe Ia measured cosmic acceleration.
Although in his model the cosmological principle is relaxed, 
it maintains almost isotropy about our worldline 
in agreement with CBR observations. 

Zimdahl {\it et al} (2000) explain an accelerated expansion 
of the present universe, suggested from observations 
of supernovae of type Ia at high redshift, by introducing an anti-frictional 
force that is self-consistently exerted on the particles of the cosmic 
substratum. Cosmic anti-friction,
which is intimately related to ``particle production'', 
is shown to give rise to an effective negative pressure of the cosmic medium. 
While other explanations for an accelerated expansion 
(cosmological constant, quintessence) introduce a component 
of dark energy besides ``standard'' cold dark matter (CDM) 
they resort to a phenomenological one-component model 
of CDM with internal self-interactions. 
They demonstrate how the dynamics of the $\Lambda$CDM model might 
be recovered as a special case of cosmic anti-friction. 
The connection with two-component models is discussed, 
providing a possible phenomenological solution to the coincidence problem. 
\subsection{Equating Vacuum Energy with Dark Matter.}
Vacuum energy being non-tangible and (excluding the Casimir effect)
so far non-measurable is idealy equated
with dark matter which is likewise.
Another way of thinking about
the quantum vacuum is that it might have large scale effects,  it 
could be just a different way of referring to dark matter - albeit 
with a different physical interpretation. 
Some recent papers which can be thought 
of as equating vacuum energy with dark matter include the following.

Spindel and Brout \cite{SBr} (1993) 
use as dynamical variable the square of the radius of the Universe, 
they solve analytically the Einstein equations in the framework of 
Robertson-Walker models where a
cosmological constant describing phenomenologically 
the vacuum energy decays into radiation. 
Emphasis is put on the computation of the entropy creation. 

Abdel-Rahman \cite{abdel} (1995) 
introduces a nonsingular closed universe 
model with continuous creation of radiation 
or matter from the vacuum.
Although primordial nucleosynthesis in the model follows 
the standard scenario it does not require the density 
of the baryonic matter to be below the critical density 
as in standard cosmology.   The model predicts a present 
a present vacuum energy comparable with matter energy.   
Its predictions for classical low red-shift cosmological 
tests agree with the standard flat model results.

Lima  and Maia \cite{limamaia} (1995)
deduce the spectrum of  a pure vacuum state, assuming
that it is formed by a kind of radiation  satisfying the 
equation of state $p=-\rho$.
Actually,  in this paper they deduce the spectrum for a large class of
massless particles satisfying the equation of state 
$p=(\gamma-1)\rho$.  Particular attention is given for the vacuum case 
($\gamma=0$)
and its thermodynamic behavior in their section V.  From a historical
viewpoint, the introduction is also of interest.   The
vacuum spectrum  deduced here is completely different of the one
appearing in the zero-point approach of stochastic electrodynamics. Some
consequences for cosmology are discussed in in their section VII.

Gr{\o}n and Soleng \cite{GSol} (1996) note that
on the scales of galaxies and beyond there is evidence for unseen dark matter.
In this paper they find the experimental limits to the density of dark matter 
bound in the solar system by studying its effect upon planetary motion.
Roberts \cite{mdrpluto} (1987) has studied how the cosmological constant
might alter the orbit of Pluto.   Van Flandern \cite{vanflandern} (1999)
has also studied possible connections between dark matter and solar system 
dynamics.

Pfenning and Ford \cite{PF} (1996) develop
quantum inequality restrictions on the stress-energy tensor 
for negative energy for three and four-dimensional 
static spacetimes. 
They derive a general inequality in terms of a sum of mode functions 
which constrains the magnitude and duration of negative energy seen 
by an observer at rest in a static spacetime. 
This inequality is evaluated explicitly for 
a minimally coupled scalar field in three and 
four-dimensional static Robertson-Walker universes. 
In the limit of vanishing curvature, the flat spacetime inequalities 
are recovered. 
More generally, these inequalities contain the effects of spacetime curvature. 
In the limit of short sampling times, 
they take the flat space form plus subdominant
curvature-dependent corrections. 

Antonsen and Bormann \cite{AB} (1998) note that
for a Friedman-Robertson-Walker spacetime in which the only contribution 
to the stress-energy tensor comes from the renormalised zero-point energy 
(i.e. the Casimir energy) of the fundamental fields the evolution of the 
universe (the scale factor) depends upon whether the universe is open, 
flat or closed and upon which fundamental fields
inhabit the space-time. 
They calculate this "Casimir effect" using the heat kernel method, 
and the calculation is thus non-perturbative. 
They treat fields of spin $0,1/2,1$ coupled to
the gravitational background only. 
The heat kernels and/or $\zeta$-functions for the various spins 
are related to that of a non-minimally coupled one. 
A WKB approximation is used in obtaining the radial part of that heat kernel. 
The simulations of the resulting equations of motion seem to exclude the 
possibility of a closed universe, $K=+1$, as these turn out
to have an overwhelming tendency towards a fast collapse - the details 
such as the rate of this collapse depends on the structure of 
the underlying quantum degrees of freedom: a
non-minimal coupling to curvature accelerates the process. 
Only $K=-1$ and K=0 will in general lead to macroscopic universes, 
and of these $K=-1$ seems to be more favourable. 
The possibility of the scale factor being a concave rather 
than a convex function potentially indicates that the problem 
of the large Hubble constant is non-existent as
the age of the universe need not be less than or equal to the Hubble time. 
Note should be given to the fact, however, that we are not able to pursue 
the numerical study to really
large times neither do simulations for a full standard model. 

\"Ozer \cite{ozer} (1999) 
shows that in the cosmological models based on 
a vacuum energy decaying as $a^{-2}$, 
where a is the scale factor of the universe, 
the fate of the universe in regard to whether 
it will collapse in future or expand forever 
is determined not by the curvature constant $k$ 
but by an effective curvature constant $k_{eff}$. 
He argues that a closed universe with k=1 may expand forever, 
in other words simulate the expansion dynamics of a flat 
or an open universe because of the possibility that $k_{eff}=0~{\rm or}-1$, 
respectively. 
Two such models, in one of which the vacuum does not interact 
with matter and in another of which it does, are
studied. He shows that the vacuum equation of state 
$p_{vac}= -\rho_{vac}$ may be realized in a decaying vacuum 
cosmology provided the vacuum interacts with matter. 
The optical depths for gravitational lensing as a 
function of the matter density and other parameters 
in the models are calculated at a source
redshift of 2. The age of the universe is discussed 
and shown to be compatible with the new Hipparcos lower limit of 11Gyr. 
He suggests the possibility that a time-varying
vacuum energy might serve as dark matter. 

Turner \cite{turner} (1999) notes that 
more than sixty years ago Zwicky made the case 
that the great clusters of galaxies are held 
together by the gravitational force of unseen (dark) matter. 
Today, he claims that the case is
stronger and more precise;
I disagree and prefer dynamical explanations.
Turner says that dark, nonbaryonic matter accounts for 
30\% +/- 7\% of the critical mass density, with baryons 
(most of which are dark) contributing only 4.5\% +/- 0.5\% 
of the critical density. 
The large-scale structure that exists in the Universe 
indicates that the bulk of the nonbaryonic dark matter 
must be cold (slowly moving particles). 
The Super Kamiokande detection of neutrino oscillations 
shows that particle dark matter exists, crossing an important threshold. 
Over the past few years a case has developed for a dark-energy problem. 
This dark component contributes about 80\% +/- 20\% 
of the critical density and is characterized by very negative pressure 
($p_X < -0.6 rho_X$). 
Consistent with this picture of dark energy and dark matter 
are measurements of CMB anisotropy that indicate that total contribution 
of matter and energy is within 10\% of the critical density. 
Fundamental physics beyond the standard model is implicated 
in both the dark matter and dark energy puzzles: 
new fundamental particles (e.g., axion or neutralino) and
new forms of relativistic energy 
(e.g., vacuum energy or a light scalar field). 
Note that here Turner is not necessarily equating vacuum energy 
with the cosmological constant.
A flood of observations will shed light on the dark side 
of the Universe over the next two decades; as
it does it will advance our understanding of the Universe 
and the laws of physics that govern it. 
\subsection{Sonoluminescence.}
Sonoluminescence was first found to occur when degassed water is irradiated 
by ultra-sound,  Frenzel and Schultes \cite{FSch} (1934).
A stable sonoluminescence can be contrived with a bubble that is trapped
at the pressure anti-node of a standing sound wave in a spherical 
or cylindrical container and that collapses and reexpands 
with the periodicity of the sound.
Barber and Putterman \cite{BP} (1991) 
note that sonoluminescence is a non-equilibrium phenomenon in which the energy
in a sound wave becomes highly concentrated so as to generate flashes of light
in a liquid.   They show that these flashes,  which comprise of over $10^5$
photons,  are too fast to be resolved by the fastest photomultiplier tubes
available.   Furthermore,  when sonoluminescence is driven by a resonant 
sound field,  the bursts can occur in a continuous repeating,  regular fashion.
These precise 'clock-like' emissions can continue for hours at drive
frequencies ranging from audible to ultrasonic.   These bursts represent an
amplification of energy by eleven orders of magnitude.
A light pulse is emittied during every cycle of the sound wave see 
Gaitan {\it et al} \cite{GCCR}(1992).
Schwinger \cite{schwinger} (1994) suggested that the mechanism responsible for
radiation in sonoluminescence is a dynamical version of the Casimir effect,
compare \S3.5.   Boundaries are an essential ingredient of the Casimir effect
and in the present case Schwinger takes them to be given by the boundary 
between  a dielectric medium (the water) and the vacuum 
(the gas inside the bubble).   Perhaps the Universe,  or more accurately
a large segment of ther Universe,  can be modeled by the vibrating bubble;
in that case the study of sonoluminescence is an analog of the study
of QFT in cosmological backgrounds that can be subject to empirical testing.
A vibrating bubble analog of the Universe might have application in the 
study of discrete red shift,  compare Roberts \cite{mdrqp} (2000).

Eberlein \cite{eberlein} (1995) explains
sonoluminescence in terms of quantum radiation by moving 
interfaces between media of different polarizability. 
It can be considered as a dynamic Casimir effect, in
the sense that it is a consequence of the imbalance 
of the zero-point fluctuations of the electromagnetic field 
during the non-inertial motion of a boundary. 
The transition amplitude from the vacuum into a two-photon state 
is calculated in a Hamiltonian formalism and turns out to be governed 
by the transition matrix-element of the radiation pressure. 
She gives expressions for the spectral density and the total radiated energy.

Chodos \cite{chodos} (1996) notes that
the phenomenon of sonoluminescence (SL), 
originally observed some sixty years ago, 
has recently become the focus of renewed interest,
particularly with the discovery that one can trap 
a single bubble and induce it to exhibit SL stably 
over a large number of acoustical cycles. 
In this work he adopts a version of the 
provocative suggestion put forward by Schwinger: 
the mechanism responsible for the radiation
in SL is a dynamic version of the Casimir effect. 
It has been known since Casimir's original work in \cite{casimir} (1948) 
that the zero-point energy of
quantum fields can be modified by the presence of boundaries, 
and that these modifications generate observable effects. For example, in
Casimir's original work, the quantum fluctuations 
of the electromagnetic field in the presence 
of a pair of uncharged, parallel, perfectly
conducting plates were shown to give rise 
to an attractive force between the plates. 

Liberati {\it et al} \cite{LVBS} (1999) investigate several 
variations of Schwinger's proposed mechanism for sonoluminescence. 
They demonstrated that any realistic version of Schwinger's mechanism 
must depend on extremely rapid (femtosecond) changes in refractive index, 
and discussed ways in which this might be physically plausible. 
To keep that discussion tractable, the technical computations in their 
earlier paper were limited to the case of a homogeneous dielectric medium. 
In their later paper they investigate the additional
complications introduced by finite-volume effects. 
The basic physical scenario remains the same, 
but now there are finite spherical bubbles, and so must decompose the
electromagnetic field into spherical harmonics and Bessel functions. 
They demonstrate how to set up the formalism for calculating Bogolubov 
coefficients in the sudden approximation, and show that qualitatively 
the results previously obtained using the homogeneous-dielectric 
(infinite volume) approximation are retained.

See also Milton \cite{milton} (2000) \S3.5.
\subsection{Quantum Cosmology.}
Quantum cosmology requires notions of a vacuum from all {\it three} 
of the previous sections. 
From the {\it first} because it is a quantum field theory;
from the {\it second},  specifically \S3.9,  because it is on curved spacetime;
and from the {\it third} because it is concerned with large scales.
Some recent papers on quantum cosmology that mention the vacuum include 
those mentioned below.

Brevik {\it et al} \cite{BMOO} (2000)
apply the background field method and the effective action formalism 
to describe the four-dimensional dynamical Casimir effect. 
Their picture corresponds to the consideration
of quantum cosmology for an expanding FRW universe 
(the boundary conditions act as a moving mirror) 
filled by a quantum massless GUT which is conformally invariant. 
They consider cases in which the static Casimir energy is repulsive 
and attractive. 
Inserting the simplest possible inertial term, they find, 
in the adiabatic (and semiclassical) approximation, 
the dynamical evolution of the scale factor and the dynamical Casimir stress 
analytically and numerically (for SU(2) super Yang-Mills theory). 
They also explore alternative kinetic energy terms.

Fink and Leschke \cite{FL} (2000) note that
in order to relate the probabilistic predictions of quantum theory 
uniquely to measurement results, one has to conceive of an ensemble of
identically prepared copies of the quantum system under study. 
Since the universe is the total domain of physical experience, it cannot be
copied, not even in a thought experiment. 
Therefore, a quantum state of the whole universe 
can never be made accessible to empirical test.
Hence the existence of such a state is only a metaphysical idea. 
Despite prominent claims to the contrary, recent developments in the
quantum-interpretation debate do not invalidate this conclusion. 

Padmanabhan and Choudhury \cite{PC} (2000) note that 
starting from an unknown quantum gravitational model, 
one can invoke a sequence of approximations 
to progressively arrive at quantum field theory (QFT) in curved spacetime,
QFT in flat spacetime, nonrelativistic quantum mechanics 
and newtonian mechanics. 
The more exact theory can put restrictions 
on the range of possibilities allowed for the
approximate theory which are not derivable 
from the latter - an example being the symmetry restrictions 
on the wave function for a pair of electrons. 
They argue that the choice of
vacuum state at low energies could be such a `relic' 
arising from combining the principles of quantum theory 
and general relativity, and demonstrate this result in a simple toy model. 
Their analysis suggests that the wave function of the universe, 
when it describes the large volume limit of the universe, 
dynamically selects a vacuum state for matter fields
- which in turn defines the concept of particle in the low energy limit. 
The result also has the potential for providing 
a concrete quantum mechanical version of Mach's principle. 
\section{Principle of Equivalence}
Misner Thorne and Wheeler \cite{MTW} page 386 (1972) formulate the 
{\it principle of equivalence} as follows
\begin{quote}
In any and every local Lorentz frame,  anywhere and anytime in the Universe,
all (nongravitational) laws of physics must take on their familiar
special-relativistic forms.
\end{quote}

One can take this view as implying that the expectation value of 
non-gravitational fields is the same in every frame.
This can be viewed as a symmetry requirement on the vacuum 
which can require that it produces a cosmological constant,  
compare Sciama's \cite{sciama} argument discussed here \S4.1\P2.

Barut and Haugen \cite{BH} (1972)
produce a theory where mass is also conformally invariant,
which might allow the requirement of a local Loretz frame to be relaxed.

Alvarez and Mann \cite{AM} (1996) 
consider possible tests of the Einstein Equivalence Principle 
for quantum-mechanical vacuum energies by evaluating the Lamb shift 
transition in a class of non-metric
theories of gravity described by the $\tau$ formalism. 
They compute to lowest order the associated red shift 
and time dilation parameters, and discuss how (high-precision)
measurements of these quantities could provide new information 
on the validity of the equivalence principle. 

Bertolami and Carvalho \cite{BertC} (2000)
show in the context of a Lorentz-violating extension of the standard model 
that estimates of Lorentz symmetry violation
extracted from ultra-high energy cosmic rays beyond 
the Greisen-Kuzmin-Zatsepin (GZK) cutoff allow for setting bounds on 
parameters of that extension. 
Furthermore, they argue that a correlated measurement of the difference 
in the arrival time of gamma-ray
photons and neutrinos emitted from active galactic nuclei 
or gamma-ray bursts may provide a signature of possible violation of Lorentz
symmetry. 
They have found that this time delay is energy independent, 
however it has a dependence on the chirality of the particles involved. 
They also briefly discuss the known settings where the mechanism 
for spontaneous violation of Lorentz symmetry in the context of
string/M-theory might take place. 

Leung \cite{leung} (2000) reviews the status of testing the principle of 
equivalence and Lorentz invariance from atmospheric 
and solar neutrino experiments.

Lyre \cite{lyre} (2000) discusses generalizing the principle of equivalence,
as does Roberts \cite{mdrsym} (1989),  who argues that the Higgs mechanism,
see \S2.10,  is not compatible with a general principle of equivalence.

Nikolic \cite{nikolic} (2000) discusses the nature of acceleration,
which by the equivalence principle is related to the gravitational field.
Whether there is a maximal acceleration is reviewed in 
Papini \cite{papini} (1995).
\section{Energy In General Relativity.}
\subsection{Does Vacuum Energy have Gravitational Effects?}
If vacuum energy is non-zero then it will have a gravitational effect via
field equations equating geometry to matter.   The vacuum energy of 
non-gravitational fields will contribute to matter;  
whether there is a contribution to the geometry side via 
``vacuum energy from geometry'' depends upon how this is viewed.
From the point of view of supersymmetric theories there is and the terms cancel
out term by term,  see \S2.6.   At this point it is necessary to address what,
if anything,  is the meaning of the semi-classical equations
\be
G_{ab}=<T_{ab}>
\ee
equation the Einstein tensor to the expected value $<>$ of the stress $T_{ab}$,
c.f. \S3.9.
One could view the lowest value of this expectation as being 
the vacuum value,  then a non-zero vacuum expectation implies 
that the Ricci-flat equations $R_{ab}=0$ never occur in nature,   
compare Roberts \cite{mdr84} (1985).
There is the question of what $<G_{ab}>$ could mean,
and this is part of the subject of quantum gravity.

Sciama's \cite{sciama} view on the relationship 
between zero-point energy and gravitation is now presented.
If an energy $\fr{1}{2}h\nu$ is ascribed 
to each mode of the vacuum radiation field,
then the total energy of the vacuum is infinite.
It would clearly be inconsistent with the original assumption 
of a background Minkowski spacetime to suppose that this energy 
produces gravitation in a manner controlled by Einstein's 
field equations of general relativity.
It is also clear that the spacetime of the real world approximates closely 
to the Minkowski state,  at least on macroscopic scales.
It thus appears that must be regularized the zero-point energy of the 
vacuum by subtracting it out according to some systematic prescription,
and this is one way of looking at QFT's on curved spaces,  see \S3.9,
and for example Baym \cite{baym} (1994) discusses how to locate renormalized 
energy in Robertson-Walker spacetimes.
At the same time,  it would expected that zero-point energy 
{\it differences} gravitate.
For example,  the (negative) Casimir energy between two plane-parallel perfect
conductors would be expected to gravitate;  otherwise,  the relativistic 
relation between measured energy and gravitation would be lost.
Similarly,  the regularized vacuum energy in a curved spacetime 
would be expected to gravitate,  where the regularization is achieved 
by subtracting out the Minkowski contribution in a systematic way.
This procedure is needed in order to obtain a pragmatically workable theory.
The difficulty with it is that existing theory does not tell which is the
fiducial state whose energy is to be set to zero.
Sciama \cite{sciama} claims that: ``It is no doubt an intelligent guess 
that one should take Minkowski spacetime as
this fiducial state,  but the awkward point is that this (or any other) 
choice is not {\it prescribed} by existing theory.  
Clearly,  something essential is missing.''

I think that Sciama is wrong,  at least for QFT's on curved spaces,
the correct procedure is to normalize a stress relative to its
Minkowski value.
Another way of looking at this problem is in terms of the cosmological
constant,  see \S4.1.   

Datta \cite{datta} (1995) discusses 
an interesting development in semiclassical gravity.
Using an improved Born-Oppenheimer approximation,  
the semi-classical reduction of the Wheeler-deWitt equation 
turns out to give important insights into the nature 
and the level of validity of the semi-classical Einstein equations (SCEE).
He shows back reactions from the quantized matter fields in SCEE  
to be completely determined by adiabatically induced U(N) gauge potentials.   
The finite energy from the vacuum polarization,  in particular,  
is found to be intimately related to the 'magnetic' type geometric gauge 
potential.   As a result the vacuum energy in a universe from a 'source-free'
flat simply connected superspace is gauge equivalent to zero,  
leading to some dramatic consequences.
\subsection{Various Approaches to Gravitational Energy.}
Gravitational energy is unusual in that it is usually negative.
This means that gravitational energy can cancel out stresses $T_{ab}$ which
obey good energy conditions.   An example of this is the imploding
scalar-Einstein solution,  Roberts \cite{mdrcounter,mdrimplode} (1989,1996)
where the positive energy of the scalar field and the negative energy of
the gravitational field cancel out.   Such a situation might happen 
for vacuum energy,  the energy of the gravitational field might cancel it out.
To gain insight into this it is necessary to study the energy of the 
gravitational field.

The energy of the gravitational field is unlike that of other fields 
as it is not usually represented by a tensor.   If a non-tensorial
expressions is chosen there exists a coordinate system in which the 
expression can be transferred to zero.   A spacetime might have a preferred 
vector field,  this extra information can allow non-tensorial expressions to 
be given an invariant meaning.   For example for asymptotically flat 
spacetimes there is a preferred vector field normal to the 3-sphere at 
infinity which allows construction of an expression for total energy measured 
at infinity.   In spacetimes which are not asymptotically flat there are 
several ways of approaching gravitational energy.   Here the two-point 
approach is investigated by looking at its form in deSitter spacetime.   
An observer might not be able to measure energy at a single point but
there is the possibility of measuring an energy difference between two points.
Two-point energy expressions have been investigated by Synge and Lanthrop,
see the next paragraph.

Gravitational energy is definable for asymptotically flat spacetime.
For spherically symmetric spacetimes there is an expression 
de Oliveria and Cheb-Terrab \cite{bi:dOCT} (1996)
which reduces to the correct expression for an 
asymptotically flat spacetime and which agrees with quasi-local expressions.
In general there are many,  not necessarily compatible,  ways of defining 
gravitational energy.   Some approaches to gravitational energy in arbitrary 
spacetime are:  
{\it one} to construct pseudo-tensors Goldberg  
\cite{bi:goldberg} (1958),   
{\it two} to construct tensors from the Riemann tensor 
such as the tensors of Bel  \cite{bi:bel} (1962) 
(see also Alder {\it et al} \cite{bi:ALN}(1977) 
Appendix E) ,  and produce 'square root' tensors from these of 
the correct dimensions Bonilla and Senovilla \cite{bi:BS} (1997) and Bergqvist 
\cite{bi:bergqvist} (1998) , 
{\it three} to construct tensors from the Lanczos
tensor Roberts \cite{bi:mdr1} (1988),  these have been used to measure the 
speed of energy transfer in gravitational waves Roberts  
\cite{bi:mdr2} (1994),   
{\it four} to construct quasi-local expressions such as those of
Brown and York \cite{bi:brown} (1993), 
Hayward \cite{bi:hayward} (1995) gives examples of these,
{\it five} general relativity
can be re-written as a tele-parallel theory and in this theory there is a 
tensorial expression for gravitational energy Maluf \cite{bi:maluf1} (1995),
{\it six} to construct Synge \cite{bi:synge1,bi:synge2} (1960/2)-Lathrop 
\cite{bi:lathrop} (1975) two-point expressions.
Other two-point approaches are those of Droz-Vincent \cite{bi:DV} (1996) 
who compares the energy of an exploding spacetime to that at the origin 
of the coordinates,  and of Isaacson \cite{bi:isaacson} (1968)
who compares the energy of gravitational waves at two separate times.
Babak and Grishchuk \cite{BG} (2000),  also study
the energy-momentum tensor for the gravitational field.
One can also ask if there is a maximum gravitational energy,
which is related,  by the equivalence principle,  see \S5,
to the question of whether there is a maximum acceleration,  
see the review of Papini \cite{papini} (1995).
For the relation of acceleration to quantum localization,
see Jaekel and Reynaud \cite{JR} (1999).

Brown {\it et al} \cite{BLY} (1998)
consider the definition $E$ of quasilocal energy stemming 
from the Hamilton-Jacobi method as applied to the canonical form 
of the gravitational action. 
They examine $E$ in the standard "small-sphere limit," 
first considered by Horowitz and Schmidt in their examination of 
Hawking's quasilocal mass. 
By the term "small sphere" they mean a cut $S(r)$, 
level in an affine radius $r$, of the lightcone belonging 
to a generic spacetime point. 
As a power series in $r$, they compute the energy $E$ 
of the gravitational and matter fields on a spacelike
hypersurface spanning $S(r)$. 
Much of Their analysis concerns conceptual 
and technical issues associated with assigning the zero-point of the energy. 
For the small-sphere limit, they argue that the correct zero-point is obtained 
via a "lightcone reference," which stems from a certain isometric embedding 
of $S(r)$ into a genuine lightcone of Minkowski spacetime. 
Choosing this zero-point, we find agreement with Hawking's quasilocal mass 
expression, up to and including the first non-trivial order in 
the affine radius. 
The vacuum limit relates the quasilocal energy directly 
to the Bel-Robinson tensor. 
They present a detailed examination of the variational principle 
for metric general relativity as applied to a ``quasilocal'' spacetime region
$M$ (that is, a region that is both spatially and temporally bounded). 
Our analysis relies on the Hamiltonian formulation of general
relativity, and thereby assumes a foliation of $M$ 
into spacelike hypersurfaces $\Sigma$. 
They allow for near complete generality in the
choice of foliation. 
Using a field-theoretic generalization of Hamilton-Jacobi theory, 
they define the quasilocal stress-energy-momentum of
the gravitational field by varying the action with respect to 
the metric on the boundary $\partial M$. The gravitational
stress-energy-momentum is defined for a two-surface B spanned 
by a spacelike hypersurface in spacetime. They examine the behavior of
the gravitational stress-energy-momentum under boosts 
of the spanning hypersurface. The boost relations are derived from the
geometrical and invariance properties of the gravitational action 
and Hamiltonian. 

Johri {\it et al} (1995) \cite{JKSE} examines
the role of gravity in the evolution of the universe is examined.
They claims that in co-moving coordinates,  calculation of the Landau-Lifshitz
pseudotensor for FRW models reveals that:\\
i) the total energy of a spatially closed universe irrespective 
of the equation of state of the cosmic fluid is zero at all times.\\
ii)the total energy enclosed within any finite volume of the 
spatially flat universe is zero at all times,\\
iii)during inflation the vacuum energy driving the accelerated expansion 
and ultimately responsible for the creation of matter (radiation) 
in the universe,  is drawn from the energy of the gravitational field.
In a similar fashion, certain cosmological models which abandon
adiabaticity by allowing for particle creation,  use gravitational
energy directly as an energy source.

Tung and Nester \cite{TN} (2000)
note that in the tetrad representation of general relativity, 
the energy-momentum expression, found by M\"oller in 1961, 
is a tensor with respect to coordinate transformations but is not a tensor 
with respect to local Lorentz frame rotations. 
This local Lorentz freedom is shown to be the same as the six
parameter normalized spinor degrees of freedom in the quadratic spinor 
representation of general relativity. 
From the viewpoint of a gravitational field theory in flat spacetime, 
these extra spinor degrees of freedom allow them to obtain a local 
energy-momentum density which is a true tensor over both coordinate 
and local Lorentz frame rotations. 
\section{Inertia.}
\subsection{Inertia in General.}
In non-quantum physics inertia would normally be approached 
through the principle of equivalence, see \S5.   
What an inertial vacuum is is discussed 
in Padmanabhan and Choudhury \cite{PC} (2000). 
Ciufolini and Wheeler \cite{CWhe} (1995) also discuss gravitation and inertia.
Meissner and Veneziano \cite{MV} (1991) discuss what can be thought of as 
trying to find which frames are inertial.

Bozhkov and Rodrigues (1995) \cite{BR} 
consider the Denisov-Solov'ov (1983) \cite{DSol}
example which some claims shows that the inertial mass is not well defined 
in general relativity.   They show that the mathematical reason 
why this is true is a wrong application of Stokes theorem.   
Then they discuss the role of the order of asymptotically flatness 
in the definition of mass.   In conclusion they present some comment on the 
conservation laws in general relativity.

Farup and Gr{\o}n \cite{FG} (1996) 
investigate if there is any inertial dragging effect associated 
with vacuum energy.
Spacetime inside and outside a rotating thin shell,  
as well as the mechanical properties of the shell,  
are analyzed by means of Israel's general relativistic 
theory of surface layers.
Their investigations generalizes that of Brill and Cohen \cite{BC} (1966)
who found vacuum solutions of Einstein's field equations 
(with vanishing cosmological constant),  inside and outside a rotating shell.
They include a non-vanishing vacuum energy inside the shell.   
They find that the inertial dragging angular velocity 
increases with increasing density of the vacuum.
da Paola and Svaiter \cite{dPS} (2000) discuss how rotating vacuums 
have inertia and their relation to Mach's principle.

Directional preferences and zero-point energy are discussed in Winterberg
\cite{winterberg} (1998),  a directional preference suggests 
a non-inertial frame.

Herrera and Mart\'inez \cite{HM} (1998) establish for slowly rotating fluids, 
the existence of a critical point
similar to the one found for nonrotating systems.   As the fluid approaches the
critical point,  the effective inertial mass of any fluid element decreases,
vanishing at that point and changing sign beyond it.   This result implies 
that the first order perturbative method is not always reliable to study
dissipative processes occurring before relaxation.   
They comment upon physical consequences
that might follow from this effect.
They use a nonstatic axisymmetric line element.

Milgrom \cite{milgrom} (1998)
has suggested in several papers that dynamics is modified to include
an acceleration constant so as to explain dynamics on galactic scales
and larger.   In this paper he suggests that the vacuum gives rise to
the acceleration constant and is also responsible for inertia.
He says that either cosmology enters local dynamics by affecting the vacuum, 
and inertia in turn, through the constant of acceleration $a_0$; 
or the same vacuum effect enters both the modified dynamics (MOND) 
through $a_0$ and cosmology (e.g. through a cosmological constant). 
He goes on to say for the vacuum to serve as substratum for inertia 
a body must be able to read in it its non-inertial motion; 
this indeed it can, by detecting Unruh-type radiation. 
A manifestation of the vacuum is also seen, even by inertial observers, 
in a non-trivial universe (marked, e.g., by curvature or expansion). 
A non-inertial observer in a nontrivial universe will see the combined effect.
An observer on a constant-acceleration ($a$) trajectory in a de Sitter 
universe with cosmological constant $\Lambda$ sees
Unruh radiation of temperature $T\propto [a^2+a_0^2]^{1/2}$, 
with $a_0=(\Lambda/3)^{1/2}$. 
The temperature excess over what an inertial observer sees, 
$T(a)-T(0)$, turns out to depend on a in the same way that MOND inertia does. 
An actual inertia-from-vacuum mechanism is still a far cry off. 

Capozziello and Lambiase \cite{CL} (1999) calculate
the inertial effects on neutrino oscillations induced by the acceleration 
and angular velocity of a reference frame. 
Such effects have been analyzed in the
framework of the solar and atmospheric neutrino problem. 

De Paola and Svaiter \cite{dPS} (2000) consider
a quantum analog of Newton's bucket experiment in a flat spacetime: 
they take an Unruh-DeWitt detector in
interaction with a real massless scalar field.
They calculate the detector's excitation rate when 
it is uniformly rotating around some fixed
point and the field is prepared in the Minkowski vacuum 
and also when the detector is inertial 
and the field is in the Trocheries-Takeno vacuum state. 
They compare these results and discuss the relations with a quantum analog 
of Mach's principle. 

Haisch {\it et al} \cite{HRP} (1994),
Haisch and Rueda \cite{HR} (1999) and 
Haisch {\it et al} \cite{HRD} (2000) note that
even when the Higgs particle is finally detected, 
it will continue to be a legitimate question to ask 
whether the inertia of matter as a
reaction force opposing acceleration 
is an intrinsic or extrinsic property of matter. 
General relativity specifies which geodesic path a free
particle will follow, but geometrodynamics has no mechanism 
for generating a reaction force for deviation from geodesic motion. 
They discuss a different approach involving the electromagnetic 
zero-point field (ZPF) of the quantum vacuum. It has been found that certain
asymmetries arise in the ZPF as perceived from an accelerating 
reference frame. In such a frame the Poynting vector and momentum
flux of the ZPF become non-zero. 
Scattering of this quantum radiation by the quarks and electrons 
in matter can result in an acceleration-dependent reaction force. 
Both the ordinary and the relativistic forms of Newton's second law, 
the equation of motion, can
be derived from the electrodynamics of such ZPF-particle interactions. 
They give conjectural arguments are given why this interaction should
take place in a resonance at the Compton frequency, 
and how this could simultaneously provide a physical basis for the de Broglie
wavelength of a moving particle. This affords a suggestive perspective 
on a deep connection between electrodynamics, the origin of inertia
and the quantum wave nature of matter. 

Modanese \cite{mod} (2000) recalls different ways to define inertial mass 
of elementary particles in modern physics, 
he then studies the relationship between the
mass of charged particles and zero-point electromagnetic fields. 
To this end he first introduce a simple model comprising a scalar field
immersed in stochastic or thermal electromagnetic fields. 
Then he sketches the main steps of Feynman mass renormalization procedure.
His approach is essentially pedagogical and in line with the standard 
formalism of quantum field theory, but he also tries to keep an open
mind concerning the physical interpretation. 
He checks, for instance, if it is possible to start from 
a zero bare mass in the renormalization
process and express the finite physical mass in terms of a cut-off. 
Finally he briefly recall the Casimir-induced mass modification of
conducting or dielectric bodies. 

Nikolic \cite{nikolic} (2000) studies the role of acceleration 
in the twin paradox. 
From the coordinate transformation that relates an accelerated and an inertial
observer he finds that, from the point of view of the accelerated observer, 
the rate of the differential lapses of time depends not only on the
relative velocity, but also on the product of the acceleration 
and the distance between the observers. However, this result does not have a
direct operational interpretation because an observer at a certain position 
can measure only physical quantities that are defined at the same
position. For local measurements, the asymmetry between the two observers 
can be attributed to the fact that noninertial coordinate
systems, contrary to inertial coordinate systems, 
can be correctly interpreted only locally. 

Rosu \cite{haretcr} (2000) introduces in a nonrigorous manner, 
the write-up of the talk extends and updates several sections of the  
review {\tt gr-qc/9406012},   last updated in 1997. 
The recently introduced glass analogy for black holes is presented,
but in order to have a more detailed picture there is a collection
from the literature some useful material related to the violations 
of the fluctuation-dissipation theorem in glass physics and stationary
driven systems. 
Nonequilibrium effective temperatures from glass irreversible 
thermodynamics are considered as useful and quite
general concepts for noninertial quantum fluctuations, 
though the analogy is not fully disentangled. 
Next, the stationary and nonstationary scalar vacuum noises is discussed 
in some detail and the radiometric nature of the Frenet invariants 
of the stationary worldlines is emphaised, 
rather than sticking to the thermal interpretation 
of the vacuum excitations as apparent for the uniformly accelerated 
quantum detector. 
The Hacyan-Sarmiento approach for calculating electromagnetic vacuum physical 
quantities is discussed next.
The application to the circular case led Mane to propose 
the identification of the Hacyan-Sarmiento zero-point radiation 
with the ordinary synchrotron radiation, but the issue remains still open. 
The spin flip synchrotron radiation in the context of Bell and
Leinaas proposal is breifly discussed.
Jackson's approach showing why this proposal is implausible is included. 
Finally, there is a short random walk in
the literature on the fluctuation-dissipation relationship. 

Salehi {\it et al} (2000) study a model for analyzing the effect 
of a principal violation of the Lorentz-invariance on the structure of vacuum.
The model is based on the divergence theory developed by Salehi (1997). 
They show that the divergence theory can be used to model an ensemble of 
particles. The ensemble is characterized by the condition that 
its members are basically at rest in the rest frame of a preferred inertial
observer in vacuum. In this way we find a direct dynamical interplay between 
a particle and its associated ensemble. They show that this
effect can be understood in terms of the interaction of a particle 
with a relativistic pilot wave through an associated quantum potential. 
\subsection{Superfluid Analogy.}
As with quantum field theory,  inertia also has superfluid analogies.

Duan \cite{duan} (1993) discusses the inertial mass of moving singularities.
Duan and \u{S}im\'anek \cite{DS} (1994)extended to finite temperature 
the theory of the inertial mass of a fluxon in the type II superconductors,  
due to the coupling between the fluxon and the lattice deformation.
They propose an ansatz for the quasi-particle fraction valid at all 
temperatures below $T_c$ and solve the associated strain field.   
For high $T_c$ supercondutors the mass is $10^5$ electron mass/cm at low 
temperature (or at least the same order of magnitude as the 
electromagnetic inertial mass)
and vanishes at $T_c$,  this resolving an outstanding problem of the 
previous theory.

Mel'nikov \cite{melnikov} (1996) considers
the dynamics of titled vortex lines in Josephson coupled layered 
superconductors in considered within the time dependent Ginzburg-Landau
theory.   The frequency and angular dependences of the complex valued vortex 
mobility $\mu$ are studied.   The components of the viscosity and 
inertial mass tensors are found to increase essentially for 
magnetic field orientations close to the layers.   
For superconducting/normal metal multilayers the frequency ($\om$) 
range is shown to exist where the $\mu^{-1}$) value depends 
logarithmically on $\om$.

Gaitonde and Ramakrishnan \cite{GRam} (1997) calculate the inertial mass 
of a moving vortex in cuprate superconductors.
\section{Relativity of Motion in Vacuum.}
Nogueira and  Maia \cite{NM2} (1995) show that for a 
self-interacting mass the scalar field in the geometry of Casimir plates 
and in $N=m+1$ spacetime dimensions the renormalized ZPA vanishes.  
So,  it is undetectable via Casimir forces.

Guendelman and Rabinowitz \cite{GRab} (1996) say...???

Nogueira and Maia \cite{NM} (1996) 
investigate a possible difference between the effective potential 
and zero-point energy. 
They define the zero-point ambiguity (ZPA) as the difference between these two
definitions of vacuum energy. 
Using the $\zeta$-function technique, 
in order to obtain renormalized quantities, 
they show that ZPA vanishes, 
implying that both of the above definitions
of vacuum energy coincide for a large class of geometries 
and a very general potential. 
In addition, they show explicitly that an extra term, 
obtained by E. Myers some years ago for
the ZPA, disappears when a scale parameter $\mu$ 
is consistently introduced in all $\zeta$-functions 
in order to keep them dimensionless.

Kardar and Golestanian \cite{KG} (1997) note that
the static Casimir effect describes an attractive force between two 
conducting plates, due to quantum fluctuations of the electromagnetic 
(EM) field in the intervening space. {\it Thermal fluctuations} 
of correlated fluids (such as critical mixtures, super-fluids, 
liquid crystals, or electrolytes) are also modified by the boundaries, 
resulting in finite-size corrections at criticality, 
and additional forces that effect wetting and layering phenomena. 
Modified fluctuations of the EM field can also account for the `van der Waals'
interaction between conducting spheres, and have analogs in the 
fluctuation--induced interactions between inclusions on a membrane. 
They employ a path integral formalism to study these phenomena for 
boundaries of arbitrary shape. 
This allows them to examine the many unexpected phenomena 
of the dynamic Casimir effect due to moving boundaries.
With the inclusion of quantum fluctuations, 
the EM vacuum behaves essentially as a complex fluid, 
and modifies the motion of objects through it. 
In particular, from the mechanical response function of the EM vacuum, 
they extract a plethora of interesting results, the most notable being: 
(i) The effective mass of a plate depends on its shape, and
becomes anisotropic, 
(ii) There is dissipation and damping of the motion, 
again dependent upon shape and direction of motion, due to emission of photons,
(iii) There is a continuous spectrum of resonant cavity modes that can be 
excited by the motion of the (neutral) boundaries.
Kardar and Golestanian \cite{KG2} (1999) discuss whether the vacuum has a
'friction'.

Jaekel \cite{jaekel} (1998) reconsider 
the question of relativity of motion because of
the existence of vacuum fluctuations.
His article is devoted to this aim with a main line 
which can be formulated as follows: 
``The principle of relativity of motion is directly related 
to symmetries of quantum vacuum''. 
Keeping close to this statement, he discusses the controversial relation
between vacuum and motion. 
He introduces the question of relativity of motion in its 
historical development before coming to the results obtained more recently.

Jaekel and Reynaud \cite{JR} (1998) 
define quantum observables associated with Einstein localization in 
spacetime. These observables are built on Poincare and dilatation generators.
Their commutators are
given by spin observables defined from the same symmetry generators. 
Their shifts under transformations to uniformly accelerated frames 
are evaluated through algebraic
computations in conformal algebra. 
Spin number is found to vary under such transformations 
with a variation involving further observables introduced 
as irreducible quadrupole momenta. 
Quadrupole observables may be dealt with as non commutative polarisations 
which allow one to define step operators increasing 
or decreasing the spin number by unity. 
\section{The Vision Thing.}
The solution suggested here to the nature of the vacuum
is that Casimir energy can produce short range effects because of boundary
conditions,  
but that at long range there is no overall effect of vacuum energy,
unless one considers lagrangians of higher order than Einstein's as vacuum 
induced.   
That such higher order lagrangians describe nature is likely 
because they occur as an approximation to most quantum gravity theories,  
and the Bach lagrangian part of higher order theory might explain various long
range effects, see Roberts \cite{galr} (1991).
Different reductions of quantum gravity theories produce different ratios
of the coupling constants of the resulting higher order lagrangian theories.
A constant ``cosmological constant'' does not exist.
A non-constant ``cosmological constant''  which is really a type of
perfect fluid might not be zero.
No original calculations are presented in support of this position.
\section{Acknowledgement.}
I would like to thank George Ellis and Ulrich Kirchner 
for reading some of the manuscript,
and Edward Witten for mentioning that different ratios of coupling constants
result from different reduction schemes.
The following people commented on the first version of the paper:
Orfeu Bertolami, 
Emilio Elizalde,
Remo Garattini,
Eduardo I.Gundelman,
Alejandro Jakubi,
Klaus Kirsten,
José Ademir Sales de Lima,
Alessandro Melchiorri,
Mordehai Milgrom,
Serguei Odintsov,
Roh S.Tung,
She-sheng Xue,
Sergio Zerbini.
The following people commented on the second version of the paper:
Boris Kastening,
Boris P.Kosyakov.
This work was supported by the Californian Institute of Physics and 
Astrophysics.
\section{Referencing Style.}
The almost alphabetical order of the first references is maintained,
the reason being so as not to necessitate rewriting my author index:\newline
http://cosmology.mth.uct.ac.za/~roberts/4hcv/4hurl/home.html.\newline
References 187,  272,  310,  345,  \& 352 are out of order,
16 small corrections are made to first version references.
References for the second version go at the end.

\end{document}